\documentclass[twocolumn]{aastex62}

\turnoffedit

\usepackage{amsmath}

\graphicspath{{./}{figures/}}

\newcommand{\msun}{\mathrm{M_{\odot}}} 
\newcommand{\msunyr}{\msun\,{\rm yr}^{-1}} 

\newcommand{\mhe}{M_{\mathrm{He}}} 
\newcommand{\mwd}{M_{\mathrm{WD}}} 
\newcommand{\mch}{M_{\mathrm{Ch}}} 
\newcommand{\mheenv}{M_{\mathrm{He,env}}}

\newcommand{\peryr}{\mathrm{yr^{-1}}} 
\newcommand{\mdotup}{\dot{M}_{\mathrm{up}}} 
\newcommand{\mdotlow}{\dot{M}_{\mathrm{low}}} 
\newcommand{\mdothe}{\dot{M}_{\mathrm{He}}} 
\newcommand{\mdotwd}{\dot{M}_{\mathrm{WD}}} 
\newcommand{\mdotedd}{\dot{M}_{\mathrm{Edd}}} 

\newcommand{\lsun}{\mathrm{L_{\odot}}} 
\newcommand{\Lhe}{L_{\mathrm{He}}} 
\newcommand{\Lwd}{L_{\mathrm{WD}}} 
\newcommand{\Ledd}{L_{\mathrm{Edd}}}

\newcommand{\rsun}{\mathrm{R_{\odot}}} 
\newcommand{\Rwd}{R_{\mathrm{WD}}} 
\newcommand{\Rcwd}{R_{\mathrm{cWD}}} 
\newcommand{\Rcirc}{R_{\mathrm{circ}}} 

\newcommand{\inipara}{(M^{i}_{\mathrm{He}}, M^{i}_{\mathrm{WD}}, \log P^{i}_{\mathrm{d}} )}
\newcommand{\iniMhe}{M^{i}_{\mathrm{He}}}
\newcommand{\iniMwd}{M^{i}_{\mathrm{WD}}}
\newcommand{\iniP}{\log P^{i}_{\mathrm{d}}}

\newcommand{\finalMhe}{M^{f}_{\mathrm{He}}}

\newcommand{\mdotw}{\dot{M}_{\mathrm{w}}}
\newcommand{\jdotw}{\dot{J}_{\mathrm{w}}}
\newcommand{\lw}{l_{\mathrm{w}}}
\newcommand{\binaryomega}{\Omega_{\mathrm{orb}}}
\newcommand{\vrl}{v_{\mathrm{RL}}}
\newcommand{\kms}{\mathrm{km} \text{ } \mathrm{s^{-1}}}
\newcommand{\vw}{v_{\mathrm{w}}}
\newcommand{\vesc}{v_{\mathrm{esc}}}

\newcommand{\betacurrent}{\beta_{\mathrm{current}}}
\newcommand{\betaexplicit}{\beta_{\mathrm{explicit}}}
\newcommand{\fbeta}{f_{\beta}}

\newcommand{\betahi}{\beta_{\mathrm{hi}}}
\newcommand{\betalo}{\beta_{\mathrm{lo}}}
\newcommand{\fbetahi}{f_{\beta}^{\mathrm{hi}}}
\newcommand{\fbetalo}{f_{\beta}^{\mathrm{lo}}}

\newcommand{\mesa}{{\tt\string MESA}}
\newcommand{\mltpp}{{\tt\string MLT++}}

\newcommand{\Rrl}{R_{\mathrm{RL}}}
\newcommand{\mdotcr}{\dot{M}_{\mathrm{cr}}}

\newcommand{\kB}{\ensuremath{k_\mathrm{B}}} 
\newcommand{\crad}{\ensuremath{a}} 
\newcommand{\mb}{\ensuremath{m_\mathrm{u}}} 
\newcommand{\Pgas}{\ensuremath{P_\mathrm{gas}}}
\newcommand{\Prad}{\ensuremath{P_\mathrm{rad}}}

\newcommand{\dif}{\ensuremath{d}}
\newcommand{\ddr}[1]{\frac{\dif #1}{\dif r}}
\newcommand{\dlndlnr}[1]{\frac{\dif \ln #1}{\dif \ln r}}

\newcommand{\epsHe}{\ensuremath{\epsilon_{\mathrm{He}}}}
\newcommand{\epsacc}{\ensuremath{\epsilon_{\mathrm{acc}}}}

\newcommand{\Tcr}{\ensuremath{T_{\rm cr}}}
\newcommand{\vcr}{\ensuremath{v_{\rm cr}}}
\newcommand{\rhocr}{\ensuremath{\rho_{\rm cr}}}
\newcommand{\rcr}{\ensuremath{r_{\rm cr}}}

\newcommand{\vph}{\ensuremath{v_{\rm ph}}}

\begin{document}

\title{Evolution of Helium Star - White Dwarf Binaries Leading up to Thermonuclear Supernovae}

\author[0000-0001-9195-7390]{Tin Long Sunny Wong}
\affiliation{Department of Astronomy and Astrophysics, University of California, Santa Cruz, CA 95064, USA}

\author[0000-0002-4870-8855]{Josiah Schwab}
\altaffiliation{Hubble Fellow}
\affiliation{Department of Astronomy and Astrophysics, University of California, Santa Cruz, CA 95064, USA}

\correspondingauthor{Tin Long Sunny Wong}
\email{twong31@ucsc.edu}

\begin{abstract}
We perform binary evolution calculations on helium star - carbon-oxygen white dwarf (CO WD) binaries using the stellar evolution code \mesa. This single degenerate channel may contribute significantly to thermonuclear supernovae at short delay times. We examine the thermal-timescale mass transfer from a 1.1 - 2.0 $\msun$ helium star to a 0.90 - 1.05 $\msun$ CO WD for initial orbital periods in the range 0.05 - 1 day.  Systems in this range may produce a thermonuclear supernova, helium novae, a helium star - oxygen-neon WD binary, or a detached double CO WD binary. Our time-dependent calculations that resolve the stellar structures of both binary components allow accurate distinction between the eventual formation of a thermonuclear supernova (via central ignition of carbon burning) and that of an ONe WD (in the case of off-center ignition). Furthermore, we investigate the effect of a slow WD wind which implies a specific angular momentum loss from the binary that is larger than typically assumed. We find that this does not significantly alter the region of parameter space over which systems evolve toward thermonuclear supernovae. Our determination of the correspondence between initial binary parameters and the final outcome informs population synthesis studies of the contribution of the helium donor channel to thermonuclear supernovae. In addition, we constrain the orbital properties and observable stellar properties of the progenitor binaries of thermonuclear supernovae and helium novae.
\end{abstract} 


\section{Introduction}
\label{sec:intro}

Type Ia supernovae (SNe Ia) are believed to originate from thermonuclear explosions of white dwarfs (WDs; e.g., \citealt{1960ApJ...132..565H}). As factories of  iron group elements, SNe Ia are important in understanding the chemical evolution of galaxies (e.g., \citealt{1983A&A...118..217G}, \citealt{1986A&A...154..279M}). Moreover, as standardizable candles, SNe Ia have played a crucial role in the discovery of an accelerating universe (\citealt{1998AJ....116.1009R}, \citealt{1999ApJ...517..565P}). Despite the important roles played by SNe Ia, there is still debate about the progenitor systems of SNe Ia \citep[for recent reviews, see e.g.,][]{2014ARA&A..52..107M, 2018PhR...736....1L}.

A scenario in which a carbon-oxygen WD (CO WD) grows up to the Chandrasekhar mass ($\mch$) by accreting mass from a non-degenerate helium star (He star) companion  may contribute to thermonuclear supernovae (TN SNe).\footnote{The precise explosion mechanism of a CO WD that reaches $\mch$
has not been definitely resolved \citep[see e.g.,][]{2017hsn..book.1275N}.
Thus we cannot unambiguously link the explosive end product of a $\mch$
WD to any particular class or subclass of observed thermonuclear events.  In particular, our stellar
evolution models cannot distinguish between SN Ia and SN Iax and so we conflate the likely end products of the helium donor channel into TN SNe.} 
This channel, which we will hereafter refer to as the helium donor channel\footnote{This is to be distinguished from
the scenario involving lower mass helium donors
in which lower accretion rates lead to the accumulation of
an unburned He shell that subsequently detonates
and leads to the explosion of a sub-Chandrasekhar mass WD
\citep[e.g.,][]{1991ApJ...370..615I,1994ApJ...423..371W}.},
has several theoretically attractive properties.
First, eliminating the thermal instabilities associated with simultaneous hydrogen and helium shell burning in the classical single-degenerate channel, the helium donor channel can offer a more efficient pathway to grow the WD up to $\mch$ (e.g., \citealt{1994ApJ...431..264I,2003A&A...412L..53Y}). Second, population synthesis studies have shown that this channel can dominate the formation of TN SNe with short delay times (e.g., \citealt{2009ApJ...699.2026R}, \citealt{2009MNRAS.395..847W}, \citealt{2014A&A...563A..83C}). Not only is the helium donor channel a favorable formation channel for short-delay time Type Ia supernovae (SNe Ia), but it may also explain the preference of the subclass Type Iax supernovae (SNe Iax), which have low ejecta velocities and lower peak luminosities, to late-type galaxies \citep{2013ApJ...767...57F}. 
Several lines of observational evidence have led to the helium donor channel being
the currently favored scenario for SNe Iax \citep[e.g.,][]{2017hsn..book..375J}\footnote{\added{Alternatively, it has been suggested that non-degenerate helium-donor systems exploding through double detonations (see Footnote 2) may also explain the rates, delay time distributions, and luminosity distribution of SNe Iax \citep[e.g.,][]{2013A&A...559A..94W}.  However, it is not clear that such explosions resemble SNe Iax in detail \citep[e.g.,][]{2011ApJ...734...38W,2019ApJ...873...84P}}}.
In particular, \citet{2014Natur.512...54M} have suggested that the blue point source found in the pre-explosion image of the SN Iax 2012Z is consistent with a non-degenerate He star of $\approx 2$ $\msun$.  In addition, helium is found in the spectra of two SNe Iax 2004cs and 2007J \citep{2013ApJ...767...57F}.\footnote{\citet{2015ApJ...799...52W} identified these events as Type IIb SNe, but see \citet{2016MNRAS.461..433F} for counterarguements.}
The discovery of the first and to date only helium nova, V445 Puppis \citep{2003A&A...409.1007A},
may also fit into this picture.
Subsequent light curve analysis suggests it is consistent with a massive $\gtrsim 1.35\,\msun$ WD retaining half of the accreted mass during the nova event \citep{2008ApJ...684.1366K}.
This hint of WD growth and efficient mass retention may 
indicate that the helium donor channel can indeed produce plausible TN SN candidates. 

\cite{1994ApJ...431..264I} first proposed that massive He stars can donate helium to a massive WD companion at a rate of $\sim 10^{-6}$ to $10^{-5}$ $\msun$ $\peryr$, which allows the helium to burn steadily on the WD surface and thus enables the WD to grow smoothly to $\mch$. This was followed up by \cite{2003A&A...412L..53Y}, who performed binary evolution calculations on a $1.6$ $\msun$ He star and a $1.0$ $\msun$ CO WD in a $0.124$ day orbit. Their calculations confirm that such a system allows thermally-stable accretion of helium onto the WD, and is an efficient channel to grow the WD to $\mch$. \cite{2009MNRAS.395..847W} then found the region in initial binary parameter space leading up to a TN SN by performing a series of binary evolution calculations.
We share the goal of identifying which binaries (in terms of the initial component masses and period) are progenitors of TN SNe and will refer to this part of parameter space as the ``TN SN region''.

\cite{2016ApJ...821...28B} pointed out the significance of fully solving the stellar structure of the WD instead of making the common point-mass treatment.  They calculated the binary evolution of a $1$ $\msun$ WD in a $3$ hour orbit with He stars of masses ranging from $1.3$ $\msun$ to $1.8$ $\msun$, and find that, for sufficiently high accretion rates, an off-center carbon ignition is initiated. Instead of a TN SN, this leads to formation of an oxygen-neon (ONe) WD; the ONe WD may subsequently undergo an accretion-induced collapse (AIC) and form a neutron star upon reaching $\mch$ \citep{2017ApJ...843..151B}. \added{\cite{2017MNRAS.472.1593W} recently reviewed} their previous parameter space calculations in \cite{2009MNRAS.395..847W}. They determine the critical mass transfer rate near $\mch$ that would lead to an off-center carbon ignition, and use this as a criterion for determining which of their previous models are off-center ignitions. They find a reduction in the TN SN region leading to a reduction of their estimated Galactic SN Ia rate through this channel from $\approx 0.3 \times 10^{-3}$ yr$^{-1}$ to $\approx 0.2 \times 10^{-3}$ yr$^{-1}$.
These rates are roughly consistent with inferred SN Iax rates \citep{2013ApJ...767...57F,2017ApJ...848...59M}.

Given its promise, the helium donor channel requires further investigations. Firstly, \cite{2017MNRAS.472.1593W} have adopted a single criterion (i.e., the mass transfer rate when the WD is near $\mch$) in detecting off-center ignitions. It is of interest to see the results of time-dependent calculations that resolve the full stellar structures of both binary components, as has been suggested by \cite{2016ApJ...821...28B}. Secondly, previous calculations have usually assumed 
that any material lost from the binary system takes the form of
a fast wind launched from the WD (i.e., that the wind velocity
is significantly above the orbital velocity and so the material carries the specific orbital angular momentum of the WD).  \cite{2016ApJ...821...28B} point out that the fast wind assumption may not always prevail.  A slow wind may gravitationally torque the binary and extract additional angular momentum, affecting the subsequent mass transfer. Therefore, the effect of angular momentum loss from the wind on the TN SN region requires further study.

This paper is organized as follows. In Section \ref{sec:accretion regimes}, we give an overview of the helium donor scenario and our basic modeling assumptions.
In Section \ref{sec:modeling and methodology}, we describe the important stellar and binary evolution controls in $\mesa$, the stopping conditions for our binary setup, and the choices of the initial binary parameters. In Section \ref{sec:fast-wind}, we show the results of grids of binary models---distributed over initial He star mass, WD mass, and binary orbital period---adopting the assumption of a fast wind. We compare with previous works in Section \ref{sec:comparison with previous works}. We relax the fast-wind assumption in Section \ref{sec:effect of enhanced angular momentum loss} and show that the TN SN parameter space does not show significant changes with enhanced angular momentum loss. In Section \ref{sec:properties of OTW} we describe the properties of the optically-thick winds we invoke in our binary models. In Section \ref{sec:discussion}, we discuss uncertainties including the effects of rotation and the accretion picture, describe the origin of the He star - CO WD systems,  
and outline the observational constraints derived from our models. We conclude in Section \ref{sec: conclusion}. 


\section{The Helium Donor Channel}
\label{sec:accretion regimes}

Our models of the helium donor channel begin with a detached He star - WD binary.\footnote{We will describe how these binaries form in Section~\ref{subsec:formation channel}.}
As the He star evolves, it eventually overfills its Roche lobe
and starts to donate mass onto the WD.
We indicate the rate at which helium is donated to the WD by
its companion He star as $|\mdothe|$.  
The WD grows at the rate the helium is donated only
when it can burn the helium at the same rate in a thermally-stable manner%
\added{\citep[e.g.,][]{1982ApJ...253..798N, 2014MNRAS.445.3239P,2018RAA....18...49W}}.
The assumptions about what happens outside of the narrow range of rates where this is possible are important in determining whether the WD can reach $\mch$ and thus in determining the ultimate fate of the binary.
In this section we discuss the different regimes in which accretion can occur and 
describe how our models answer the critical question of how much of the
transferred He is retained on the WD.

\subsection{The Red Giant Regime and $\mdotup$}
\label{subsec:red giant regime}

Above the maximum stable accretion rate (hereafter the upper stability line $\mdotup$), the WD cannot burn helium as fast as it is accreted.
This occurs because there exists a maximum luminosity for a shell-burning star. The core-mass luminosity relation \citep{1970AcA....20...47P} says that the luminosity of a shell-burning star is primarily dependent on the core mass. This can be understood in the context of hydrostatic equilibrium -- in shell burning stars, the pressure due to the envelope is negligible, and the core mass is dominant in setting up the condition for hydrostatic equilibrium \citep{2012sse..book.....K}. Since nuclear burning depends sharply on the temperature, the luminosity, which largely derives from nuclear burning, is then related to the core mass through hydrostatic equilibrium. For accreting WDs, however, the luminosity derives not only from nuclear burning of the accreted material, but also from the gravitational potential energy released when the accreted material settles from the surface to the base of the envelope. As both the nuclear burning rate and the ``accretion luminosity'' depend on the accretion rate, this gives rise to a maximum stable accretion rate dependent on the core mass \citep{2007ApJ...660.1444S}.
The calculations by \cite{1982ApJ...253..798N} show that the upper stability line for helium accretion is
\begin{equation}
\mdotup = 7.2\times10^{-6} \left(\frac{\mwd}{\msun} - 0.60 \right) \text{ $\msun$ $\peryr$}, 
\end{equation}
which is valid for CO WDs of mass $0.75$ $\msun \leqslant \mwd \leqslant 1.38$ $\msun$. The value of $\mdotup$ scales positively with $\mwd$, since the equilibrium temperature at the burning shell increases with the core mass and allows for nuclear burning at a higher rate.

For $|\mdothe| > \mdotup$ the WD is not able to burn material as fast it is donated, so material piles up in the envelope, inflating it to red giant dimensions.
Typically, a mass loss prescription that allows the WD to dispose of the excessive mass and circumvent the formation of a common envelope is invoked \citep[e.g.,][]{2003A&A...412L..53Y,2009MNRAS.395..847W,2015A&A...584A..37W}.
Physically, this may correspond to the suggestion by \cite{1996ApJ...470L..97H} that an optically-thick wind can result\footnote{Their calculations were applied to hydrogen accretors, but an analogy can be and has been made to helium accretors.} (called the ``accretion wind'').
As the WD expands, its envelope cools and gradually becomes radiation-dominated as the iron opacity bump traps the outgoing photons, resulting in a strong radiation-driven wind. In this scenario, the WD accretes from its companion through an equatorial accretion disk and loses the excessive mass from the system through a bipolar outflow (e.g., \citealt{2001ApJ...558..323H}).

This picture indicates that the WD grows at an effective rate of $\mdotup$.
Therefore in practice, the wind is often implemented simply by removing material at a rate given by the amount that $|\mdothe|$ is in excess of $\mdotup$.
Our work follows this optically-thick wind scenario, though in implementation it mirrors the approach of \cite{2016ApJ...821...28B} by removing mass from the system when the WD model expands (see Section~\ref{subsec:mesa setup}), rather than using a form of $\mdotup$ prescribed in advance.

One of the goals of this work is to critically examine many of the assumptions
made in this regime.  We discuss and
compare past approaches in more detail in Section~\ref{sec:comparison with previous works}.
We consider the specific angular momentum carried by the mass loss in Section \ref{sec:effect of enhanced angular momentum loss}. 
We explore the physical plausibility of the optically-thick wind in Section \ref{sec:properties of OTW}.

\subsection{The Helium Nova Regime and $\mdotlow$}
\label{subsec:helium nova regime}

Below the minimum stable accretion rate (hereafter the lower stability line $\mdotlow$), the helium shell is thermally unstable and undergoes a series of helium flashes.
This thermal instability in the burning shell happens when a temperature perturbation causes the nuclear burning rate to increase faster than the cooling rate either by expansion work or radiative cooling (e.g., \citealt{2007ApJ...663.1269N}, \citealt{2007ApJ...660.1444S}). For low accretion rates, the thin envelope leads to a less efficient cooling by expansion work and is hence thermally unstable. The thermal content of the envelope, which determines the equation of state, may also come into play. For an envelope with a lower thermal content, pressure has a lower dependence on the temperature, making cooling by expansion work negligible. In general, a lower mass accretion rate below the lower stability line leads to a stronger helium flash. Like $\mdotup$, $\mdotlow$ itself increases with $\mwd$, since a stronger surface gravity leads to a higher shell temperature and hence burning rate, driving the envelope mass lower and therefore less thermally stable for a given accretion rate. 

For $|\mdothe| \leqslant \mdotlow$, the existence of helium flashes can also lead to the ejection
of mass from the system.  It is then necessary to understand the mass retention efficiency (the ratio of mass that remains on the WD to the total mass transferred over a nova cycle) to determine how the WD grows in mass.   
The helium flash regime is not a focus of our work.  Therefore, once the WD enters the He flash regime 
instead of following our models through the flashes, we terminate the simulations 
and report the required average retention efficiency for the WD to grow to $\mch$.
These values can then be compared to previous results characterizing the helium nova retention efficiency as a function of $\mwd$ and $\mdothe$ \citep[e.g.,][]{2004ApJ...613L.129K, 2014MNRAS.445.3239P,2017A&A...604A..31W}.

\subsection{Summary}

To sum up, the growth of the WD mass is determined by the following regimes: if $|\mdothe| \geqslant \mdotup$, the WD effectively accretes at roughly $\mdotup$, and the excess is lost as a wind; if $|\mdothe| \leqslant \mdotlow$, the WD undergoes helium flashes and the exact growth rate depends on the mass accumulation efficiency over a nova cycle; if the mass transfer rate is in the stable regime, the WD accretes at exactly the donor mass transfer rate $|\mdothe|$.


\section{Modeling and Methodology}
\label{sec:modeling and methodology}

In this section we describe the stellar and binary evolution controls, as well as the initial models in our calculations. Since the parameter space involves numerous binary systems, we stop the binary runs when the outcome of the binary system is clear. We describe the stopping conditions here.


\subsection{{Stellar and binary evolution with \tt\string MESA}} 
\label{subsec:mesa setup}

We evolve a CO WD and a He star of various masses in a binary using version {\tt\string 10108} of Modules for Experiments in Stellar Astrophysics (\mesa; \citealt{2011ApJS..192....3P,2013ApJS..208....4P,2015ApJS..220...15P,2018ApJS..234...34P}). We use \mesa\ to evolve the stellar structures of both stars as well as the binary parameters self-consistently, until the outcome of the mass transfer episode from the He star is clear. We describe the important controls in the {\tt\string binary} module as follows. 

We start the evolution with a He ZAMS star between $1.1$ $\msun$ and $2.0$ $\msun$ and a CO WD of $0.90-1.05$ $\msun$. Prior to the He star leaving the He ZAMS, the binary orbit decays slightly solely through emission of gravitational waves. We do not consider the effects of magnetic braking.

As the He star finishes core helium burning, it expands and fills up its Roche lobe. Mass transfer onto the WD then ensues. For the mass loss from the He star, we adopt the {\tt\string Ritter} mass loss scheme \citep{1988A&A...202...93R}, which accounts for the finite pressure scale height of the donor near its Roche limit. We solve for the mass loss using the implicit scheme in \mesa, which accepts the computed mass loss at the start of a time step, $\dot{M}_{\mathrm{RLOF}}$, only if the computed mass loss at the end of the time step, $\dot{M}_{\mathrm{end}}$, has a relative change less than some threshold $\xi$:
\begin{equation}
\left| \frac{\dot{M}_{\mathrm{end}} - \dot{M}_{\mathrm{RLOF}}}{\dot{M}_{\mathrm{end}}} \right| \leqslant \xi,
\end{equation}
and we take $\xi = 1 \times 10^{-4}$. 

For $|\mdothe| \geqslant \mdotup$, some mass is lost from the vicinity of the WD and carries off some angular momentum from the system. We must compute the system mass loss rate, $\beta \mdothe$, where $\beta$ is the fraction of the mass transfer rate $\mdothe$ lost from the system.
To do so, we use a prescription that takes advantage of the tendency of the WD expand to red giant dimensions. The value of $\beta$ is 0 when the WD radius ($\Rwd$) is within $2$ $\Rcwd$, two times the radius of a cold WD of the same mass, but $\beta$ gradually increases to $1$ when $\Rwd$ reaches $10$ $\Rcwd$.
Generally, we want these transition radii to be somewhere between the
cold WD radius and the Roche lobe radius.
As noted by \cite{2016ApJ...821...28B}, the expansion of the WD at the upper stability line occurs so sharply as $\mdotwd$ increases that it does not matter which radius one chooses to implement the wind mass loss.\footnote{We choose a fixed physical radius, while \cite{2016ApJ...821...28B} choose a fraction of the Roche radius.  Because we explore longer period systems, we found it numerically advantageous to not allow the WD to develop a large envelope during the calculation.}
This procedure effectively holds the growth rate of the WD, $\mdotwd = (1-\beta)\mdothe$, at $\mdotup$.

For the determination of the exact value of $\beta$, we adopt an implicit scheme similar to the one described above. In other words, we require that the fractional change in the computed system mass loss between the start and end of the time step to vary less than $\xi = 1 \times 10^{-4}$ . This is important because near the upper stability limit, the WD expands so rapidly that the time step size may have an effect on the computed value of $\beta$ in an explicit scheme. The implicit scheme we adopt allows us to self-consistently calculate $\beta$ and is described in more detail Appendix~\ref{sec:mass loss prescription}.

The characterization of $\mdotup$ by rapid increase of $\Rwd$, is indeed consistent with the statement that above the upper stability line the WD expands to red giant dimensions. However, whether the wind mass loss occurs at the onset of expansion, or whether efficient wind mass loss can happen at all, is itself another issue. For example, \cite{2003A&A...412L..53Y} have adopted a wind mass loss that scales not only with $\Rwd$, but with the WD luminosity $\Lwd$ too, and the upper stability line defined as such is different from ours. We adopt the optically-thick wind theory as a plausible physical scenario for mass loss at mass transfer rates above the upper stability line.  We stress that the particular values of WD radii to implement the mass loss in our prescription do not carry physical significance. Our assumption that a wind will carry all the excess mass above $\mdotup$, defined by rapid expansion of the WD, is convenient for calculations.  We will discuss the physical possibility of such a wind via wind calculations and energetic arguments in Sections \ref{sec:properties of OTW} and \ref{subsubsec: the accretion picture}. 

\added{Our $\mesa$ models also include a super-Eddington wind scheme for the WD.  This only active when the WD exceeds the Eddington luminosity while undergoing helium flashes (generally at the onset of accretion), and so  does not affect the upper stability line.  We discuss its effect in Section~\ref{subsec:fiducial grid}, but it is of minor importance since the focus of this study is on the phase of thermally-stable mass transfer.}

The important controls for the stellar models during the binary evolution are described below.\footnote{The complete list of controls is available to the reader as our \mesa\ input files are posted online at \url{https://doi.org/10.5281/zenodo.2630887}.} For the He star, we use the ``predictive mixing" scheme of $\mesa$ which iteratively finds the location of the convective boundary (described more in detail in Section \ref{subsec:comparison with Brooks}). This change is important during the HeMS when a convective core exists. Equally important in modelling the convective core is the use of OPAL Type 2 opacities \citep{1996ApJ...464..943I}, which accounts for enhanced carbon and oxygen abundances due to He burning. We also artificially enhance the efficiency of convection in
near-Eddington, radiation-dominated regions by reducing the excess of the temperature gradient over the adiabatic temperature gradient, in order to avoid numerical difficulties associated with the iron opacity bump in the most stripped He star models (discussed more in Appendix \ref{uncertainties resulting from stellar evolution controls}). The corresponding controls are 

\begin{list}{}{}
\setlength{\itemsep}{-5pt}
\item \tt\string predictive\_mix(2) = .true.
\item predictive\_zone\_type(2) = `burn\_He'
\item predictive\_zone\_loc(2) = `core'
\item predictive\_superad\_thresh(2) = 0.01
\item predictive\_avoid\_reversal(2) = `he4'
\end{list}

\begin{list}{}{}
\setlength{\itemsep}{-5pt}
\item \tt\string okay\_to\_reduce\_gradT\_excess = .true.
\item gradT\_excess\_lambda1 = -1
\item gradT\_excess\_max\_logT = 6
\end{list}

\begin{list}{}{}
\setlength{\itemsep}{-5pt}
\item \tt\string use\_Type2\_opacities = .true.
\item Zbase = 0.02
\end{list}




For the WD, we also use Type 2 opacities. We note that for sufficient spatial resolution of the burning shell, we adopt ${\tt \string mesh\_delta\_coeff = 0.4 }$ which yields $\gtrsim 3000$ zones during the accretion ($\sim 400$ zones are around the He-burning shell).



\subsection{Stopping Conditions} 
\label{subsec:stopping conditions}

To save computation time, we evolve our models until one of the following conditions is met:

1. \textbf{Center Ignition. } When $\mwd$ approaches $\mch$, compression of the core to higher densities may lead to center carbon ignition. A thermonuclear runaway then happens. We detect the runaway by comparing the rate of non-nuclear neutrino cooling, $\epsilon_{\nu}$ and the rate of carbon burning, $\epsilon_{\mathrm{cc}}$. When $\epsilon_{\nu} \leqslant \epsilon_{\mathrm{cc}}$, we assume that thermal equilibrium can no longer be maintained by having neutrino cooling carry the energy produced by carbon burning, and that a runaway reaction occurs. The result is likely to be a TN SN.  The observational manifestation of Chandrasekhar-mass core carbon ignitions has not been definitively theoretically established, in part due to uncertainties related to the existence of the detonation-to-deflagration transition during the explosion. Thus these core ignitions might be either normal SNe Ia 
\citep[in the case of delayed detonations, e.g.,][]{2005ApJ...623..337G, 2008A&A...478..843B, 2013MNRAS.429.1156S}
or SNe Iax \citep[in the case of pure deflagrations, e.g.,][]{2013MNRAS.429.2287K,2014ApJ...789..103L}.

2. \textbf{Off-center Ignition. } If the WD accretes at high accretion rates (near $\mdotup$) for a prolonged period, compressional heating in the shell (i.e., the
region of the off-center temperature peak that develops) may proceed faster than in the core. As a result, the WD shell may reach conditions for an off-center carbon ignition. A slow carbon flame propagates to the center and the likely outcome is a ONe\footnote{However, see \citet{2018arXiv181108638W} who suggest in a closely-related circumstance that this may lead to burning beyond ONe.} WD which undergoes accretion-induced collapse into a neutron star \citep{1985ApJ...297..531N}. We detect off-center ignition using the same conditions as in center ignition, but we can distinguish the two either by examining whether $\mwd$ is significantly sub-Chandrasekhar, or by examining the mass coordinate of maximum carbon burning.  

3. \textbf{Center/Off-center Ignition. } In very few cases, we find that both the core and the shell reach the line where $\epsilon_{\mathrm{cc}} = \epsilon_{\nu}$. That is, we find models very close to the boundary in parameter space between a center ignition and an off-center ignition. While the occurrence or the final product of a hybrid center/off-center ignition is not clear, we label these systems to emphasize that they are lying near the boundary between a center ignition and an off-center ignition given the uncertainties. 

4. \textbf{Helium Flashes. } When $|\mdothe|\leqslant\mdotlow$, the helium accreted onto the WD is thermally unstable and leads to helium flashes. We then terminate the binary run since evolving through a full helium flash cycle is computationally expensive. We report the minimum required retention efficiency for the WD to grow to $\mch$, given the remaining He star envelope mass ($\mheenv^{f}$) and WD mass ($\mwd^{f}$) at termination:
\added{%
\begin{equation}
    \text{min. efficiency} = \frac{\mch - \mwd^{f}}{\mheenv^{f}}.
\end{equation}
}


5. \textbf{Detached Double WD Binary. } It may happen that the He donor exhausts its envelope and underfills its Roche lobe again. In this case we would expect that a detached double WD binary would result
\added{\citep{2013MNRAS.429.1425R,2018MNRAS.473.5352L}}. If both the WDs are CO WDs, they may merge following orbital decay by gravitational waves and contribute to the double-degenerate channel of SNe Ia \citep[e.g.,][]{1984ApJS...54..335I,1984ApJ...277..355W,2010ApJ...709L..64G,2011ApJ...737...89D}.

6. \textbf{Mass Transfer Runaway. } Depending on the prescription of angular momentum lost from the system, and the binary mass ratio, a mass transfer runaway may occur -- further mass and angular momentum loss may lead to even greater loss. In reality we would expect such a system to form a common envelope, or the WD may merge with the core of the He star.


\subsection{Initial Binary Parameters}
\label{subsec:initial binary parameters}

We compute grids of models by varying the initial He star mass ($\iniMhe$), WD mass ($\iniMwd$), binary period ($\iniP$), and degree of wind angular momentum loss from the system. Our fiducial parameter grid is with a $1.0$ $\msun$ CO WD, where we compute models evenly distributed in donor mass (for $\iniMhe$ from $1.0$ $\msun$ to $2.0$ $\msun$) and in logarithmic initial period (for $\iniP$ from $-1.3$ to $0.0$ in days). The shortest period corresponds to the limit where the He star donor fills up its Roche lobe at He ZAMS. The other parameter space limits are determined such that the TN SN region is well enclosed. 

In addition to the grid with initial WD mass of $1.0$ $\msun$, we also compute grids with initial WD masses of $0.90$ $\msun$, $0.95$ $\msun$ and $1.05$ $\msun$. As $\iniMwd$ decreases, the parameter space shrinks as the WD needs to accrete much more mass to reach $\mch$. A WD mass of $0.90$ $\msun$ is roughly the lowest WD mass where a TN SN outcome is still likely.  For $\iniMwd \geqslant 1.05$ $\msun$, the WD is likely a hybrid carbon-oxygen-neon (CONe) or an ONe WD (e.g., \citealt{2007A&A...476..893S}). It is uncertain whether such WDs can contribute to TN SNe. An ONe WD growing up to $\mch$ is likely to undergo accretion-induced collapse and form a neutron star. Therefore, we do not consider $\iniMwd$ above $1.05$ $\msun$. 

The initial models are made to approximate the previous common envelope episode(s) these He star - CO WD binaries have undergone. For the He star, we create He ZAMS stars with \mesa. The He stars have solar metallicity, that is, {\tt\string Y=0.98} and {\tt\string Z=0.02}. We scale up the mass fraction of $^{14}\mathrm{N}$ to the equilibrium value of the CNO cycle, since the He star has previously undergone hydrogen burning. The CO WD models are created by stripping the envelope of a He star. We evolve a He star and a WD in a binary just as in our grid setup, since we know that for long periods and large donor mass, the He star eventually depletes its envelope and forms a degenerate CO core. We then use part of the {\tt\string MESA} test suite {\tt\string make\_co\_wd} to strip more mass off the CO core through a stellar wind. The CO core is allowed to cool for 10 Myr. Although this is not exactly the evolutionary channel the CO WD comes from, the stripping of a He star in any case suffices to model the formation of the CO WD. We test various combinations of periods and donor masses through this method to produce the CO WD models of masses $0.90$ $\msun$, $0.95$ $\msun$ and $1.0$ $\msun$ to be used in the grid models. However, since this method produces a hybrid CONe WD for a mass of $1.05$ $\msun$ -- which reinforces the fact that $1.05$ $\msun$ is the boundary between CO WD and ONe WD -- we artificially scale up the $1.0$ $\msun$ model to create our $1.05$ $\msun$ CO WD. 

It may be of concern whether the initial conditions in the WD may affect the final outcome. While the carbon/oxygen ratio at the core may affect the temperature and density at which carbon ignites near $\mch$, the initial core temperature has little effect on carbon ignition in our case. The high WD accretion rates of $\sim 10^{-6}$ $\msunyr$ allows fast convergence of the core density-temperature trajectory to a common attractor with little dependence on initial conditions, as shown by \cite{2016ApJ...821...28B}.

Finally, we adopt the fast wind assumption in the fiducial grids to be presented in Section~\ref{sec:fast-wind}. \added{As in previous work \citep[e.g., ][]{2003A&A...412L..53Y,2009MNRAS.395..847W}, this assumes} that the WD wind carries the specific angular momentum of the WD itself:
\begin{equation}
 \frac{\dot{J}_{\mathrm{w}}}{\dot{M}_{\mathrm{w}}}  = \left( \frac{q}{1+q} \right)^{2} a^{2} \Omega_{\rm orb},
\end{equation}
where $\dot{J}_{\mathrm{w}}$ and $\dot{M}_{\mathrm{w}}$ are the orbital angular momentum and mass loss rates from the system, $q = \mhe/\mwd$ is the mass ratio, $a$ is the semimajor axis, and $\Omega_{\mathrm{orb}}$ is the orbital angular frequency. 

\begin{figure}
\fig{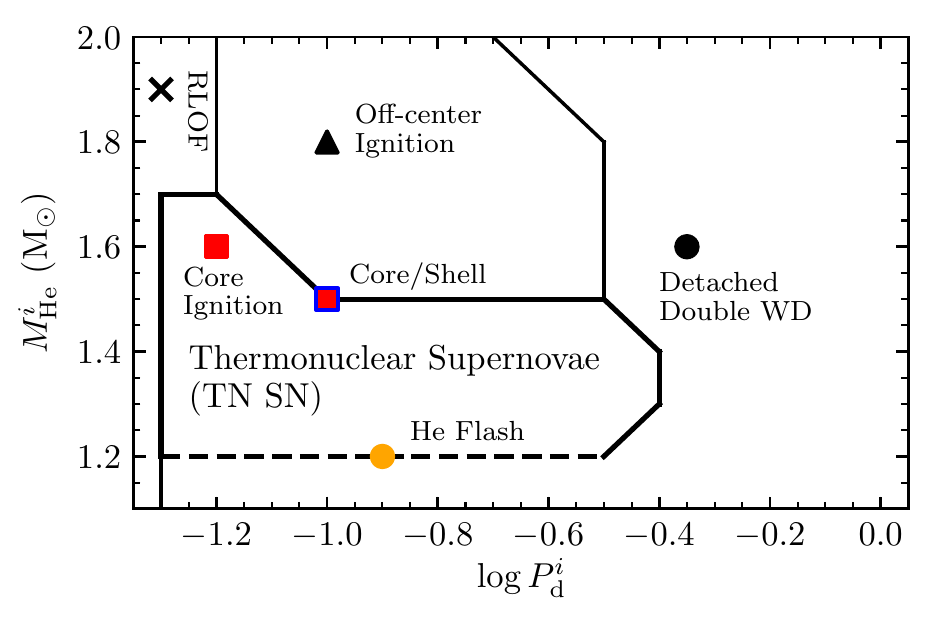}{\columnwidth}{}
\caption{Schematic result of a set of binary evolution models.  By identifying the outcome of each of our grid of models, we divide the initial parameter space into the set of outcomes described in Section~\ref{subsec:stopping conditions}.
\added{The lower boundary of the TN SN region is dashed to indicate that our models do not directly find the boundary between the He flash systems that produce TN SNe and those that eventually become detached double CO WD binaries.}
\label{fig:legend}}
\end{figure}


Figure~\ref{fig:legend} illustrates the schematic result of one of these sets of model grids.\footnote{The boundaries approximately, but not exactly, correspond
to the results from the case for $\mwd = 1.0\,\msun$ shown in Figure~\ref{fig:grids}c.}
Our calculations partition the parameter space into various outcomes described in Section~\ref{subsec:stopping conditions}, with our particular interest being in the TN SN region. Beyond the left boundary of the TN SN region, the He star is Roche lobe-filling at He ZAMS, and these systems labelled ``RLOF'' in Figure~\ref{fig:legend} and marked with an X are unlikely to have been formed.


\section{Fast Wind Results}
\label{sec:fast-wind}

In this section we describe the results of our binary calculations.
Throughout, we keep the wind angular momentum loss fixed at the fast wind limit.
We first choose a few cases to illustrate the binary calculation itself,
then we describe the TN SN region.


\subsection{The Mass Transfer History} 
\label{subsec:the mass transfer history}

\begin{figure*}
\gridline{
\fig{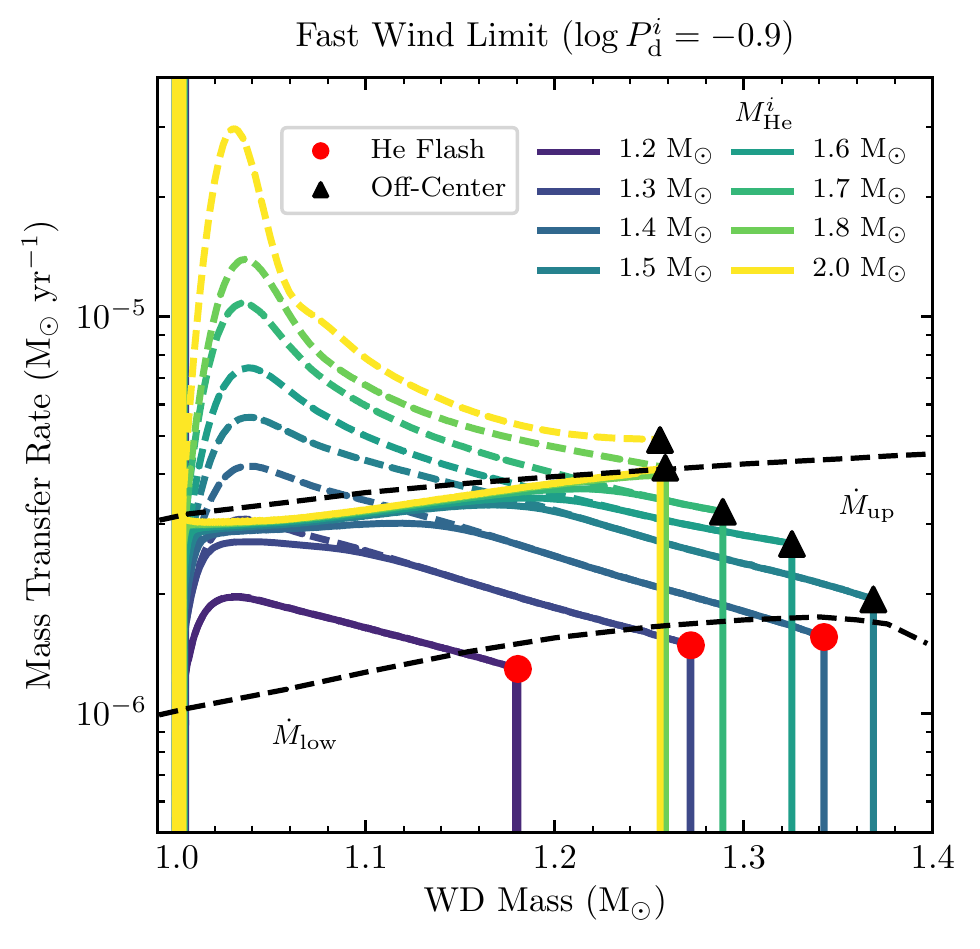}{0.5\textwidth}{(a)}
\fig{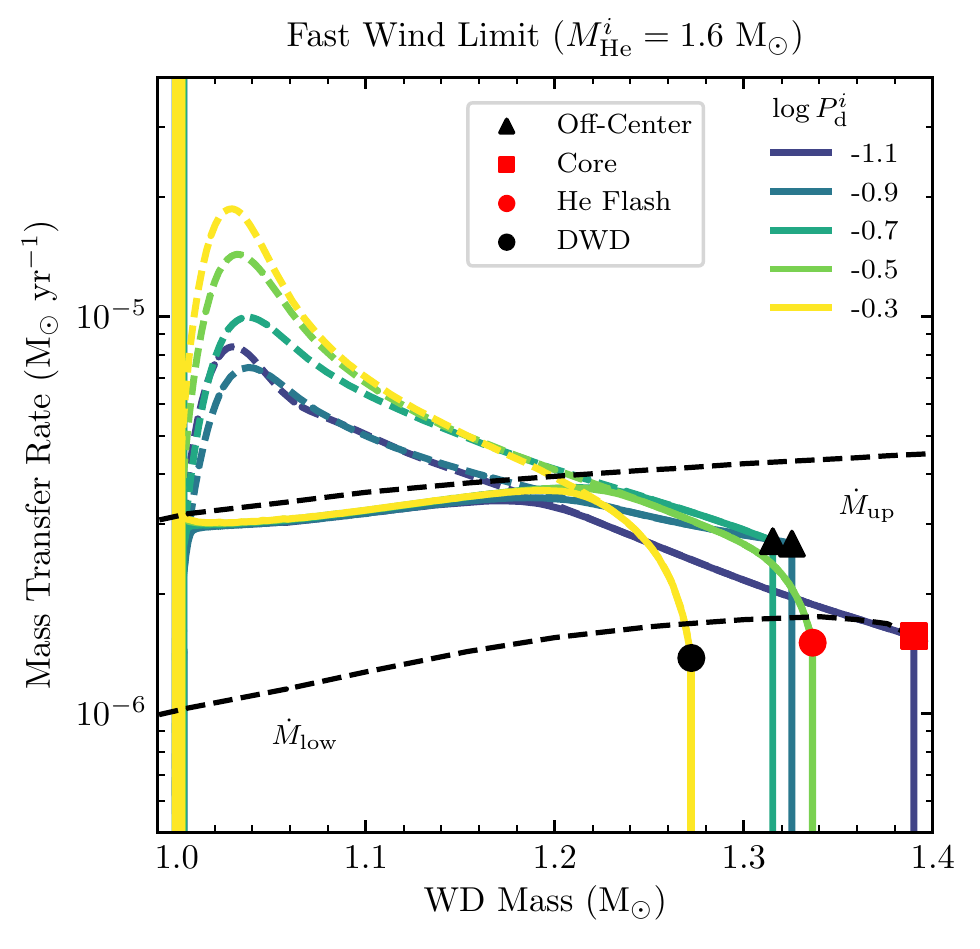}{0.5\textwidth}{(b)}
          }
\caption{ The mass transfer history for models at fixed period (Panel a; $\iniP=-0.9$) and fixed donor mass (Panel b; $\iniMhe = 1.6$) adopting the fast wind limit. 
The dashed lines show the mass loss rate of the He star, $\mdothe$, 
and the solid lines show the accretion rate on the WD, $\mdotwd$.
As mass transfer begins $|\mdothe|$ increases due to evolutionary expansion of the He star and peaks, while later $|\mdothe|$ decreases as the donor structure adjusts to the mass loss and expansion of the binary.
We assume an optically-thick wind is driven when $|\mdothe| \geqslant \mdotup$ (upper dashed black line), which then holds $\mdotwd \approx \mdotup$.
The symbols at the end of each track indicate the stopping condition
of each run, with the red square and blue triangle indicating core and off-center carbon ignition, respectively.  For systems where $|\mdothe| \leqslant \mdotlow$ (lower dashed black line), we have either 
a detached double WD binary (black circle) or if the WD begins to undergo helium flashes, we halt the calculation and denote this by a red filled circle. 
The $\mdotup$ and $\mdotlow$ curves are from \citet{{2016ApJ...821...28B}}.
\authorcomment1{We altered the aspect ratio of the plot and included fewer lines in order to enhance its readability.}
\label{fig:fiducial_case}
}
\end{figure*}

To demonstrate the mass transfer history leading up to the corresponding final outcome of the binary, we show a subset of the binary calculations in
Figure \ref{fig:fiducial_case}. Panel (a) shows a set at fixed period and varying He star mass, $\inipara$ = (1.1 -- 2.0, 1.0, -0.9), while panel (b) shows as set at varying period but fixed donor mass, $\inipara$ = (1.6, 1.0, -1.2 -- -0.3).

Mass transfer initiates as a consequence of both orbital decay by gravitational waves and evolutionary expansion of the He star. As the He star, evolved from the He ZAMS, exhausts helium in the core and proceeds to helium shell burning, it rapidly expands and overfills its Roche lobe. Mass transfer then proceeds on the thermal timescale of the He star, yielding a typical mass transfer rate of $\sim 10^{-6}-10^{-5}$ $\msun$ $\peryr$. The WD accretes from the He star and grows in mass.

Initially, as $|\mdothe|$ is still low and the WD is cold, matter accreted onto the WD is cold and dense, leading to a few cycles of helium flashes, which explain the very high $|\mdotwd|$ -- the WD is in fact losing mass due to the inclusion of a super-Eddington wind. The strength of the helium flash decreases with each cycle as the thermal content of the WD surface increases and $|\mdothe|$ increases further. Afterwards, $|\mdothe|$ (colored, dashed lines) enters the stable regime or even rises above $\mdotup$.  In this case, $|\mdotwd|$ (colored, solid lines) is effectively limited to $\mdotup$, and we assume the remainder of the donated mass is lost in a fast wind carrying the specific angular momentum of the WD. 

The final outcome of each system is indicated by the symbol at the end of its track.  The outcome shifts as the mass transfer history changes.  We clearly see 
that increasing $\iniMhe$ and $\iniP$ generally leads to higher values of  $|\mdothe|$, but that the trends in the outcome are more complex.

Panel (a) of Figure~\ref{fig:fiducial_case} shows that with increasing $\iniMhe$, off-center carbon ignition in the WD is more favored. This results from the fact that a more massive donor is able to sustain high $|\mdothe|$ for a longer period of time. In general, for a more massive donor, either $\mdotwd=\mdotup$ for a longer time, or $\mdotwd=|\mdothe|$ tends to be higher within the steady accretion regime. Either of these leads to higher accretion rate onto the WD, favoring off-center carbon ignition in high mass donors. This is certainly the case for the most massive donors (1.8 - 2.0 $\msun$).
For less massive donors (1.5 - 1.7 $\msun$), $|\mdothe|$ eventually falls within the stable regime, but the generally high accretion rates throughout the accretion episode still leads to an off-center carbon ignition. The WD mass at which the off-center carbon ignition happens is higher for a lower $\iniMhe$, because the lower $|\mdothe|$ leads to less compressional heating, delaying the evolution of the shell to carbon ignition. 

Conversely, low $\iniMhe$ mean lower mass transfer rate on average. The WD may accrete for a while -- or even not at all -- at $\mdotup$, and the drop in $|\mdothe|$ leads to accretion in the stable regime and eventually in the helium flash regime. The lower $\iniMhe$ is, the higher the helium flash retention efficiency is required to reach $\mch$.  This is because the WD does not grow too much further in mass during the stable accretion. However, the mass retention efficiency depends on $|\mdothe|$, and may even be negative for very low $|\mdothe|$.  Recall that we stopped our evolutionary calculations at the onset of the He flashes, so the retention efficiencies must come from other 
calculations that follow WDs though many flashes.

Panel (b) of Figure~\ref{fig:fiducial_case} shows that an off-center ignition is more favored with increasing $\iniP$. For a given $\iniMhe$, longer periods give rise to a larger donor Roche radius and a larger $|\mdothe|$ can occur when the donor overfills its Roche lobe. This means $|\mdothe|$ is higher initially. The higher compressional heating caused by high $|\mdothe|$ is why the outcome shifts from a core ignition at $\iniP = -1.2,-1.1$ to an off-center ignition at $\iniP$ from $-1.0$ to $-0.5$. (we describe this in more detail in the next subsection.)
Ultimately for even larger $\iniP$, the formation of a detached double WD binary is favored. Longer periods lead to higher initial $|\mdothe|$, so the donor envelope is stripped more efficiently; as the WD can only accrete at most at $\mdotup$, the very low accretion efficiency by the WD may cause the donor to exhaust its envelope before the WD can grow up to $\mch$. Then a detached double WD binary is formed, as in the two longest period systems.

\subsection{Core/Shell Competition}
\label{subsec:core shell competition}

\begin{figure*}
\gridline{\fig{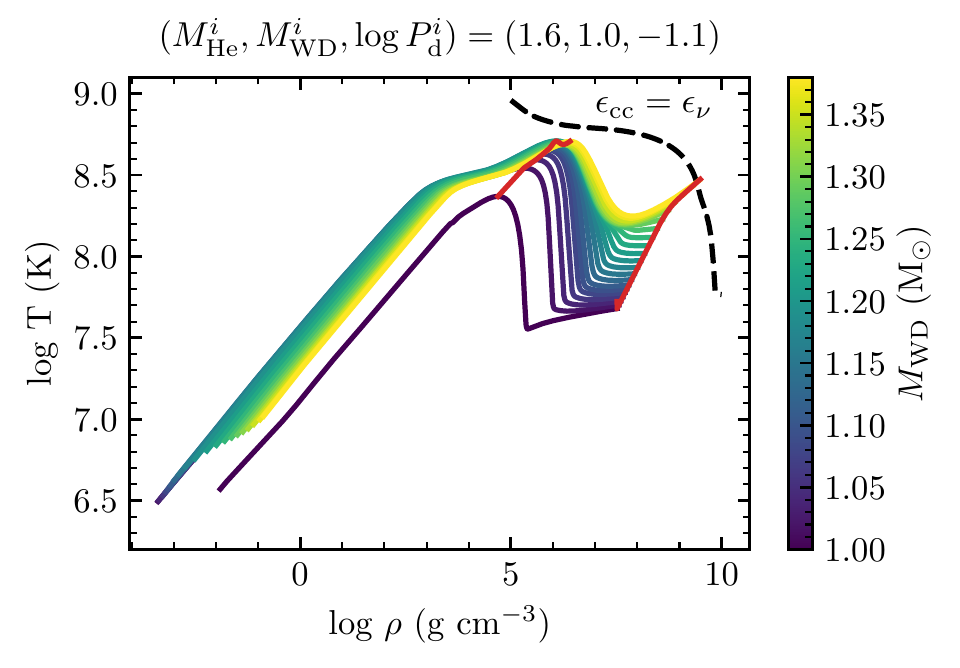}{0.5\textwidth}{(a)}
          \fig{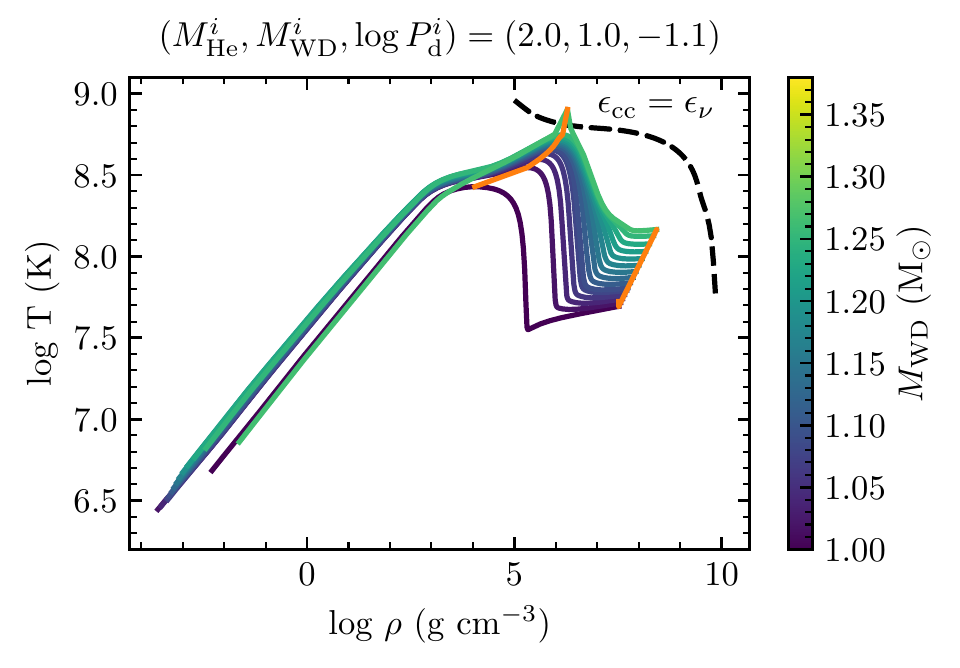}{0.5\textwidth}{(b)}
          }
\gridline{\fig{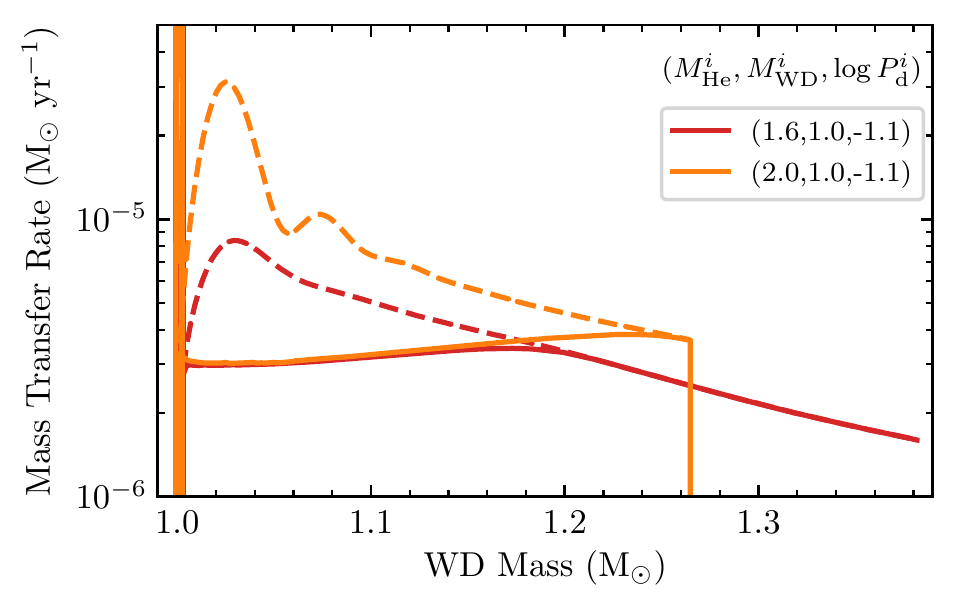}{0.5\textwidth}{(c)}
          }
\caption{Thermal evolution of the WD during accretion. Panel (a) shows the model $\inipara = (1.6,1.0,-1.1)$ which eventually undergoes central carbon ignition; Panel (b) shows the model $\inipara = (2.0,1.0,-0.9)$ which eventually undergoes off-center carbon ignition.  These panels plot the WD density-temperature profile at different WD masses. The red (Panel a) and orange (Panel b) lines track the evolution of the core (right) and the shell (left), one of which will eventually cross the black dashed line where the rate of carbon burning is equal to the thermal neutrino losses. Panel (c) shows the evolution of the mass transfer rates with $\mwd$. Note that the WD in Panel (b) has a higher accretion rate at all times (solid line), and hence ignites off-center due to stronger compressional heating in the shell than in the core. \label{fig:core_shell}}
\end{figure*}

\cite{2016ApJ...821...28B} have brought to attention the core/shell competition in the WD, in which the mass accretion history determines whether a carbon ignition occurs at the center or off-center.  
Here we describe the physics behind the thermal evolution in the core and the shell.

The mass accretion rate on the WD, $\mdotwd$, determines the energy generation rate and subsequent heat distribution within the WD. Energy is generated in the burning shell via stable helium burning and by the release of gravitational potential energy as each Lagrangian shell is buried deeper inside the WD and compressed while the WD increases in mass. The local (Lagrangian) compression rate leading to the release of gravitational energy originates from two sources (see equation~(6) of \citealt{1982ApJ...253..798N}). One arises due to the increase in density at a fixed fractional mass coordinate $q$ while the WD increases in mass; the other arises from the compression to higher densities of the shell itself as it moves inwards to lower $q$ \citep{1982ApJ...253..798N}. A temperature peak is driven at high accretion rates because this ``compressional heating'' proceeds faster near the surface than at the center for high accretion rates -- the timescale for compressional heating is faster than the timescale for heat transport  \citep{1982ApJ...253..798N}. Therefore, for higher accretion rates the WD shell evolves more rapidly to higher temperature and density \citep{2016ApJ...821...28B}. An off-center carbon ignition is thus more likely. 

In Figure \ref{fig:core_shell}, we show the evolution of the WD density-temperature profile, for two cases of accretion. Both panels (a) and (b) start with a $1.0$ $\msun$ WD accreting from a He star companion in an initial orbital period (in days) of $\iniP = -1.1$. Panel (a) has a $1.6$ $\msun$ He star, whereas Panel (b) has a $2.0$ $\msun$ He star. We show the corresponding mass transfer rates in Panel (c). Since in Panel (a) the donor has a lower envelope mass, as mass transfer proceeds $|\mdothe|$ falls into the stable regime, whereas the Panel (b) WD always accretes at $\mdotup$. Due to the higher mass accretion rate, the Panel (b) WD experiences stronger compressional heating in the shell than in the core. As the shell evolves to higher temperature and densities, carbon is eventually ignited off-center. On the contrary, the Panel (a) WD is able to grow up to $\mch$ and undergo central carbon ignition. Figure \ref{fig:core_shell} illustrates the point that a higher mass accretion rate favors an off-center carbon ignition, so properly resolving the WD stellar structure is needed in order to investigate the TN SN region. Our fiducial grid, to be described in the following section, showcases our time-dependent binary runs resolving both components.


\subsection{The Fiducial Grid}
\label{subsec:fiducial grid}

As the fiducial grid, we run models evenly distributed in $\iniMhe$ and $\iniP$ space, using $\iniMwd=1.0$ $\msun$ and a fast wind assumption. The corresponding mass transfer history for each model is similar to the ones shown in Figure~\ref{fig:fiducial_case}. Here we describe the general trends in the outcome across the parameter space.  Figure~\ref{fig:legend} shows a schematic version the outcomes, while Figure~\ref{fig:grids}, panel (c) shows the detailed outcome for each binary calculation in the fiducal grid.

The left-most boundary of the TN SN region is determined by the condition that the He star not be Roche-filling at He ZAMS. The shortest period that the He star can still fit in its Roche lobe is $\iniP=-1.3$, except for models with $\mhe=1.8-2.0$ $\msun$. 
The rest of this period may be so tight that the He star, while still helium-burning at the core \added{(case BA mass transfer)}, may expand due to evolution, transfer mass in the He flash regime, adjust and be detached repeatedly.
\added{The super-Eddington wind present in our models effectively keeps the accumulation efficiency near zero during the He flashes and so the WD experiences little growth in mass during this phase.}
Some particular models may experience He flashes that cause numerical problems in $\mesa$ which is why some models are missing from the grid. The models that run through eventually transfer mass at the stable regime as the He star exhausts its core helium \added{(case BB mass transfer)}, although the donor mass at the start of the stable mass transfer may be reduced from its mass at He ZAMS. 

The upper and right boundaries of the TN SN region comes from the occurrence of off-center carbon ignitions in the WD, or formation of detached double WD binaries. As mentioned, higher $\iniMhe$ and $\iniP$ lead to higher accretion rates and favor off-center ignitions. These will likely lead to a mass-transferring He star with an ONe WD companion which may undergo accretion-induced collapse near $\mch$ \citep{2017ApJ...843..151B}. Even longer $\iniP$ strip the He donor so efficiently that the donor becomes detached again. With longer periods more time has elapsed between He star - WD binary formation and donor RLOF, therefore the donor is more evolved at the start of RLOF. As a result the CO core of the He donor grows more, so that the donor may become a more massive WD when it becomes detached again. The less massive remnants may become a second CO WD. The subsequent orbital decay through gravitational waves may lead to a double CO WD merger and hence to TN SN through the double-degenerate channel. The more massive remnants may become an ONe WD and the final outcome of such a CO + ONe WD merger may also be an interesting transient event \citep{2018arXiv181100013K}. 

For lower $\iniMhe$ systems, $|\mdothe|$ eventually enters the He flash regime. Following evolution through the helium flashes is tractable only by time-dependent, multi-cycle calculations, so in Figure~\ref{fig:grids} we use the colorbar to report the required retention efficiency for systems that begin to flash. Referring to the low-mass 1.1-1.2 $\msun$ donors in Panel (a) of Figure \ref{fig:fiducial_case} and Panel (c) of Figure \ref{fig:grids}, we see that the required efficiency is near unity, but the fact that they have low $|\mdothe|$ means that the helium flashes will have very low retention efficiency. These low mass donors are unlikely candidates as systems that will grow the WD up to $\mch$, and hence define the lower boundary of the TN SN -- systems below this boundary will ultimately become detached \replaced{double WD binaries.}{double CO WD binaries \citep{2018MNRAS.473.5352L}. We do not determine the minimum $\iniMhe$ that can still contribute to TN SNe since we do not evolve the WD through the He flashes; in Figure~\ref{fig:legend} we draw the lower boundary at systems with 60\% required efficiency since it broadly agrees with the lower boundaries of \citet{2009MNRAS.395..847W} and \citet{2017A&A...604A..31W}.} Previous works have attempted to calculate the mass retention efficiency of helium flashes as a function of $\mwd$ and $\mdotwd$ \citep[e.g.,][]{2004ApJ...613L.129K,2014MNRAS.445.3239P,2017A&A...604A..31W}, which we have briefly discussed in Section \ref{subsec:helium nova regime}.

In between the boundaries for off-center carbon ignition, detached double WD binary, and low retention efficiency helium flash, is the region of central ignition (likely TN SN progenitors). These are systems with $|\mdothe|$ low enough to avoid strong compressional heating in the shell and thus an off-center carbon ignition, or exhausting the donor envelope, but high enough to avoid helium flashes with low retention efficiency. Near the short period end, there is a trend for the center/off-center ignition boundary to move to higher $\iniMhe$. This is because compared with long periods, high mass donors at short periods become Roche-filling at a less evolved stage with lower core mass, and in general avoid very high $|\mdothe|$ so as to cause an off-center ignition in the WD. 


\subsection{A Different Donor Mass}
\label{subsec:a different donor mass}

\begin{figure*}
\gridline{
\fig{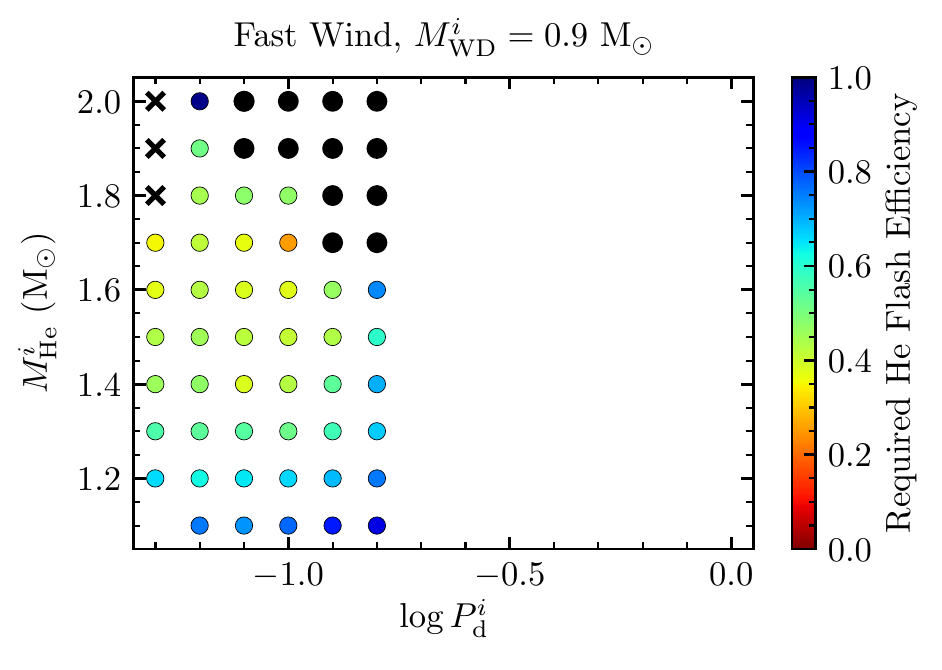}{0.5\textwidth}{(a)}
\fig{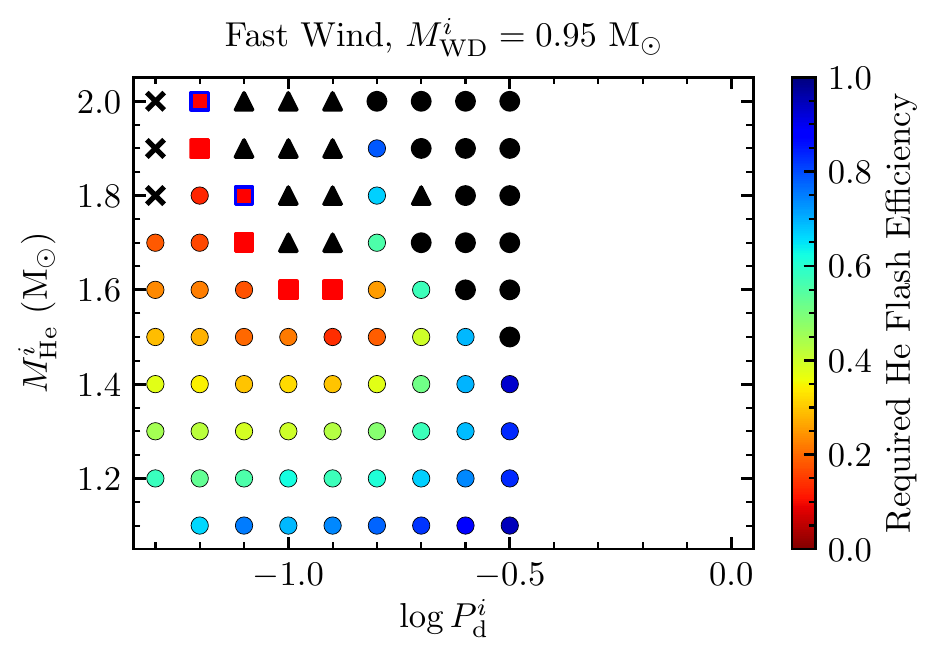}{0.5\textwidth}{(b)}
          }
\gridline{
\fig{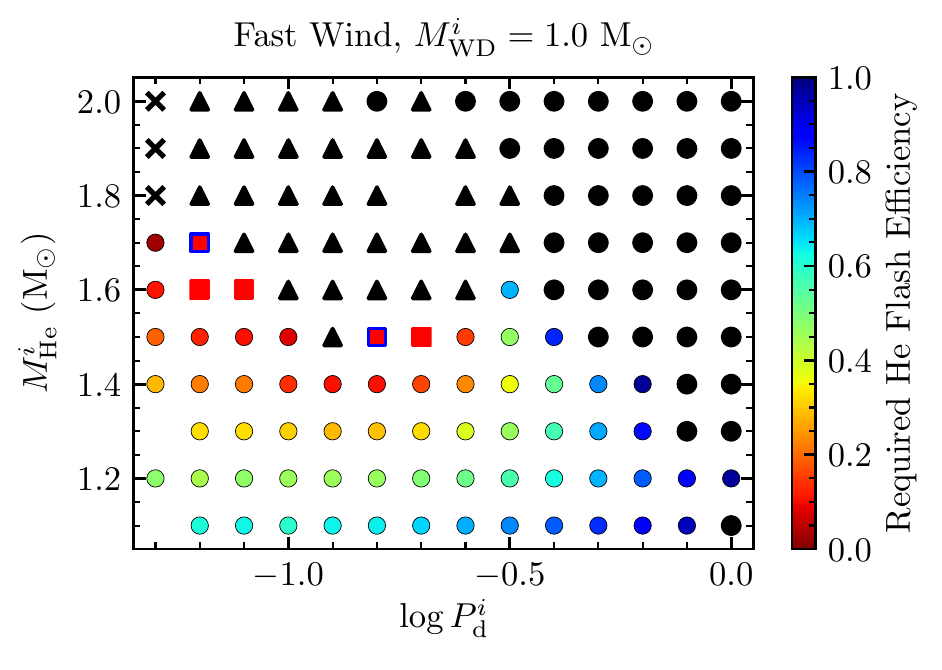}{0.5\textwidth}{(c)}
\fig{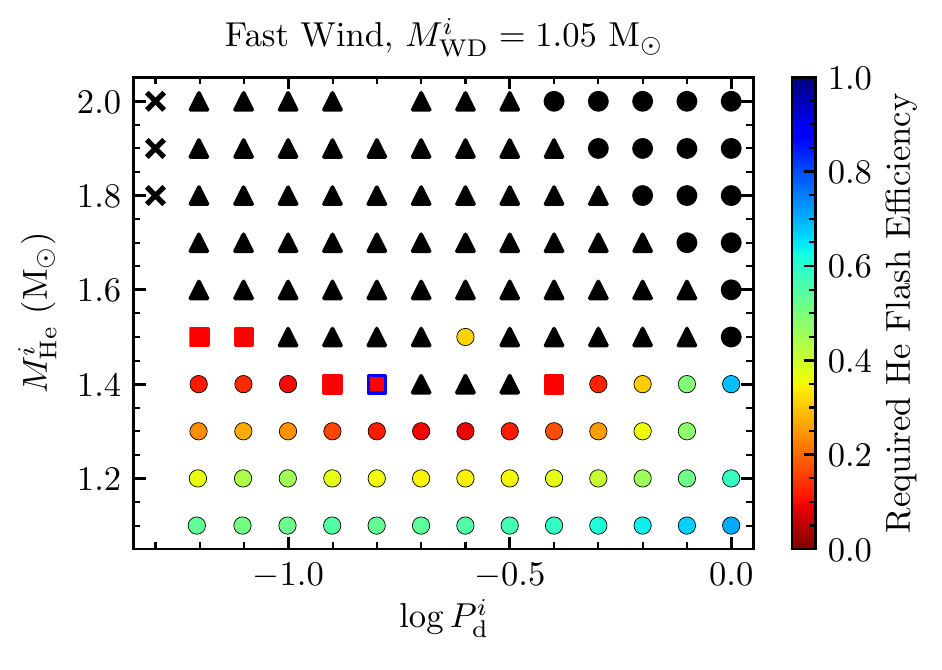}{0.5\textwidth}{(d)}
          }
\caption{ The fate of the He star - CO WD binary as a function of $\iniMhe$ and $\iniP$, with each panel representing a different initial WD mass.  
As a guide to interpreting these plots, the reader is referred to Figure~\ref{fig:legend},
which is a schematic version of panel (c).
The red squares represent systems where the WD undergoes a core ignition through direct accretion. The colored circles represent the systems where the WD undergoes helium flashes; we color code by the required retention efficiency for the WD to grow to $\mch$. The red squares with a blue edge represent systems where both core and shell ignitions are detected, representing systems located at the core/shell ignition boundary in the parameter space. The black triangles represent systems where the WD experiences a shell ignition and will likely form an ONe WD. The black circles indicate systems likely to form a detached double WD binary; \added{systems with high required retention efficiencies are also likely to produce detached double WDs.  Our work does not determine the actual retention efficiency during the He flashes, so does not directly indentify the mininum He star donor mass required for a TN SN through this channel.} The crosses indicate systems where the He star is Roche-filling at He ZAMS. The TN SN region grows to longer $\iniP$ but lower $\iniMhe$ as $\iniMwd$ increases.
\label{fig:grids}}
\end{figure*}

The fiducial grid employs a $1.0$ $\msun$ WD as the accretor, but it is also interesting to see how the parameter space changes with $\iniMwd$. Figure \ref{fig:grids} shows the results of several grids run with a different $\iniMwd$. 

In general, the parameter space shrinks with lower initial WD mass.  The most significant change is at the long period end, where the regime for forming detached double WD binaries starts at a shorter period (a shift of $\approx 0.3$ in $\iniP$) for the $0.95$ $\msun$ grid (panel b) compared with the $1.0$ $\msun$ grid (panel c).  The long period binaries tend to have higher $|\mdothe|$ initially, which the WD cannot accept fully due to the upper stability limit, and hence lower overall accretion efficiency. As the donor is stripped of its envelope rapidly, the question then becomes whether the WD can grow up to $\mch$ before the donor envelope is exhausted.  This is simply more difficult for lower $\iniMwd$.

On the short period end of the $0.95$ $\msun$ grid, we see a slight shift of the core-ignition regime into the parameter space with high mass donors. This may be attributed to the lower value of $\mdotup$ for lower $\mwd$, such that the WD growth rate is lower during the time before $|\mdothe|$ falls below $\mdotup$ and the WD enters the stable accretion regime. The lower accretion rate leads to weaker compressional heating in the shell and allows the WD to avoid an off-center ignition. Therefore, a lower $\iniMwd$ shifts the boundary between center and off-center ignitions to a higher $\iniMhe$ in the parameter space, and vice versa.

To summarize, the parameter space for TN SNe is restricted to lower donor masses due to off-center ignition for a higher initial WD mass, but broadens to include longer period systems by outracing the stripping of the donor envelope.

\section{Comparison with Previous Works}
\label{sec:comparison with previous works}

 Now having described the results of our fast wind grids in Section~\ref{sec:fast-wind}, we discuss and compare with the results of previous works. 

\subsection{Comparison with \citet{2016ApJ...821...28B}}
\label{subsec:comparison with Brooks}

The most direct comparison we can make is with \citet{2016ApJ...821...28B}
who also used \mesa\ and who provided a starting poing for our work. A major difference between the two studies is our use of {\tt\string MESA}'s predictive mixing capability. \cite{2018ApJS..234...34P} emphasized the importance of self-consistently locating convective boundaries such that $\nabla_{\rm rad}$ and $\nabla_{\rm ad}$ are equal on the convective side of the boundary. They implemented a scheme, called ``predictive mixing'', that served to satisfy this constraint. This has a significant effect on the extent of the convective core during core He burning (see their section 2.4). This leads to differences in the stellar structure of the donor and hence mass transfer rates.

\begin{figure}
\fig{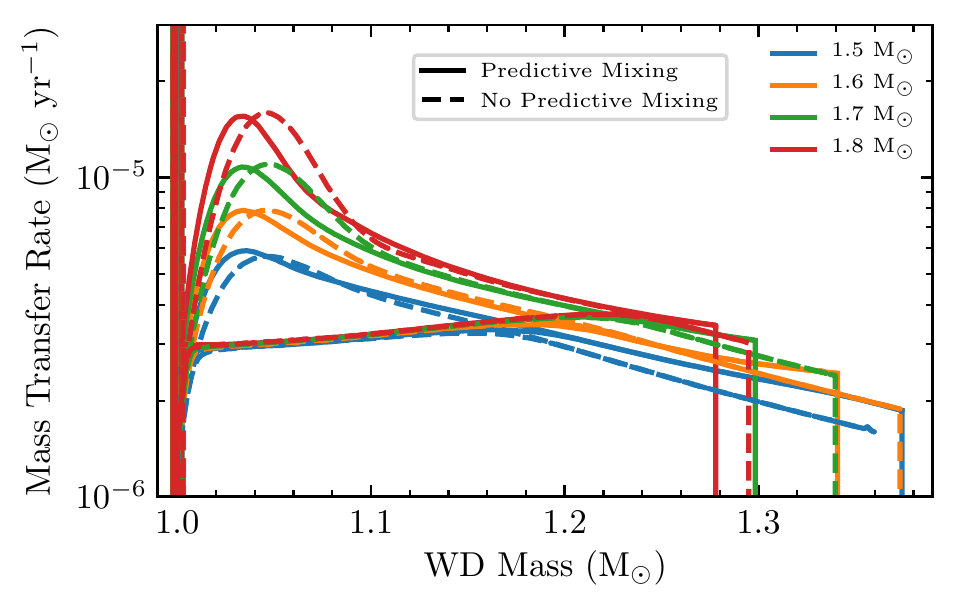}{\columnwidth}{}
\caption{Comparison of two runs with a 1.0 $\msun$ WD at an initial orbital period 0.125 days with $\iniMhe$ ranging from 1.5 to 1.8 $\msun$, one iteratively solving for the convective boundary of the donor (shown as solid lines) and the other without this ``predictive mixing'' capability (shown as dashed lines). The runs with ``predictive mixing'' have higher $|\mdothe|$ initially, as a result of the different donor convective core size on the HeMS.
\label{fig:pred_mix}}
\end{figure}

We illustrate that the use of the predictive mixing scheme for the He donor leads to a slightly different binary evolution in Figure \ref{fig:pred_mix}. The self-consistent determination of the convective boundary 
leads to a larger convective core.  This has several effects.
First, it produces a larger carbon core after core helium exhaustion 
and thus the helium envelope mass available for mass transfer to the WD is smaller.  Second, because the core burning lifetime is longer and we begin at the He ZAMS, the binary separation by the time mass transfer happens is slightly smaller, as gravitational waves have had more time carry away orbital angular momentum. Finally, as the donor has a slightly different structure, mass transfer and subsequently the binary evolution takes a slightly different path.

\begin{figure}
\fig{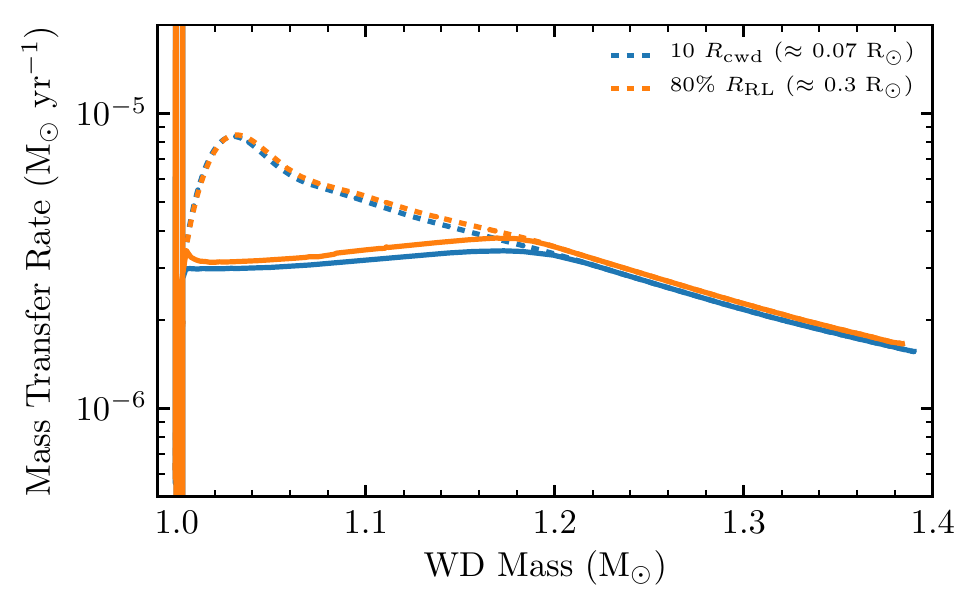}{\columnwidth}{}
\caption{Comparison of two runs at $\inipara=(1.6,1.0,-1.1)$, one using a larger radius of $80\%$ $\Rrl$ for initiating the wind mass loss (orange) and the other a smaller radius of 10 $\Rcwd$ (blue). The difference in the effective upper stability line is small and both WD models reach $\mch$.
\label{fig:radius}}
\end{figure}

There are also slight differences in our wind mass loss prescriptions. We both implement a wind mass loss when the accreting WD is at the upper stability line, but whereas \cite{2016ApJ...821...28B} limit $\Rwd$ to less than 60\% of the WD Roche radius $\Rrl$, we limit $\Rwd$ to a slightly more compact configuration, 10 $\Rcwd$.  Our implementation leads to a slightly lower $\mdotup$, since the transition to a He red giant does not happen at a infinitely sharp mass accretion rate, and our prescription chooses the lower end of this transition. Figure \ref{fig:radius} compares two runs at $\inipara=(1.6,1.0,-1.1)$, with one limiting the radius to 10 $\Rcwd$, and the other to 80\% $\Rrl$.  This illustrates that our choice of limiting radius in the wind prescription does not lead to significant differences in the wind mass loss and mass transfer rate.

At an initial orbital period of 3 hr, \cite{2016ApJ...821...28B} find the transition between core and shell ignitions is around $\mhe \approx 1.7\,\msun$.
In our calculations, this transition is somewhat lower, around $\mhe \approx 1.5\,\msun$.  Figure~\ref{fig:pred_mix} illustrates that, in terms of final WD mass, models run with predictive mixing appear like models with $\mhe$ lower by $\approx 0.1\,\msun$ run without predictive mixing.  This partially explains the shift.
Based on their posted inlists, we believe that \cite{2016ApJ...821...28B} also included magnetic braking, which means that the orbits shrank slightly before mass transfer began, making their initial period effectively shorter than 3 hr.
This also goes in the correct direction to explain the change, as shrinking the period by 0.1 dex increases the transition mass by $\approx 0.1\,\msun$. For the case of lower mass donors, comparing the $\mhe=1.3\,\msun$ and $1.4\,\msun$ models in Panel (a) of Figure \ref{fig:fiducial_case} with the equivalent models in Figure 3 of \cite{2016ApJ...821...28B}, we see that the models start experiencing strong helium flashes at similar WD masses, $\mwd \approx 1.27$ $\msun$ and $1.35\,\msun$, respectively.  The $|\mdothe|$ at which the strong helium flashes start is slightly higher in our models, which may be due to differences in the accretion histories and in the adopted opacities.  Together, these minor differences appear to account for most of the difference between our results and \cite{2016ApJ...821...28B}.
We emphasize that overall the agreement is good, which is to be expected given the similarity of our approaches.

\subsection{Comparison with \citet{2003A&A...412L..53Y}}
\label{subsec:comparison with Yoon and Langer}

\begin{figure}
\fig{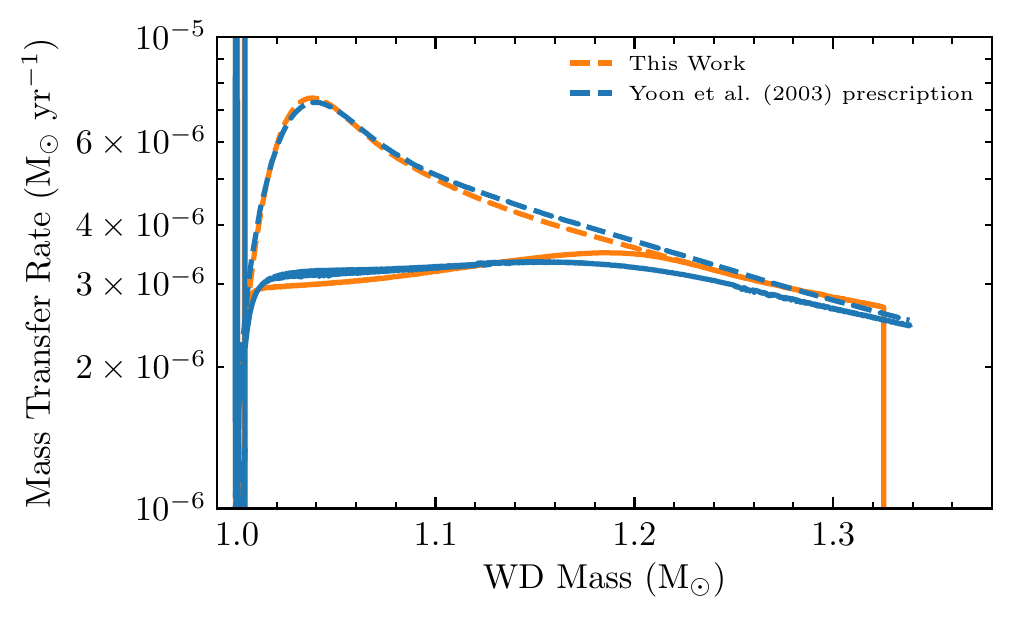}{\columnwidth}{}
\caption{A comparison of our work and a binary run adopting the \citet{2003A&A...412L..53Y} prescription. Both are run at $\inipara=(1.6,1.0,-0.9)$. Our mass accretion rates are similar, except that due to a dependence on $\Lwd/\Ledd$, the \citet{2003A&A...412L..53Y} prescription gives rise to some wind mass loss when the WD is massive. 
The difference in mass transfer history does not affect the outcome and
both models shown experience an off-center ignition before reaching $\mch$.
\label{fig:yoon}}
\end{figure}

\cite{2003A&A...412L..53Y} computed the mass transfer between a 1.6 $\msun$ zero age main sequence He star and a 1.0 $\msun$ WD initially at an orbital period of 0.124 days. Gravitational wave losses are included in the initial orbital decay. The WD is treated as a point mass until $|\mdothe|$ is above $10^{-6}$ $\msun$ $\peryr$, at which point a ``heated'' WD model is used to approximate the heating by the initial helium flashes. The WD is eventually able to grow up to $\mch$ and experience a central ignition.

The most similar model in our grid has the same binary component masses with $\iniP = -0.9$. Instead of a core carbon ignition found by \cite{2003A&A...412L..53Y}, we find an off-center carbon ignition at about $\mwd \approx 1.32$ $\msun$. We examine the differences by running a \mesa\ model with $\iniP = -0.9$ adopting the mass loss prescription of \cite{2003A&A...412L..53Y}. We show the results of comparing this with our standard model in Figure \ref{fig:yoon}. The Yoon-like model experiences an off-center ignition at $\approx 1.34$ $\msun$, similar to our standard model.

The mass transfer histories of both models are very similar. As expected, the donor mass transfer rates are almost identical, with a slightly different accretion retention fraction due to the wind mass loss prescriptions adopted. In particular,  \cite{2003A&A...412L..53Y} have adopted a wind mass loss with the form $\mdotw = 10^{-2} \Rwd \Lwd / G \mwd (1-\Gamma)$. This form is based on dimensional arguments (modifying the gravitational potential to account for the radiation pressure), and normalized to fit the mass loss rates observed for Wolf-Rayet stars. On the other hand, we implement a mass loss algorithm that limits the WD radius to a rather compact 10 $\Rcwd$.

The dependence of the mass loss prescriptions on different stellar parameters is the main cause in the slight difference between the two models shown in Figure \ref{fig:yoon}. Whereas we implement a mass loss only when the WD experiences radial expansion, the \cite{2003A&A...412L..53Y} prescription also has mass loss even when the WD is quite compact but instead has a high luminosity close to the Eddington limit due to the accretion. This can be clearly seen when the WD is quite massive. In general, the \cite{2003A&A...412L..53Y} prescription leads to slightly lower accretion rates,
which would slightly favor a core ignition.
As can be seen in Figure \ref{fig:yoon}, the difference in WD accretion rates between our prescription and the \citet{2003A&A...412L..53Y} prescription is not significant as both these models experienced an off-center ignition. Thus, instead of a difference in mass loss prescription,
the reason why we find an off-center ignition where 
\cite{2003A&A...412L..53Y} find a core ignition 
may be due to different donor models, e.g., the use of $\mesa$'s predictive mixing capability in our work. Moreover, the $\inipara=(1.6,1.0,-0.9)$ case is located at the boundary between center and off-center ignitions in the parameter space, therefore the final outcome is sensitive to the binary evolution prescription.


\subsection{Comparison with \citet{2017MNRAS.472.1593W}}
\label{subsec:comparison with Wang 2017}

\cite{2009MNRAS.395..847W,2017MNRAS.472.1593W} also study the parameter space for SN Ia via the helium donor channel.  They use Eggleton's stellar evolution code to evolve He star - WD binaries.  They model the WD as a point mass, but have developed a simple prescription to account for the occurrence of an off-center carbon ignition in the WD.  Here we compare our results to theirs.

Several details differ in the mass transfer histories of the models computed by \cite{2009MNRAS.395..847W} and those in our work. Such differences are reasonable in light of the different WD core thermal profiles, He donor stellar models, exact values of the accretion regime, etc., being used in our works. In particular, \cite{2009MNRAS.395..847W} have used the upper stability line of \cite{1982ApJ...253..798N}, which is slightly higher than the effective upper stability line in our calculations.

More importantly, it is informative to compare the TN SN regions found in our works. In order to find the off-center ignition models in the entire parameter space, \cite{2017MNRAS.472.1593W} examined the mass transfer histories of the models in \cite{2009MNRAS.395..847W}. If the models have mass transfer rates higher than a single critical value $\mdotcr$ when the WD is near $\mch$, that particular model is determined to experience an off-center ignition. The value of $\mdotcr$ is determined by computing a grid of models, where WDs of $\iniMwd= 0.6-1.35$ $\msun$ accrete at different constant rates. The accretion rate above which WD models experience an off-center ignition (which will happen before the WD reaches $\mch$) is then the critical mass transfer rate $\mdotcr$. In the work of \cite{2017MNRAS.472.1593W}, the value of $\mdotcr$ is $\approx 2.05\times10^{-6}$ $\msun$ $\peryr$. Of course, time-dependent mass transfer simulations will show that the WD does not accrete at a constant rate, so the occurrence of an off-center ignition depends on the mass accretion history. As a result, our grid presents a non-negligible, further correction to the upper boundary of the TN SN region, due to accounting for the time-variability of the mass transfer rate. Comparing our fiducial grid (Panel (c) of Figure \ref{fig:grids}) with Figure 7 of \cite{2017MNRAS.472.1593W}, we find that the upper boundary of our grid for $\iniMwd=1.0$ $\msun$ is generally lower than that of \cite{2017MNRAS.472.1593W} by $\iniMhe \approx 0.1 - 0.2$ $\msun$.

\begin{figure}
\fig{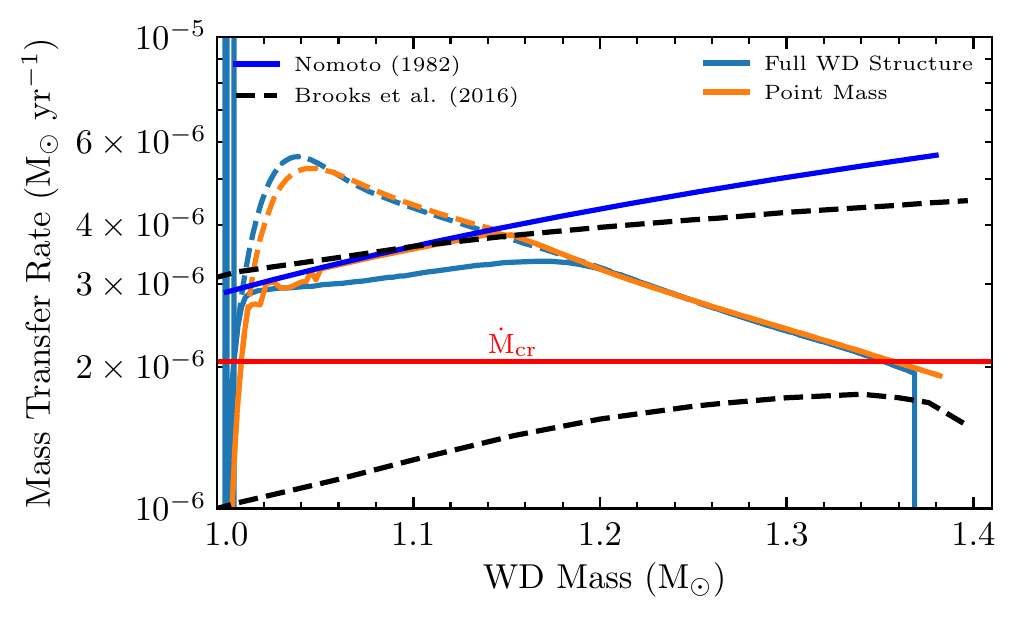}{\columnwidth}{}
\caption{A demonstration of why the $\mdotcr$-prescription of \citet{2017MNRAS.472.1593W} may fail to account for some systems undergoing shell ignitions. Two binary runs are performed at $\inipara=(1.5,1.0,-0.9)$, one from our work (blue) and other adopting the \citet{2009MNRAS.395..847W} prescription (orange). The latter run does not resolve the WD structure, and since $|\mdothe| \leqslant \mdotcr$ as the WD nears $\mch$ the $\mdotcr$-prescription regards this system as a core ignition system, whereas our work resolves the WD structure and suggests this system to be a shell ignition system. We also note that the $\mdotup$ of \citet{1982ApJ...253..798N} may be as much as 10\% above ours.
\label{fig:point_mass_wang}}
\end{figure}

To demonstrate the importance of time-dependent calculations, in Figure \ref{fig:point_mass_wang} we compare two simulations of the system $\inipara = (1.5,1.0,-0.9)$, which is an off-center ignition system in our work but a central ignition system in \cite{2017MNRAS.472.1593W}. One system is taken from our prescription fully resolving the WD. The other takes the WD as a point-mass but with the upper stability line given by \cite{1982ApJ...253..798N}. In both cases, we can observe that the WD accretes at $\mdotup$ for some time, until $|\mdothe|$ falls back into the stable accretion regime. If the occurrence of the off-center carbon ignition is not tracked, when the WD nears $\mch$ the mass transfer rate may eventually fall below the $\mdotcr$ found by \cite{2017MNRAS.472.1593W}. As a result, the lower $\iniMhe$ off-center ignition systems that we have found will be missed by the $\mdotcr$-prescription since $|\mdothe|$ eventually falls below $\mdotcr$. We note that the $\mdotup$ of our prescription is lower than that of \cite{1982ApJ...253..798N}, up to the 10\% level. If we were to adopt the \cite{1982ApJ...253..798N} $\mdotup$, our upper boundary of the TN SN region would have been even lower. 

However, a second cause may be responsible for the difference in the TN SN region upper boundaries. A point-mass calculation shows that, for a $\iniMhe$ slightly higher than in Figure \ref{fig:point_mass_wang}, say $\iniMhe=1.6$ $\msun$, the $\mdotcr$-prescription would also have agreed that the WD will ignite off-center. The only other reason why our grid does not agree with \cite{2017MNRAS.472.1593W} on this model, lies in differences in stellar models. 

Moreover, this difference in the upper boundaries found by us and by \cite{2017MNRAS.472.1593W} varies in degree depending on $\iniMwd$. Comparing our grid of $\iniMwd=0.90$ $\msun$ with Figure 8 of \cite{2017MNRAS.472.1593W}, we find very similar upper boundaries because the off-center ignitions are not important. Instead, the low value of $\iniMwd$ requires further depletion in the donor envelope to grow up to $\mch$. The WD accretes below $\mdotup$ for a longer time, so the compressional heating in the WD shell is less important. The conditions for an off-center ignition are therefore unfavorable. \deleted{Thus, we expect that as $\iniMwd$ increases, the difference in the upper boundaries found by us and by \cite{2017MNRAS.472.1593W} grows. }

We may compare the other boundaries as well. The left boundary is determined by the condition that the He donor is not Roche-filling at the He ZAMS. Comparing our fiducial grid of $\iniMwd=1.0$ $\msun$ with that of \cite{2017MNRAS.472.1593W}, we find that our left boundary is slightly larger by $\lesssim 0.1$ in $\iniP$. This discrepancy is likely to stem from differences in our stellar evolution codes\replaced{, but negligible given that post-common envelope binary systems emerging at very short periods are rare.}{. But whether this is negligible depends on the formation probability distribution of the CO WD-He star binaries -- a higher common envelope ejection efficiency $\alpha \lambda$ used by population synthesis would predict a lower formation rate of short period systems than long period systems (see Section \ref{subsec:formation channel}).} 

\added{At the shortest period ($\iniP=-1.3$ in our grids), we find that the systems undergo case BA then BB mass transfer, which agrees qualitatively with \cite{2009MNRAS.395..847W} (their Case 4 calculations). However, the super-Eddington
wind triggered in our models leads to less growth during case BA mass transfer than in the models of \cite{2009MNRAS.395..847W}. }

The bottom and right boundaries are determined by the systems that undergo helium flashes following stable accretion. In our grids, we compute the required mass retention efficiency, given $\mwd$ and $\mheenv$ when the helium flashes start, for the WD to grow to $\mch$, and contour the grids by setting the required efficiency to be greater than 60$\%$. \cite{2009MNRAS.395..847W} and subsequently \cite{2017MNRAS.472.1593W} follow through the evolution of the WD in successive helium flashes by adopting the mass retention efficiencies computed by \cite{2004ApJ...613L.129K} under the optically-thick wind framework. Thereby, the bottom and right boundaries of \cite{2017MNRAS.472.1593W} may be more thorough by virtue of following through the accretion through helium flashes. Nonetheless, given the uncertainties regarding the helium flash retention efficiency, it is sufficient to observe that our bottom and right boundaries do not show significant deviation from those of \cite{2017MNRAS.472.1593W}. 

\section{The Effect of Enhanced Angular Momentum Loss}
\label{sec:effect of enhanced angular momentum loss}

Previous work, and the models in Section~\ref{sec:fast-wind}, have adopted the assumption that mass is lost from the binary through a fast isotropic wind.  However, a slow wind may gravitationally torque the binary, leading to additional angular momentum loss.  In this section, we investigate the effect of enhanced angular momentum loss on the mass transfer histories and the TN SN region. 


\subsection{Parametrization of Angular Momentum Loss} 
\label{subsec:parametrization of angular momentum loss}

\cite{1999ApJ...522..487H} investigated the specific angular momentum by carried by a spherically symmetric wind blown from a star in a binary. They ejected a number of test particles from the surface of the mass-losing star, at $0.1$ times the inner Roche lobe radius of the star. They evolved the trajectory of the test particles in the co-rotating frame under the Roche potential and Coriolis force, and computed the specific angular momentum carried by the test particles that manage to escape. They found that when the wind speed is on the order of the binary orbital speed, $a \binaryomega$, the wind gravitationally torques the binary and extracts more angular momentum. They found the angular momentum parameter $\lw$,
which is defined as
\begin{equation}
    \left(\frac{\jdotw}{\mdotw}\right) = \lw a^2 \binaryomega~,
\end{equation}
varies as
\begin{equation}
\lw = \max \left\{ 1.7 - 0.55 \left( \frac{\vrl}{a \binaryomega}\right)^{2}, \left( \frac{q}{1+q} \right)^{2} \right\},
\end{equation}
where $\vrl$ is the radial velocity of the wind at the Roche lobe of the mass losing star, and the limiting value of $1.7$ was cited from previous restricted three-body problem (\citealt{1975A&A....43..309N}, \citealt{1976PASJ...28..593N}) and two-dimensional (equatorial plane) hydrodynamical results \citep{1984MNRAS.206..673S}. 
\citet{2016ApJ...821...28B} used the results of \citet{1999ApJ...522..487H} to suggest that wind velocities $\gtrsim 1000\,\kms$ were required to justify the fast wind assumption (see their Figure 4).
As noted by \cite{1993ApJ...410..719B}, at slow wind speeds complex trajectories result, and therefore a hydrodynamical approach likely needs to be adopted. Therefore, we view the use of results for $\vrl/a \binaryomega \la 2$ from \cite{1999ApJ...522..487H} with some caution.

\cite{2005A&A...441..589J} performed a three-dimensional hydrodynamic calculation in the co-rotating frame for the case where the mass-losing component fills half of its Roche lobe, for various initial wind speeds and mass ratios. They also conclude that slow wind speeds can significantly shrink the binary orbit. However, their conclusion is that the specific angular momentum carried by a wind outflow is smaller than that found by \cite{1999ApJ...522..487H}; the functional dependence of the wind specific angular momentum on the ratio of wind radial velocity at the Roche lobe $\vrl$ to binary orbital speed $a \binaryomega$, is also different. For the case of $q = 1$, they find that the wind specific angular momentum is
\begin{equation}
\lw = 0.25 + \frac{0.12}{\vrl/a \binaryomega + 0.02}~.
\label{eq:janahara}
\end{equation}
The 0.25 represents the fast wind limit of $[q/(1+q)]^2$ for $q = 1$.
The binaries we consider typically have
$0.5 \lesssim q \lesssim 2$, so we make the rough approximation
that Equation~\eqref{eq:janahara} continues to hold.
We then separately apply the fast wind limit
(i.e., that $\lw$ cannot fall below $[q/(1+q)]^2$) to this expression.

\begin{figure*}
\gridline{\fig{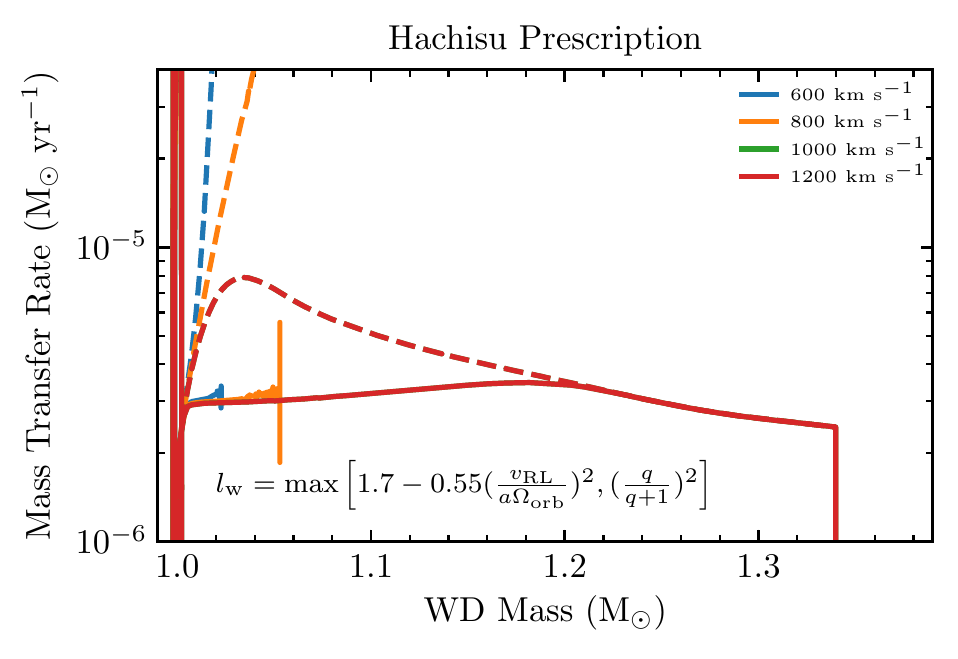}{0.5\textwidth}{(a)}
          \fig{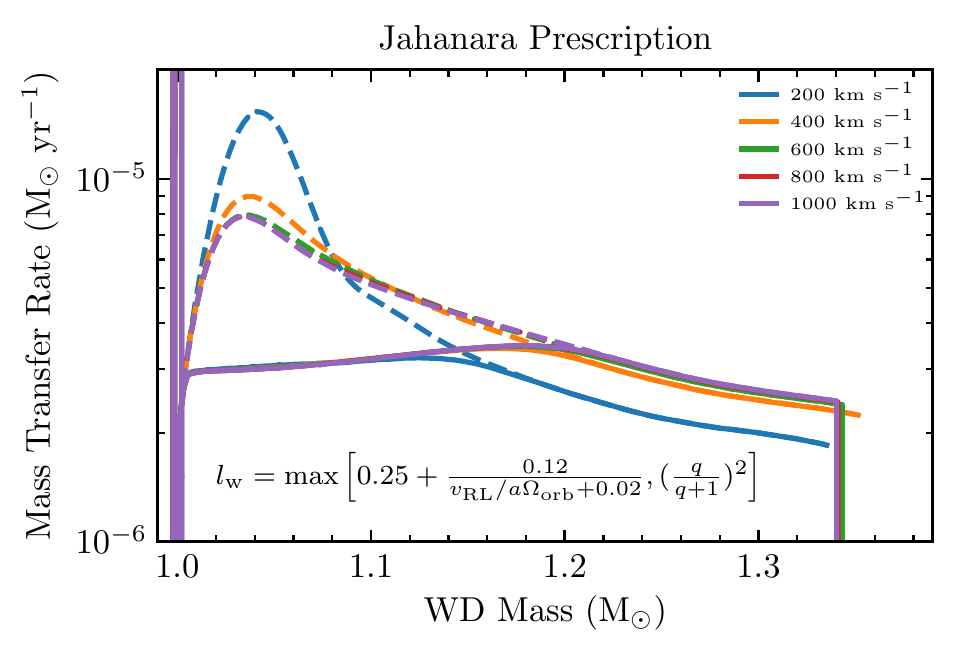}{0.5\textwidth}{(b)}
          }
\caption{Two plots showing the mass transfer histories of binary runs adopting the Hachisu prescription (a) and the Jahanara prescription (b). For the system $\inipara=(1.6,1.0,-0.9)$ where $a\binaryomega\approx600$ $\kms$, a mass transfer runaway occurs for a wind speed (measured radially at the Roche radius) of $\vw\lesssim 900$ $\kms$ assuming the Hachisu prescription, and a much lower $\vw<200$ $\kms$ assuming the Jahanara prescription. \label{fig:hachisu_jahanara}}
\end{figure*}

We perform binary calculations with both the Hachisu and Jahanara prescriptions, using $\inipara$ $= (1.6,1.0,-0.9)$. We vary the assumed radial wind speed at the Roche lobe $\vrl$ (where the binary orbital speed for this system is $a \binaryomega \approx 600$ $\kms$), and the results are shown in Figure \ref{fig:hachisu_jahanara}. Panel (a) shows the calculations adopting the Hachisu prescription, and we find a bifurcation at a wind speed of $\approx 900$ $\kms$, above which a mass transfer runaway and subsequently a merger will likely result. In Panel (b), the Jahanara prescription only leads to a noticeable change in the mass transfer history at a wind speed of $\approx 200$ $\kms$, below which we estimate that a mass transfer runaway will likely result. We note here that both test-particle and hydrodynamic calculations would likely suggest that mass loss in the red-giant regime through the RLOF scenario (corresponding to $\vrl \approx 0$), as briefly mentioned by \cite{2016ApJ...821...28B}, would lead to a mass transfer runaway. 

However, when investigating the effect of enhanced wind angular momentum loss on the TN SN region, we prefer to be agnostic about the physical mechanism regarding the wind angular momentum loss. We have chosen to parametrize this via a variant of the $\gamma$ formalism \citep{2000A&A...360.1011N}. Instead of using the total change in binary angular momentum and binary mass, we use the angular momentum and mass loss rates, and parametrize the angular momentum loss with $\gamma$ as follows
\begin{equation}
\frac{\jdotw}{J} = \gamma \frac{\mdotw}{M}, 
\end{equation}
which corresponds to 
\begin{equation}
\lw = \gamma \frac{q}{(1+q)^{2}},
\end{equation}
so the fast wind assumption corresponds to $\gamma = q$.

\begin{figure*}
\gridline{\fig{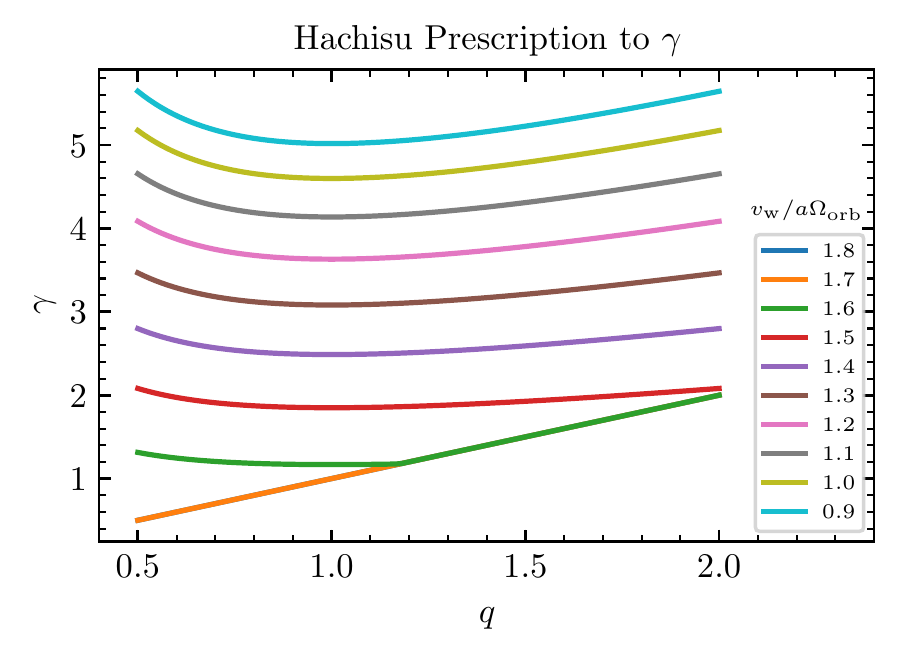}{0.5\textwidth}{(a)}
          \fig{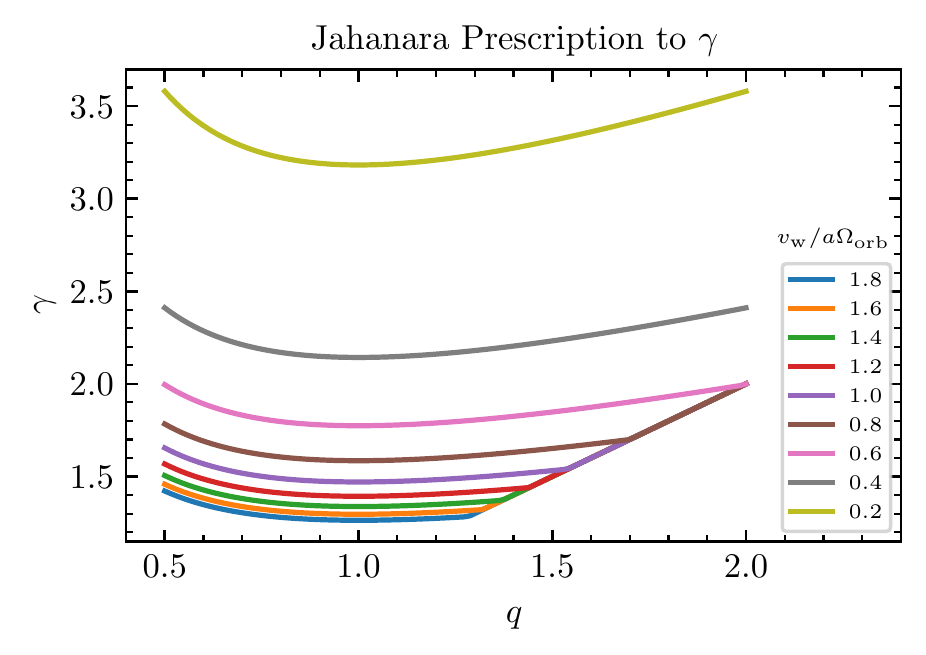}{0.5\textwidth}{(b)}
          }
\caption{The values of $\gamma$ as a function of mass ratio $q$ and ratio of wind speed over orbital speed $\vw/a\binaryomega$, assuming the Hachisu (a) and Jahanara (b) prescriptions. The straight line cutoff at the bottom is due to the fast wind limit. The limit $\vw=0$ corresponds to $\gamma\approx 8$ and $\gamma\approx 25$ for the Hachisu and Jahanara prescriptions respectively. 
\label{fig:to_gamma}}
\end{figure*}

In Panels (a) and (b) of Figure \ref{fig:to_gamma}, we provide the value of $\gamma$ as a function of mass ratio $q$, given a certain ratio of wind speed over binary orbital speed $\vw/a \binaryomega$. That is, given a mass ratio and value of $\vw/a \binaryomega$, we find the value of wind angular momentum parameter $\lw$ assuming either the Hachisu or Jahanara prescriptions, and then invert to find the corresponding value of $\gamma$.  Similarly,
if future work develops a new prescription, its effective value of $\gamma$
can be computed and then compared with our results.

\subsection{The Effect of Enhanced Wind Specific Angular Momentum Loss on the Mass Transfer History} \label{subsec:the effect of gamma}

Now we examine the effect of additional wind angular momentum loss on the mass transfer for a given period and donor mass. We illustrate this by performing binary calculations with $\inipara$ $= (1.6,1.0,-0.9)$, while varying the angular momentum loss parameter $\gamma$. 

\begin{figure}
\gridline{\fig{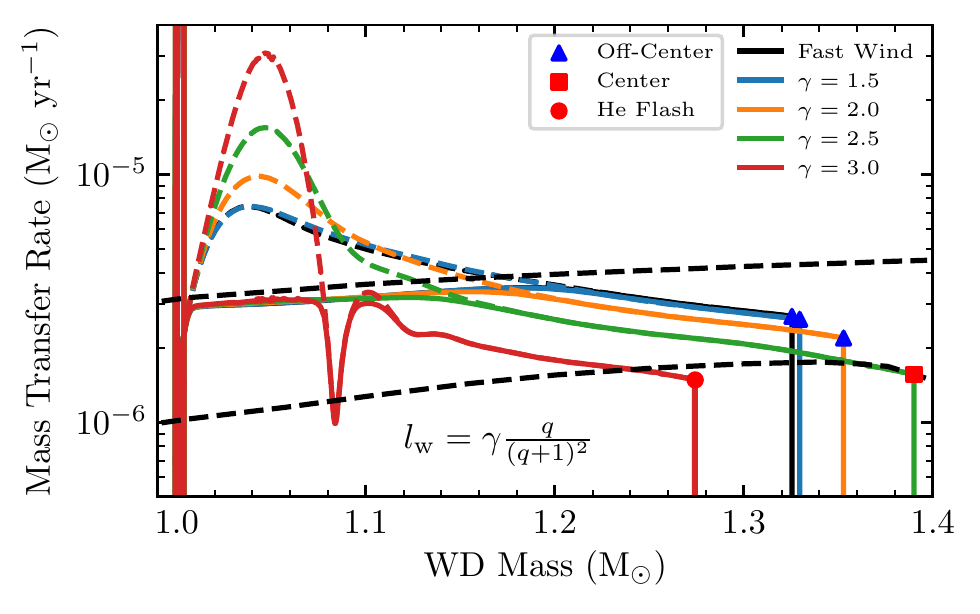}{\columnwidth}{}
        }
\caption{The mass transfer histories of runs at $\inipara=(1.6,1.0,-0.9)$ of various values of $\gamma$, ranging from 1.5 to 3.0. As $\gamma$ increases, $|\mdothe|$ increases initially but is lower at later times. A core ignition is thus favored at higher $\gamma$. Also, at the largest values of $\gamma$ shown, the rapid mass transfer throws the donor envelope out of thermal equilibrium, leading to a time-dependent adjustment of $|\mdothe|$.
\label{fig:gamma}}
\end{figure}

Figure \ref{fig:gamma} shows the results of several values of $\gamma$. The base reference is the fast wind case, where the WD undergoes an off-center carbon ignition. The evolution of the $\gamma=1.5$ case is almost identical to that of the fast wind case, since the fast wind case implies a value of $\gamma=q$, and during the early phase of mass transfer, where wind mass loss and wind angular momentum loss peak, the mass ratio is very close to $q=1.6$. 

As the value of $\gamma$ increases, the specific angular momentum carried by the wind increases, leading to an increase in the peak mass loss rate.
This has several consequences on the mass transfer in the binary. First,
the required mass loss rate may exceed that able to be launched in a wind (see Section~\ref{sec:properties of OTW}); a common envelope may form when the wind-driving process is inefficient. On the other hand, if a wind is successfully launched despite the larger $\mdotw$, then the WD still accretes at $\mdotup$, but the donor is left with less mass to transfer at later times due to this rapid stripping at the beginning. The donor is left with less envelope mass, leading to lower $|\mdothe|$. In other words, higher wind angular momentum loss leads to higher $|\mdothe|$ initially and lower $|\mdothe|$ at later times. Since the WD accretes at $\mdotup$ anyways, on average the WD accretes at a lower rate for a higher wind specific angular momentum. From previous discussion we see that this means less compressional heating in the envelope and a core ignition becomes more favorable. Another possibility is, however, that the donor envelope is effectively stripped that the donor underfills its own Roche lobe again. Then we will obtain a detached double WD binary. 

In addition, when the wind carries high specific angular momentum, for example, $\gamma=3$, then the donor may encounter difficulty adjusting its thermal structure to the rapid mass loss. When the mass transfer timescale comes close to, or is even shorter than, the donor's Kelvin-Helmholtz timescale, the donor envelope may be thrown out of thermal equilibrium. Then we observe time-dependent behavior in the donor. When the donor is out of thermal equilibrium, it may only be able to adjust its thermal structure after its envelope mass has been reduced by mass transfer, after which it may overfill its Roche lobe again. This interplay between mass transfer and thermal adjustment is observed in our models for the donors at the shorter periods and with higher masses. The effect of the mass transfer variability due to the donor's thermal response can be seen in the $\gamma=3$ case, where the donor mass transfer rate may at times drop below $\mdotup$. In general this leads to lower compressional heating, and favors a core ignition. However, as noted before, it is also likely that the donor will eventually be stripped of its envelope and form a detached double WD binary. 

\subsection{The Effect of Enhanced Angular Momentum Loss on the TN SN Region}

\begin{figure*}
\gridline{\fig{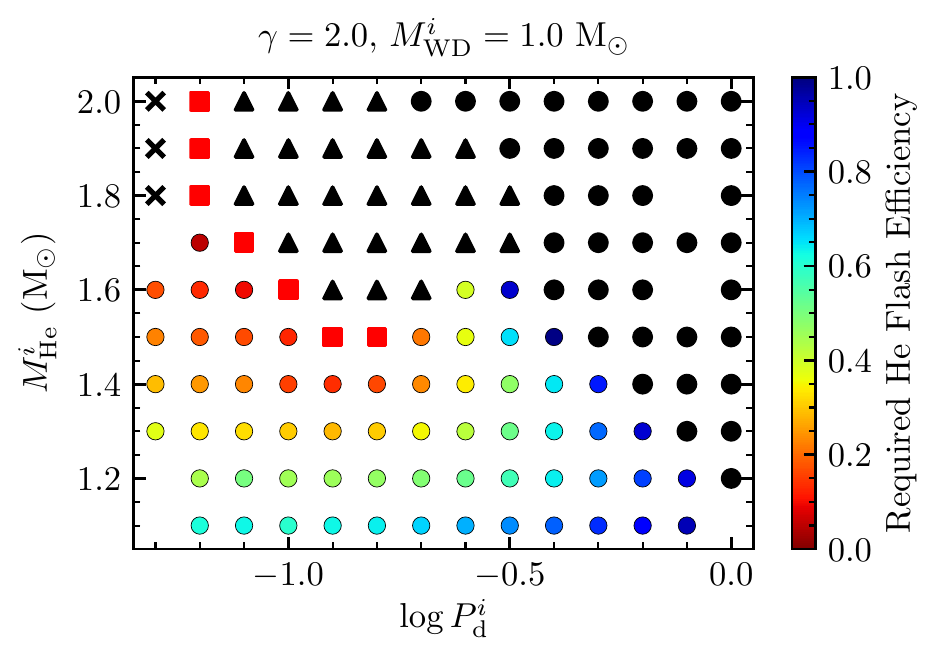}{0.5\textwidth}{(a)}
          \fig{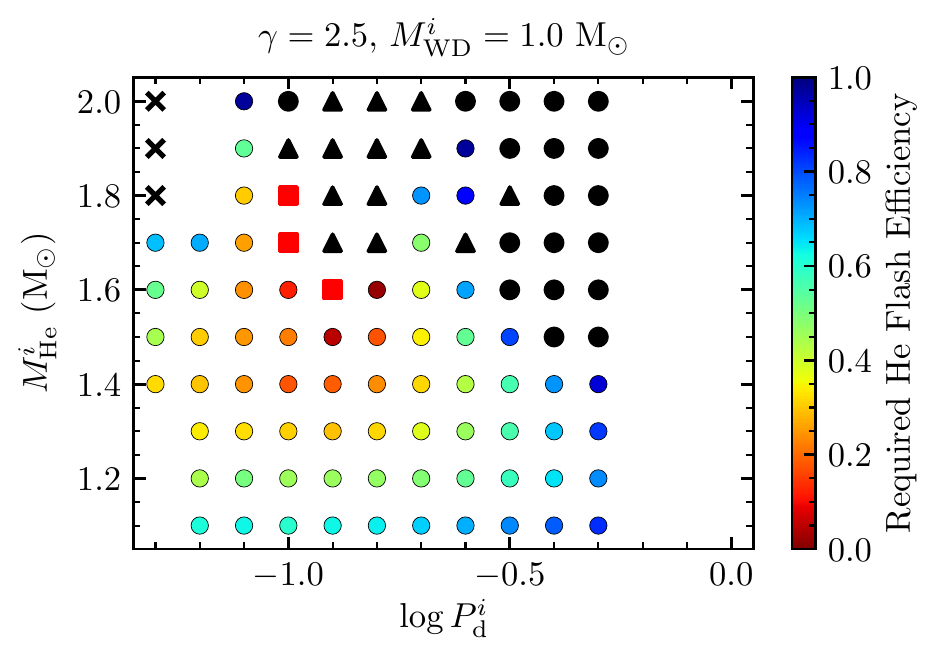}{0.5\textwidth}{(b)}
          }
\gridline{\fig{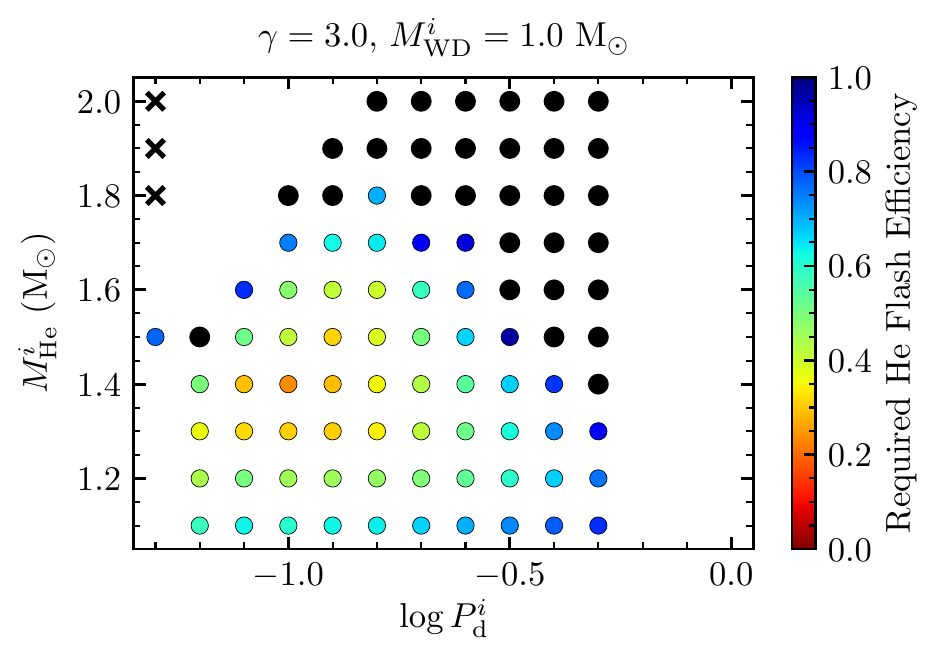}{0.5\textwidth}{(c)}
          }
\caption{ Similar to Figure \ref{fig:grids}, but with a fixed $\iniMwd=1.0$ and different values of $\gamma$. We observe that the TN SN region grows at $\gamma=2.0$ compared to that at the fast wind limit, but shrinks for larger values of $\gamma$ due to more systems experiencing mass transfer runaways.  The empty spots on the top left corner are systems undergoing mass transfer runaways, which the energy budget shows will likely end up in an common envelope. We have not run through the models at the bottom left corner since these systems do not experience wind mass loss. \label{fig:gamma_grids}
}
\end{figure*}

Now we move on to describe the effect of additional wind angular momentum loss on the TN SN region.  With greater angular momentum loss from the system, the peak mass transfer rate is higher, as explained previously. This has several global effects on the parameter space which we show via grids run at different $\gamma$ in Figure~\ref{fig:gamma_grids}.

As is observed in the $\gamma=2$ grid (panel a), the boundary between core and off-center carbon ignitions moves to higher donor mass at the shorter periods (compared to the fiducial Figure~\ref{fig:grids}, panel c). This is the result of a mass transfer variability due to the donor's thermal response. The lag between mass transfer depleting the donor envelope and the donor envelope's thermal adjustment to mass loss leads to large variations in $|\mdothe|$, but on average contributes to lower $\mdotwd$ and thus avoids an off-center ignition in the WD. 

However, for even stronger angular momentum loss ($\gamma = $ 2.5 \& 3, panels b \& c), the short period and high mass donor region leads to $|\mdothe|$ so high that it is likely that either a mass transfer runaway and hence a common envelope occurs, or the donor is rapidly stripped of its envelope to form a detached double WD binary. 

The same can be said for the long period regions. The regime for detached double WD binary slightly broadens with wind specific angular momentum, due to greater mass loss from the donor as a result of additional angular momentum loss.

While the regime for helium flashes is in general unchanged since wind mass loss is insignificant, the TN SN region slightly broadens (for $\gamma=2$) but then shrinks (for $\gamma = 2.5$ \& $3$) as $\gamma$ goes up. In fact, the missing systems in the top left corner of the $\gamma=2.5$ \& $3$ grids are likely systems undergoing mass transfer runaways. A calculation of the energy and momentum budgets shows that these systems are unlikely to sustain very high wind mass loss rates, and thus may end up in a common envelope. If the wind specific angular momentum goes up even more, it is likely that all systems on the grid will form a common envelope, for which the final outcome is unclear but
seems unlikely to be a TN SN.

Nevertheless, simply by observing the change from the fast wind grid through the $\gamma = 3$ grid, we may see that the parameter space for core ignitions, if a common envelope is not formed, remains relatively unchanged -- the only boundary affected is, as expected, the upper boundary where wind mass loss occurs. The upper boundary shifts by a model or two, but does not lead to a qualitative change. This is because a change of $\approx 0.1$ $\msun$ in $\iniMhe$ is sufficient to introduce a change in the WD accretion rate affecting the occurrence of off-center ignition. Therefore, either strong angular momentum loss leads to the formation of a common envelope for all systems, or even moderate angular momentum loss can only lead to slight shifts in the TN SN region.

\section{Properties of Optically-Thick Winds}
\label{sec:properties of OTW}

Throughout this paper, we invoke the presence of an optically-thick wind (OTW) that removes any donated mass in excess of $\mdotup$ from the binary system.  This wind mass loss rate was allowed to be arbitrarily high.  In Section~\ref{sec:budget}, 
we compute the required energy and momentum needed for the wind to be launched and compare this to the properties of observed OTWs in Wolf-Rayet stars.  In Section~\ref{sec:kh}, we provide some estimates of the structure and properties of these OTWs by formulating simple steady-state wind solutions following the approach of \cite{1994ApJ...437..802K}.  
In Section~\ref{sec:otw-solns},
we comment on the likelihood of wind launching in our models based on these constraints.

\subsection{Energy and Momentum Budget}
\label{sec:budget}

Energy and momentum conservation constrain the occurrence of mass loss from the binary. The kinetic energy of the wind must be provided by the luminosity of the WD, possibly with the help of the orbital energy of the binary if the wind torque is significant. For now we will assume the fast wind limit such that the wind does not torque the binary as it leaves the system. Then, we can find the required efficiency factor, $\eta$, for converting radiative power in the luminosity of the WD to the kinetic power of the wind from the equality $\mdotw \vw^{2} = 2 \eta \Lwd$. Adopting a wind velocity of $\vw = 1000\,\kms$ we have
\begin{equation}
\eta \approx 0.03
\left(\frac{\mdotw}{10^{-5}\,\msun\,\rm yr^{-1}}\right)
\left(\frac{\Lwd}{5\times10^4\,\lsun}\right)^{-1}~.
\label{eq:eta}
\end{equation}
For these representative fiducial parameters, powering the wind requires only a few percent of the luminosity of the WD.

\begin{figure*}
\gridline{\fig{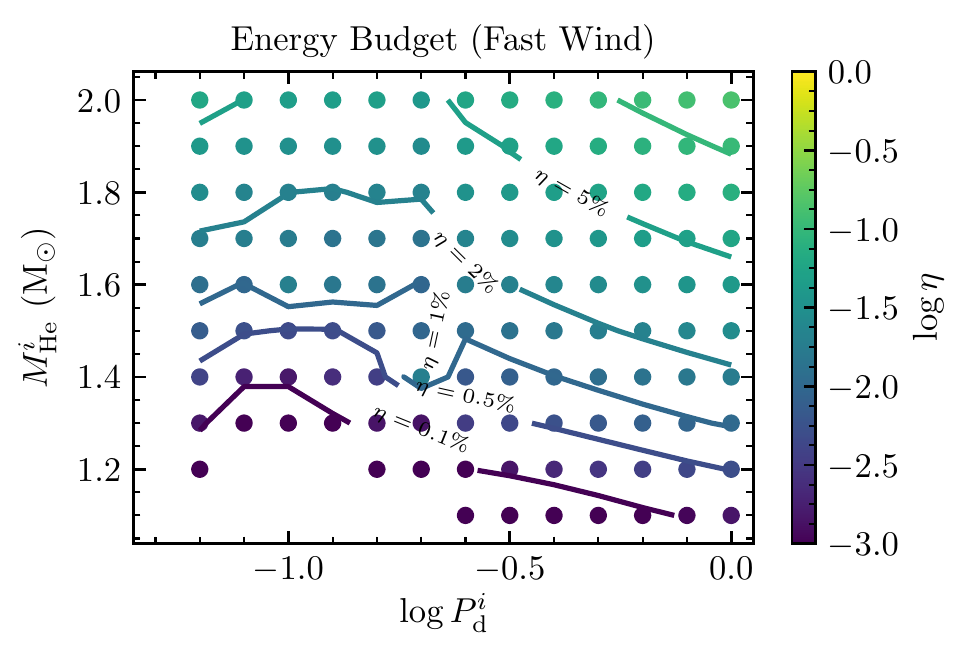}{0.5\textwidth}{(a)}
          \fig{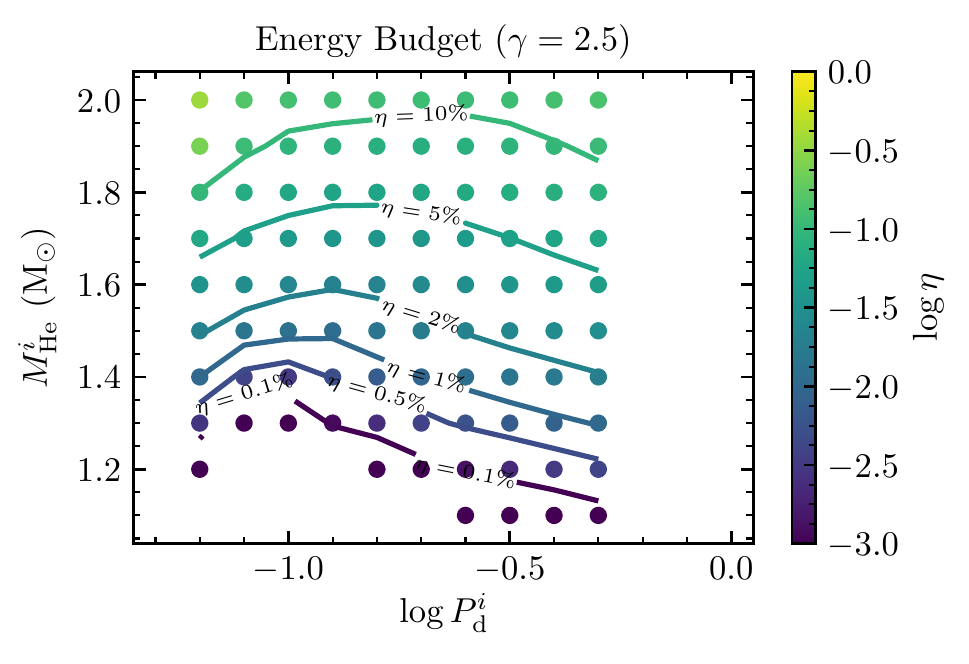}{0.5\textwidth}{(b)}
          }
\gridline{\fig{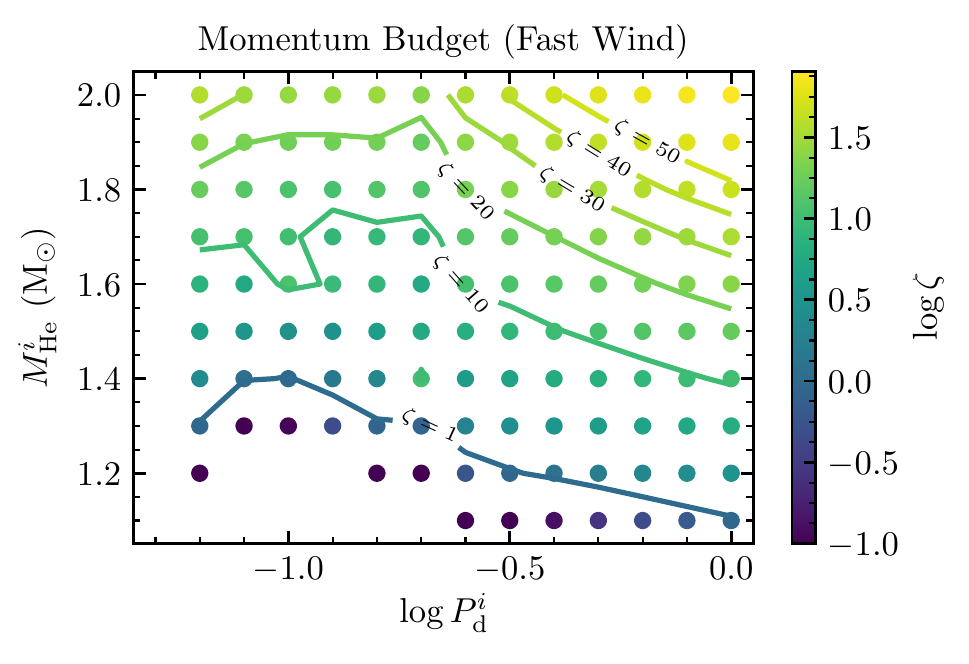}{0.5\textwidth}{(c)}
          \fig{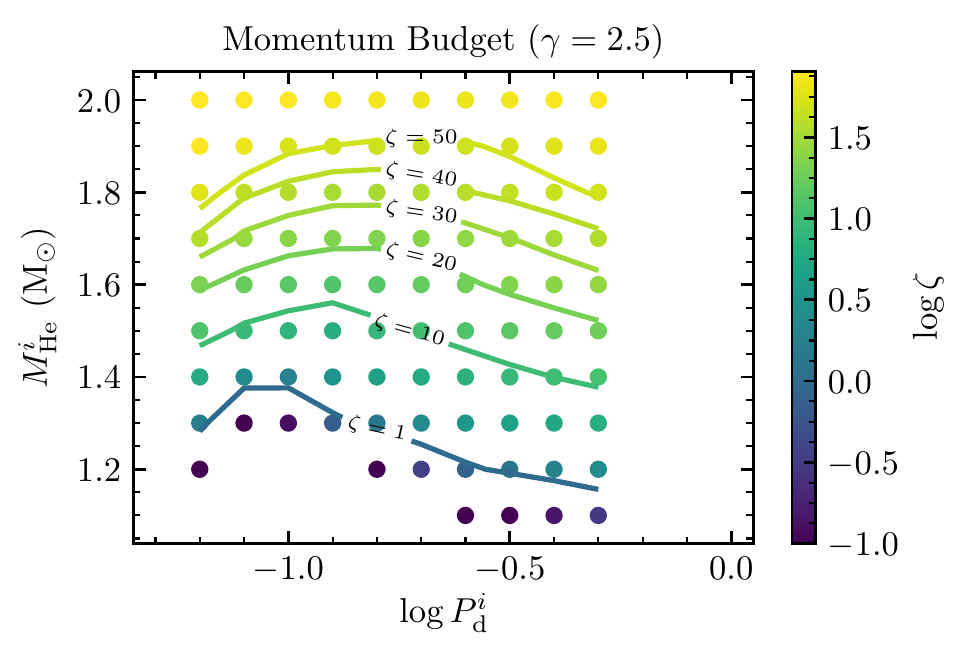}{0.5\textwidth}{(d)}
          }
\caption{Plots showing the energy and momentum budgets of the binary runs. Panels (a) and (b) compare the energy budgets of the fast wind grid and the $\gamma=2.5$ grid assuming a wind speed of $\vw=1000$ $\kms$; in some systems of the $\gamma=2.5$ grid the maximum wind kinetic energy may be as high as 10\% of $\Lwd$. Panels (c) and (d) compare the momentum budgets. We view the systems with $\zeta>50$ unlikely drivers of a wind, based on the observed limits of Wolf-Rayet stars. 
\label{fig:eta_zeta}}
\end{figure*}

In Figure \ref{fig:eta_zeta} we show the maximum value of $\eta$ during the mass transfer, for each binary model in the fast wind grid (panel a) and the $\gamma=2.5$ grid (panel b). We find that in order to drive a wind of wind speed $\vw = 1000$ $\kms$, for the fast wind grid at most a $\approx 10\%$ minimum energy transfer efficiency is required, whereas some systems in the $\gamma=2.5$ grid require a $\approx 30\%$ minimum energy transfer efficiency. The systems with required efficiency of tens of percent will likely face a tight energy constraint and may become inefficient in driving a wind. For the fast wind grid, this occurs mostly for the high mass donor and long period systems.  For the $\gamma=2.5$ grid, the high mass donors at very short periods also face the same constraint.  However, under the assumption of a successful wind, these systems all form detached double WD binaries.  Therefore, while a failed wind might suggest instead a common envelope, this difference does not directly affect our identification of which systems undergo a core ignition.

However, the value of $\eta$ in Equation~(\ref{eq:eta}) is sensitive to our choice of $\vw$.  The fiducial wind speed of $1000\,\kms$ is consistent with the fast wind assumption (of order the orbital speed).  In Section~\ref{sec:kh} we will use our OTW models to further justify this choice: because the wind is launched from the iron bump, the wind launching radius has a much lower escape velocity than the surface of the WD.  If instead, the wind were launched near the burning shell, or approximately $\Rcwd$ ($\approx 0.008\,\rsun$) then the escape speed would be $\vesc = \sqrt{G \mwd / \Rcwd} \approx $ 7000 $\kms$ for a 1 $\msun$ WD.    This would imply that the systems with $\log \eta \gtrsim -1.7$ in Figure~\ref{fig:eta_zeta} would not be energetically able to drive a wind.  The high mass systems still face stringent energy constraints on wind-driving, but again, either they face the fate of common envelope, or assuming successful wind-driving, the fate of an off-center ignition in the WD.

We can also ask whether $\Lwd$ can supply sufficient momentum to the wind to drive the outflow. 
In this case we can define the required momentum efficiency factor, $\zeta$, from the equality $\mdotw \vw c = \zeta \Lwd$.  
Again adopting a wind velocity of $\vw = 1000\,\kms$ we have
\begin{equation}
\zeta \approx 10
\left(\frac{\mdotw}{10^{-5}\,\msun\,\rm yr^{-1}}\right)
\left(\frac{\Lwd}{5\times10^4\,\lsun}\right)^{-1}~.
\end{equation}
In this case, the required momentum transfer efficiency for the fiducial parameters is significantly greater than unity.  This then requires the presence of multiple scattering in order to extract sufficient momentum from the radiation field.  The winds in Wolf-Rayet stars often exhibit $\zeta \sim 10$, where this can be physically explained by wind launching at an optical depth $\tau \sim \zeta$ \citep[][and references therein]{2002A&A...389..162N}. 
Thus values of $\zeta \gg 1$ are consistent with our assumption of an OTW, in which the acceleration region occurs near the iron-bump at relatively high optical depth.

Some Wolf-Rayet stars have reported momentum efficiencies $\approx 50$
\citep{1995A&A...299..151H}, though inferred mass loss rates
may now be a factor of a few lower after accouting for clumping
\citep[e.g.,][]{1998A&A...335.1003H, 2014ARA&A..52..487S}.
On this basis, allowing values of $\zeta$ up to 50 in our mass loss prescription leads to only a few binary systems that would be deemed inefficient in driving a wind outflow, and thus likely enter a phase of common envelope evolution.
Figure \ref{fig:eta_zeta} shows the maximum value of $\zeta$ during the mass transfer, for each binary model in the fast wind grid (panel c) and the $\gamma=2.5$ grid (panel d).
The systems that approach or exceed $\zeta = 50$ are the highest mass donors, which assuming successful wind-driving would most likely lead to an off-center ignition in the WD or form a detached double WD binary.
Therefore, our assumptions about the momentum efficiency do not affect our conclusions about core ignitions
unless we restrict $\zeta \la 10$.

However, some past work does indirectly enforce a restrictive constraint on $\zeta$ in the binary evolution
\added{\citep[e.g.,][]{2000A&A...362.1046L, 2013A&A...558A..39T}}.
Recall that the Eddington mass transfer rate can be defined by asking when the rate of energy release of the accreted material (via both the liberation of gravitational potential energy and nuclear burning) reaches the (electron-scattering) Eddington luminosity \cite[e.g.,][]{2013A&A...558A..39T}.  For helium accretion on a WD this is $\mdotedd \sim 3 \times 10^{-6}\,\msunyr$.  Note that this is roughly an order of magnitude larger than for hydrogen accretion because of the lower specific nuclear energy release and the lower electron scattering opacity.
For hydrogen accretion, WDs happen to have the interesting property that $\vesc c / \epsilon_{\rm nuc} \sim 1$ \citep{2000A&A...362.1046L}.  In our case for helium accretion and a wind velocity below the escape velcocity of the WD surface, we similarly have $\vw c / \epsHe \sim 1$.  These quantities being of order unity implies that when $\mdotw \sim \mdotedd$, the wind momentum is of order the photon momentum, that is $\zeta \sim 1$.  Based on arguments along these lines, some past work has assumed that material cannot be efficiently lost from the system if $\mdotw > 3 \mdotedd$,
and thus above this mass transfer rate a common envelope results \citep{2000A&A...362.1046L, 2013A&A...558A..39T}.
In contrast, in our work we impose no cap on $\mdotw$.  Physically, we emphasize that this is equivalent to the assumption that $\zeta \gg 1$ is allowed via mulitple scattering.

\subsection{Wind Equations}
\label{sec:kh}

OTW solutions have been calculated in the context of hydrogen and helium nova outbursts by \citet{1994ApJ...437..802K,2004ApJ...613L.129K}.  We follow their approach in solving the equations for a spherically-symmetric, steady-state wind.  The continuity equation is
\begin{equation}
  \dot{M} = 4 \pi r^2 \rho v~,
  \label{eq:continuity}
\end{equation}
and the momentum equation is
\begin{equation}
  v\ddr{v} + \frac{1}{\rho} \ddr{P} + \frac{GM}{r^2} = 0 ~.
  \label{eq:momentum}
\end{equation}

We assume that the material has the equation of state of an ideal gas
plus radiation, so the pressure is
\begin{equation}
  P = P_{\mathrm{gas}} + P_{\mathrm{rad}} = \frac{\rho \kB T}{\mu \mb} + \frac{1}{3} \crad T^4 ,
\end{equation}
and the enthalpy is
\begin{equation}
  w = u + P = \frac{5}{2}\frac{\rho \kB T}{\mu \mb} + \frac{4}{3} \crad T^4 .
\end{equation}
Energy conservation implies
\begin{equation}
  L + \dot{M} \left(\frac{1}{2}v^2 + w - \frac{GM}{r}\right) = \Lambda ~
\end{equation}
where $\Lambda$ is a constant

We assume that energy transport via convection is unimportant, and so
the temperature gradient is set by radiative diffusion,
\begin{equation}
  \dlndlnr{T} = - \frac{3 \kappa \rho L}{16 \pi \crad c r T^4} ~.
  \label{eq:luminosity}
\end{equation}
The velocity gradient can be derived by taking the derivative of
Equation~(\ref{eq:continuity}) and combining it with Equation~(\ref{eq:momentum})
which gives
\begin{equation}
  \dlndlnr{v} = \frac{\frac{2\Pgas}{\rho} - \frac{GM}{r} + \left(\Pgas+ 4 \Prad \right) \dlndlnr{T}}
  {v^2 - \frac{\Pgas}{\rho}}~.
  \label{eq:crit-point}
\end{equation}
For a transonic solution, the numerator and denominator must
simultaneously vanish.  Therefore, this condition defines two
constraints at the critical point.

\begin{figure}
  \includegraphics[width=\columnwidth]{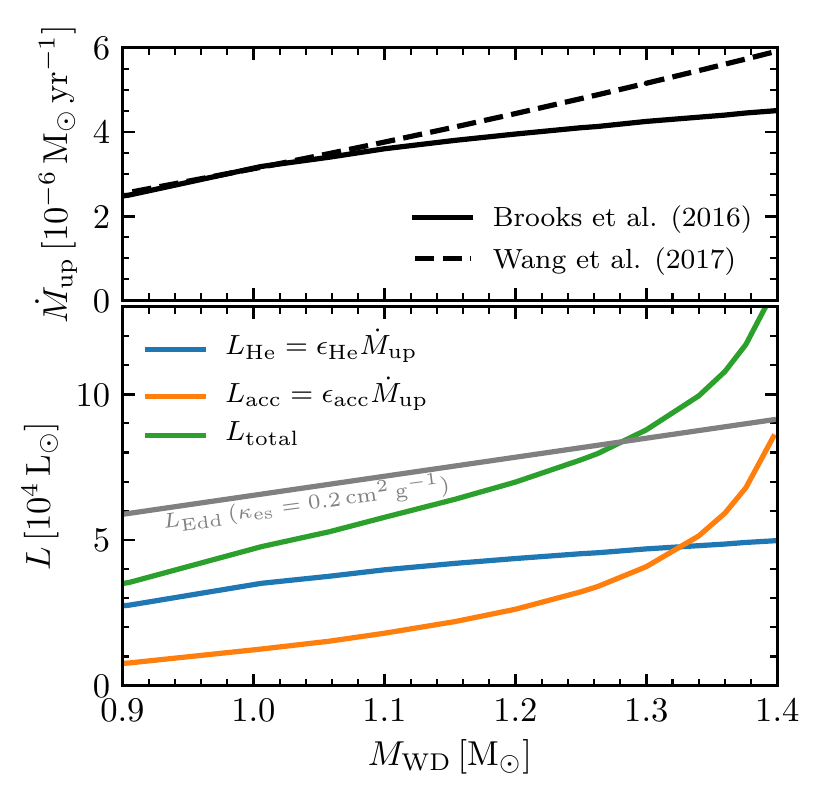}
  \caption{Relationship between the mass of the WD and the luminosity assumed in the OTW calculation.  The upper panel compares the assumed values of $\mdotup$.  Curves in the lower panel use the \citet{2016ApJ...821...28B} value. The lower panel shows the contribution to the total luminosity from helium burning and accretion. The grey line shows the electron-scattering Eddington luminosity for helium. \label{fig:mdot-and-L}}
\end{figure}

In the nova wind case, the goal is to construct a sequence of steady-state wind solutions
that connect the mass loss rate to the envelope mass.  
However, in this case, we already know the wind mass loss rate, as it assumed to be $\mdotw = |\mdothe|- \mdotup$.
We also know the luminosity, as this is set by the energy release of the material retained on the WD.  Therefore we can write
\begin{equation}
  L = \left(\epsHe + \epsacc\right) \mdotup
\end{equation}
The first term is the specific energy release from helium burning.  We
use the formula given in \citet{2002RvMP...74.1015W},
\begin{equation}
  \epsHe = \left(5.85 + 2.86 X_{\rm O}\right) \times 10^{17}\rm\,ergs\,g^{-1}\,s^{-1}~,
\end{equation}
where we take the final mass fraction of $^{16}{\rm O}$ to be
$X_{\rm O} \approx 0.3$.  The specific energy of accretion is
$\epsacc = G \mwd / \Rcwd$.  For the (cold) WD radius we use the
fitting formula from \citet{2000MNRAS.315..543H}.  The lower panel of
Figure~\ref{fig:mdot-and-L} shows how these luminosities change with
WD mass.  For $\mwd \gtrsim 1.3\,\msun$, the accretion luminosity begins to play a
dominant role and the total luminosity approaches the Eddington limit.  Note however,
that in the binary evolution models, WDs with these masses do not generally have OTWs (see Figure~\ref{fig:fiducial_case}).  Thus, the relevant luminosities are generally sub-Eddington (with respect to electron scattering) and dominated by energy release from helium burning.

Therefore, given an $\mwd$ and an $\mdotw$, we can find 
the desired wind solution via the following procedure.  First,
we make a guess for the temperature at the critical point, \Tcr.
Then, we use the vanishing of the denominator of
Equation~(\ref{eq:crit-point}) to calculate \vcr.  Using the known value of
$\mdotw$, we use Equation~(\ref{eq:continuity}) to eliminate \rcr\ in favor of
\vcr\ and \rhocr.  The numerator of Equation~(\ref{eq:crit-point}) must
also be zero at the critical point, and so we numerically solve for
the value of \rhocr\ that satisfies this constraint.  We then know all
the relevant values at the critical point.

The next step will be to integrate outwards until we reach the
photosphere, which is defined by $\tau = \kappa \rho r \approx 2.7$
\citep[see Appendix A in][]{1994ApJ...437..802K}.  Then, at the photosphere, we check if the radiative
luminosity (defined via Equation~\ref{eq:luminosity}) matches the blackbody
luminosity ($L_{\rm BB} = 4 \pi r^2 \sigma T^4$). We iterate on \Tcr\
until this condition is satisfied.

In solving these equations, we make use of the \mesa\ opacities, which
in practice are provided by OPAL \citep{1996ApJ...464..943I}
at solar metalicity ($Z = 0.02$, abundance pattern from \citealt{1998SSRv...85..161G}).
Once performed, this
procedure gives us the full structure of the wind between the critical
point and the photosphere.

It is worthwhile to remember that this model has made a number of
significant simplifications.
We assume a spherically symmetric wind.  This neglects the gravitational influence of the companion (which is negligible
far inside the Roche lobe) and the flow of mass donated by the companion (which presumably has
significant influence in the vicinity of the orbital plane at essentially all radii).  For more on this latter point, see Section~\ref{sec:discussion}.
Another caveat of the wind models used here is that energy transport via convection is not accounted for (Equation \ref{eq:luminosity} assumes only radiative diffusion).  The iron group opacity bump may lead to a 
convectively unstable region, and thus for a significant
convective luminosity roughly coinciding with the acceleration region.
Section 6.4 in \citet{1994ApJ...437..802K} discusses this point
in more detail, but importantly finds that the presence small convective
regions does not significantly affect the overall wind structure.\footnote{This conclusion too has its caveats, as it is based on
one-dimensional mixing length theory.  In this region, the convective eddy velocity will most likely be comparable to the local adiabatic sound speed which may drive shocks and lead to an inhomogeneous medium, at which point the assumptions that underpin MLT are breaking down.}
The treatment of radiation in the diffusion approximation
is also manifest in the momentum equation (Equation~\ref{eq:momentum}).
Near the critical point (at relatively high optical depth) the CAK-type line force is negligible, but will eventually become dominant at some larger radius (see Section 2.3 in \citealt{2002A&A...389..162N}).
Fully addressing the structure of this wind would require 3D calculations with coupled hydrodynamics and co-moving frame radiative transfer,
far beyond this scope of the current work.

\subsection{Wind Solutions}
\label{sec:otw-solns}

\begin{figure}
  \includegraphics[width=\columnwidth]{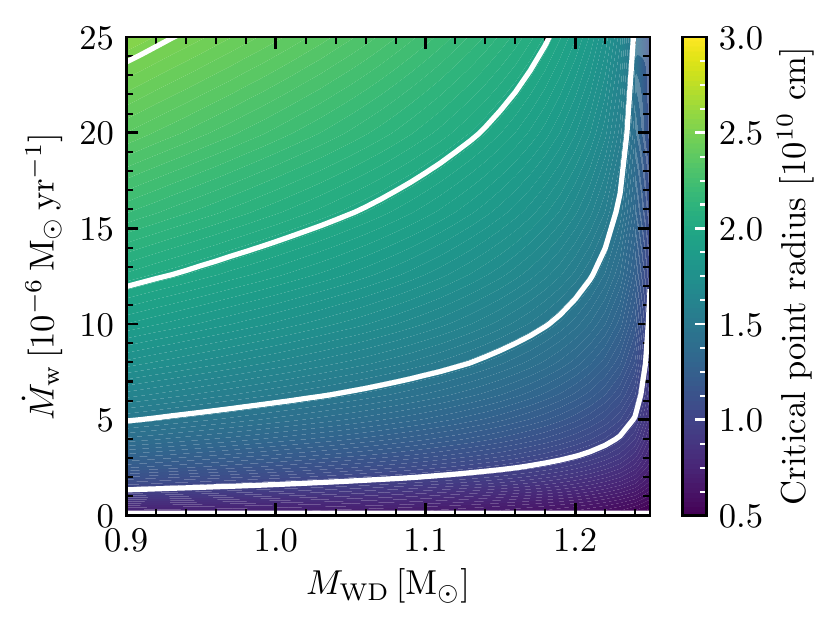}
  \caption{Location of critical point in OTW solutions.  White contours match the locations of the major ticks in the color bar.  }
  \label{fig:otw-rcr}
\end{figure}

\begin{figure}
  \includegraphics[width=\columnwidth]{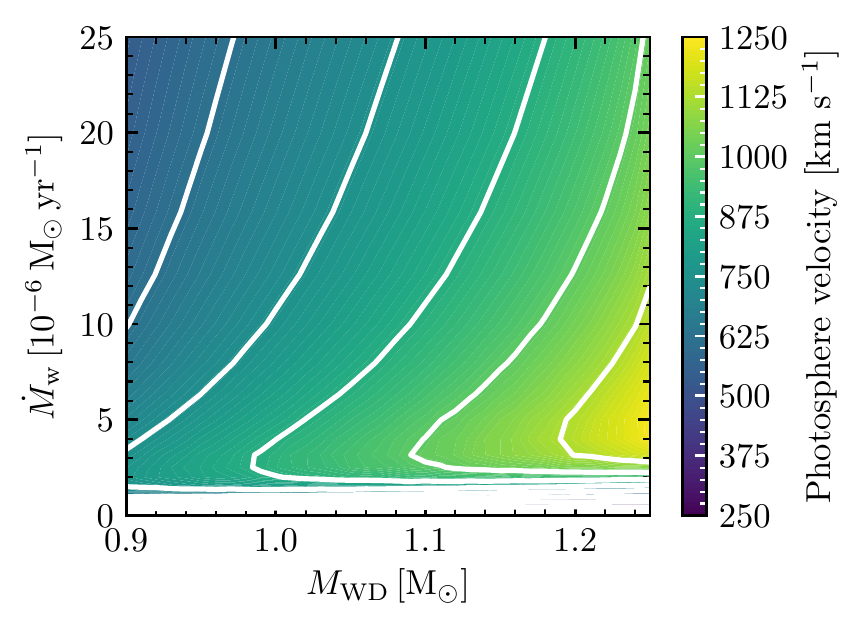}
  \caption{Velocity at OTW photosphere. White contours match the locations of the major ticks in the color bar.}
  \label{fig:otw-vph}
\end{figure}

\begin{figure}
  \includegraphics[width=\columnwidth]{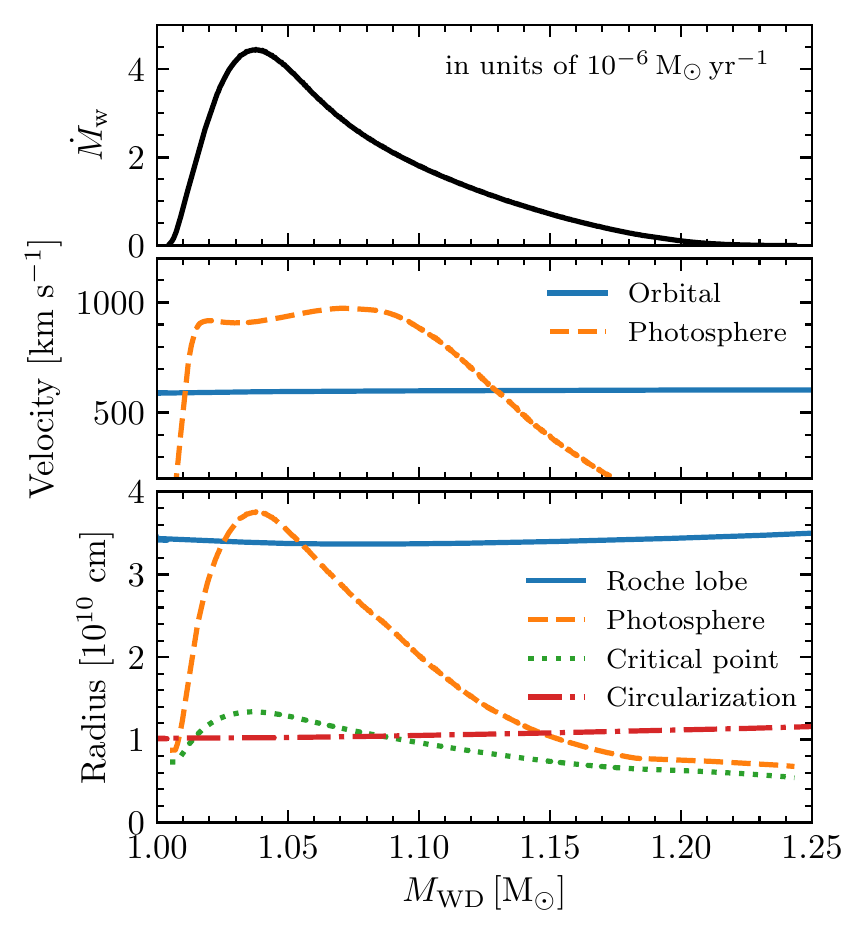}
  \caption{Key velocities and radii for the binary model with $\inipara = (1.6, 1.0, -0.9)$. \textit{Upper panel:} The wind mass loss rate enforced by our modeling assumptions.  \textit{Middle panel:} Orbital velocity and and photosphere velocity from matching OTW models.  \textit{Lower panel:} Roche lobe radius and circularization radius, along with critical point
  and photosphere radius from matching OTW models.}
  \label{fig:otw-evolution}
\end{figure}

In what follows, we focus on two quantities given by the wind solutions. 
First, we consider the radius of the critical point
(\rcr).  If this value is outside the Roche lobe, then the effectively
single star framework in which this wind solution was derived clearly
breaks down.  Figure~\ref{fig:otw-rcr} shows the values of this
quantity for a range of $\mwd$ and $\mdotw$.  It increases as each of these parameters increases, but is characteristically $\approx 10^{10}$ cm.

Second, assuming the wind is launched, then we are also interested in its  velocity in order to understand if it satisfies the conditions
for a fast wind.
The velocity at the photosphere (\vph) is beyond the acceleration
region and thus we take it to be roughly representative
of the terminal speed of the wind. 
We note that the true terminal speed may be 
modified beyond this estimate by further action of the gravitational force of the
the stars or by CAK-type forces on lines in the wind.
Figure~\ref{fig:otw-vph}
shows the values of the velocity at the photosphere
in the OTW models over a range of $\mwd$ and $\mdotw$.  Generally, \vph\, increases
with increasing  $\mwd$ and decreases with increasing $\mdotw$.
The rapid decrease in velocity at low values of $\mdotw$ corresponds to the approach towards hydrostatic solutions.  Over most of the parameter space,
characteristic values are $\approx 500 - 1000\,\kms$.

We place this in context in Figure~\ref{fig:otw-evolution} by showing
the key velocities and radii values for a fiducial binary model.
The upper panel shows the wind mass loss rate enforced by our \mesa\ calculations.
With the values of $\mwd$ and $\mdotw$,  we use Figures \ref{fig:otw-rcr} 
and \ref{fig:otw-vph} to infer the velocities and radii at
the critical point and the photosphere and show these in
the other panels.
For reference, we also show the orbital velocity and the Roche lobe
and circularization radius, which are directly set by the binary properties.
In the middle panel, we observe that the wind velocity generally exceeds the orbital velocity,
thus reinforcing our fast wind assumption.  In the lower panel, we see that the critical point radius is well within the Roche lobe, indicating that
the presence of the binary companion does not disrupt the wind launching.
However, at the peak mass transfer rate, the photosphere of the OTW model
is inferred to lie at a radius beyond the Roche lobe.  This does indicate that (at least for the short period systems) the relevance of the spherically symmetric
outflow solutions in our OTW models begins to break down.
Similarly, we see that the circularization radius
is in a similar location to the critical point, indicating the
likely complexity of the flow in the equatorial plane.  We will
discuss this more in the following section.

But while caveats apply, the OTW models that we construct,
when applied to our simulated systems,
appear to be generally consistent with the idea
that material can be accelerated
within the Roche lobe of the WD
to velocities in excess of the orbital
velocity of the binary.


\section{Discussion}
\label{sec:discussion}

We briefly discuss some of the uncertainties
associated with our modeling assumptions 
\added{ of solar metallicity stars in Section~\ref{subsubsec:metallicity} and}
of non-rotating,
spherically-symmetric WDs in Sections~\ref{subsubsec:rotation} and \ref{subsubsec: the accretion picture}.
In Section~\ref{subsec:formation channel} we discuss the formation of He star - WD binaries.
In Section~\ref{subsec: observational constraints}, we describe how our models
fit in with observed systems and observational constraints on TN SNe progenitor systems.

\added{
\subsection{Effects of Metallicity}
\label{subsubsec:metallicity}
Metallicity may have an effect on the helium donor channel, but we do not explore that in this work where all models assume $Z = 0.02$. The optically-thick wind is accelerated by the iron bump opacity, so the wind efficiency may be lower for lower metallicity \citep{1997ApJS..113..121K}, with a minimum metallicity $Z\approx0.002$ for the wind to occur \citep{1998ApJ...503L.155K}. \cite{2010A&A...515A..88W} also found that the TN SN region broadens to higher $\iniMhe$ and longer $\iniP$ for higher metallicity, which leads to a lower minimum $\iniMwd$. Overall, they found the TN SN rates are higher with shorter delay times for higher metallicity.
}


\subsection{Effects of Rotation}
\label{subsubsec:rotation}

In our binary calculations we have evolved both components as non-rotating models. In reality, sources of torque will likely enter into the binary interaction, with consequences for the stellar structures of both components, orbital angular momentum evolution, and possibly the final outcome of the system. Here we describe the possible effects that may enter if rotation is accounted for. 

When rotation is accounted for, the angular momentum evolution of the system and each component becomes complicated. In the case of double WD systems, the WD spins may be both an important drain and source of the orbital angular momentum \citep{2007ApJ...655.1010G}, but it is unclear how this would affect the stability of the He star-WD systems here. However, it is likely that the WD will spin up from the accretion of high specific angular momentum material, up to critical rotation (e.g., \citealt{2000A&A...362.1046L}). The angular momentum profile of the WD is still currently under debate, subject to the rotational instabilities at work. Some previous studies have suggested that either only uniform rotation or differential rotation may be attained (e.g., \citealt{2004A&A...425..217Y}, \citealt{2004ApJ...615.444S}, \citealt{2008ApJ...679..616P}), whereas recently \cite{2017ApJ...834...93G} have suggested both are possible assuming active baroclinic instability. 

Rapid rotation has important implications on the stellar structure of the WD. The transport of angular momentum into the WD interior may provide additional support through the centrifugal force and lead to a larger WD radius. Previous studies have shown that under differential rotation, lower central densities are attained at the conventional Chandrasekhar mass, and so the WD may accrete up to much higher mass, up to $\approx 2.0$ $\msun$ (e.g., \citealt{2004A&A...425..217Y}). Only when the WD spins down can its central density reach carbon ignition, leading to a super-Chandrasekhar event in the spin-up/spin-down scenario (e.g., \citealt{2011ApJ...738L...1D}, \citealt{2011ApJ...730L..34J}). More importantly, rotationally-induced chemical mixing may lead to different helium shell burning conditions. \cite{2004A&A...425..217Y} have studied the accretion of helium onto a CO WD at mass transfer rates in the helium flash regime. They have found that the rotationally-induced chemical mixing leads to a larger helium burning zone, and the enhanced transport of helium into the core leads to stronger energy release through the reaction $^{12}\mathrm{C}(\alpha,\gamma) ^{16}\mathrm{O}$. In addition, the lower density at the burning shell supported by the centrifugal force helps lift the degeneracy. As a result of the larger geometric thickness, lower degeneracy, and higher temperature at the burning shell, the strength of the helium flashes is greatly reduced.

In summary, even the qualitative effects of including rotation on the TN SN region are unclear.
Rotation may require the WD to grow to a larger mass to reach a core ignition, thus requiring systems
that can transfer more helium or begin from more massive WDs.  Alternatively, the higher helium flash retention fraction attainable may allow for more efficient growth, partially or totally cancelling the other effect.


\subsection{The Accretion Picture}
\label{subsubsec: the accretion picture}

In this study we have assumed that a radiation-driven wind will be blown from the WD as the WD expands to red-giant dimensions. However, it remains to be elucidated how the mass transferred is partially accreted and the rest lost through a wind in a realistic three-dimensional picture. In addition, it is unclear whether a direct-impact accretion may result when the WD expands. We do not plan to resolve these issues altogether, which likely requires three-dimensional simulations, but we describe the unresolved issues here.

To our knowledge, all works on the helium donor, Chandrasekhar-mass WD channel, have assumed that a wind carries away the excess mass from the WD once it expands (e.g., \citealt{2003A&A...412L..53Y}, \citealt{2009ApJ...701.1540W}, \citealt{2016ApJ...821...28B}). This is reasonable given that the only other alternative is a common envelope event \citep{1982ApJ...253..798N}. However, it is unclear what the flow structure would look like.  The optically-thick wind calculations are generally made assuming spherical symmetry (e.g., \citealt{1994ApJ...437..802K}).
\cite{2017arXiv171101529K} have proposed that in a steady state, the WD may accrete through an accretion disk and a bipolar, optically-thick wind may blow from the WD. Observations of the helium nova V445 Pup suggest a highly collimated outflow \citep{2009ApJ...706..738W}.  Extending the one-dimensional results to three-dimensions in order to study the bipolar nature of the wind and the influence of the companion may be important and will require additional work. 

\begin{figure}
\gridline{\fig{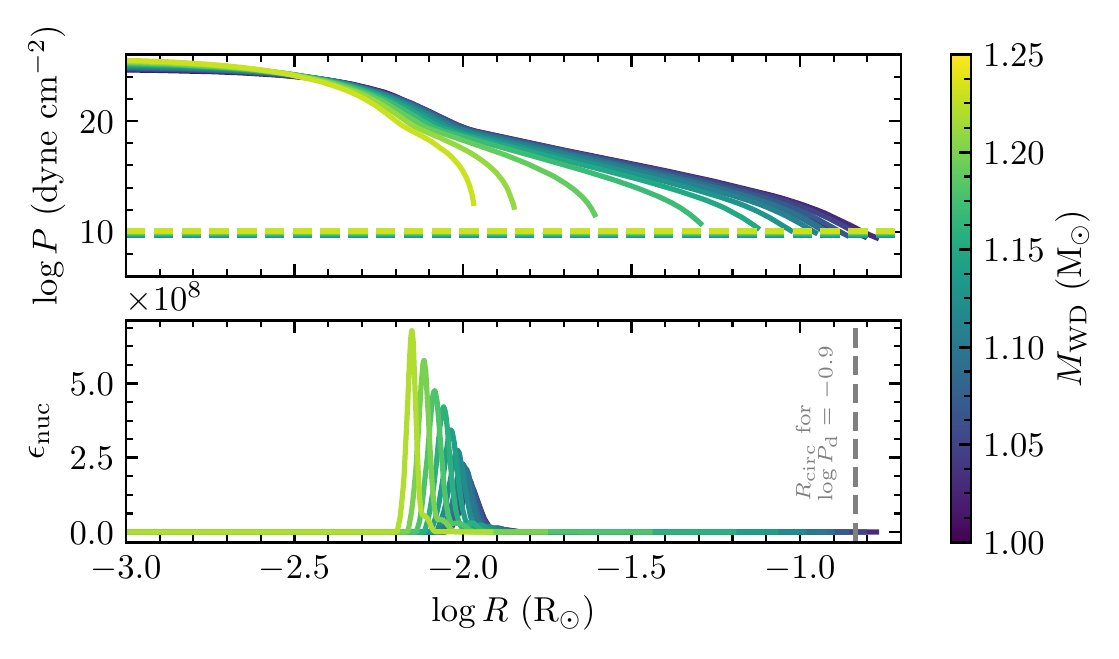}{\columnwidth}{}
		}
\caption{Calculations exploring whether a disk can be formed. The top panel shows the WD pressure profiles as a function of radius in solid lines, and the horizontal dashed lines show the estimated values of the ram pressure of the accretion stream. The bottom panel shows the nuclear burning rate $\epsilon_{\mathrm{nuc}}$, where the peak denotes the helium-burning shell, and the circularization radius for a period of $\log P_{\mathrm{d}} = -0.9$, $\Rcirc \approx 0.15$ $\rsun$. This shows that the accreted material will likely form a disk despite the rapid expansion of the WD envelope. 
\label{fig:ram_pressure}}
\end{figure}

There is, in addition, the question of whether an accretion disk can always be formed. In general, when the WD radius $\Rwd$ is smaller than the circularization radius $\Rcirc$ (defined by the Keplerian radius material would have carrying the specific angular momentum of the inner Lagrange point), a Keplerian disk will likely be formed. The disk will transport material to the WD surface with specific angular momentum equal to $\sqrt{G\mwd\Rwd}$. But when $\Rwd>\Rcirc$, one question is how deep inside the WD envelope the accreted material would settle, as determined by the ram pressure of the accreted material. In Figure \ref{fig:ram_pressure}, we allow one of our WD models to expand up to 80$\%$ of its Roche radius, and plot the pressure profiles of the WD at different epochs. We also estimate the ram pressure of the incoming material, given by $P_{\mathrm{ram}} = \rho v^{2}$. We estimate $\rho v \sim (\rho c_{\mathrm{s,iso}})_{\mathrm{L1}} $ by mass continuity, where the density around L$_{\mathrm{1}}$ is given by $\dot{M} \Omega^{2}/c^{3}_{\mathrm{s,\mathrm{L1}}}$ from \cite{1975ApJ...198..383L}, and $c_{\mathrm{s,\mathrm{L1}}} $ is taken from the conditions at the outermost zone of the He star. The other $v$ term is estimated as the free-fall velocity $\sqrt{G \mwd/ \Rwd}$ onto the WD. Since the radius for pressure equilibrium is mostly at a smaller radius than the circularization radius, it seems possible that the accretion stream will penetrate the envelope and still form an accretion disk. 

Under the accretion picture being considered here, the accretion disk will likely be embedded in an inflated envelope/wind structure with high entropy. Can heat exchange between the envelope and the disk alter the disk structure? The disk thermal timescale \citep{1981ARA&A..19..137P} is roughly  $t_{\mathrm{th,disk}} \approx \alpha^{-1} t_{\phi} = \alpha^{-1} \sqrt{R^{3}/GM}
\approx 500$ s taking $\alpha=0.1$, $R=0.1$ $\rsun$ and $M=\msun$. The photon diffusion timescale through the deeper parts of the envelope is $t_{\mathrm{diff}} \approx R^{2} \rho \kappa/c \approx 10^{4}$ s taking $\rho \approx 10^{-4}$ g cm$^{-3}$ and $\kappa \approx 0.2$ cm$^{2}$ g$^{-1}$. At larger radii ($\gtrsim 0.5\,\rsun$), outside the acceleration region of the wind, $r^{2} \rho = \mdotw/4 \pi \vw$ and so $t_{\mathrm{diff}} \approx \mdotw \kappa/ 4 \pi \vw c \approx 30$ s using $\mdotw\approx 10^{-5}$ $\msun$ $\mathrm{ yr}^{-1}$ and $\vw \approx 10^{7}$ cm s$^{-1}$. Comparing these timescales, we suggest that heat exchange between the disk and the wind/envelope may be rapid near the outer edge of the disk, where $t_{\mathrm{diff}} < t_{\mathrm{th,disk}}$, perhaps inflating the outer disk, but the inner disk should remain intact.

\subsection{Formation of He star - CO WD systems}\label{subsec:formation channel} 

Understanding the formation of He star - CO WD binaries informs the contribution of the helium donor channel to TN SNe, particularly in population synthesis calculations. This particular combination of binary components requires that at least one common envelope episode is involved. 

\cite{2012NewAR..56..122W} describe three scenarios leading to the formation of a He star - CO WD binary. Scenario A starts with a subgiant or red giant branch (RGB) primary and a main sequence (MS) secondary, where the primary undergoes Roche lobe overflow (RLOF) episodes to form a CO WD primary with a subgiant/ RGB secondary. After a dynamically unstable RLOF and a common envelope episode, a He star - CO WD binary emerges. In Scenario B, the CO WD - MS binary comes initially from a early asymptotic giant branch (EAGB) primary with a MS secondary. The EAGB-MS binary undergoes a dynamically unstable RLOF and a common envelope to form a helium red giant (He RG) - MS binary, and forms a CO WD - MS binary after a stable RLOF. In Scenario C, the He star - CO WD comes directly from a common envelope resulting from a dynamically unstable RLOF between a thermally-pulsing asymptotic giant branch (TP-AGB) and a helium-core burning star. 

\cite{2014A&A...563A..83C} also describe three scenarios for the formation of a He star - CO WD binary, while providing the contribution by each channel. Scenarios $\mathrm{A_{He}}$ and $\mathrm{B_{He}}$ of \cite{2014A&A...563A..83C} are very similar to Scenario A of \cite{2012NewAR..56..122W}. The only difference lies in when the He star starts transferring mass to the CO WD --- a He MS in Scenario $\mathrm{A_{He}}$ of \cite{2014A&A...563A..83C} and a more evolved He star in Scenario $\mathrm{B_{He}}$ of \cite{2014A&A...563A..83C} and Scenario A of \cite{2012NewAR..56..122W}. Each of $\mathrm{A_{He}}$ and $\mathrm{B_{He}}$ contributes 48\% of all TN SN progenitors according to \cite{2014A&A...563A..83C}. The last 4 \%, Scenario $\mathrm{C_{He}}$ is similar to Scenario C of \cite{2012NewAR..56..122W}. Note however that \cite{2014A&A...563A..83C} find that a He MS star may also donate mass to the CO WD and contribute to the TN SN region, whereas in our investigation the He star is more evolved, undergoing helium shell-burning as a He subgiant.

\begin{figure*}
\gridline{\fig{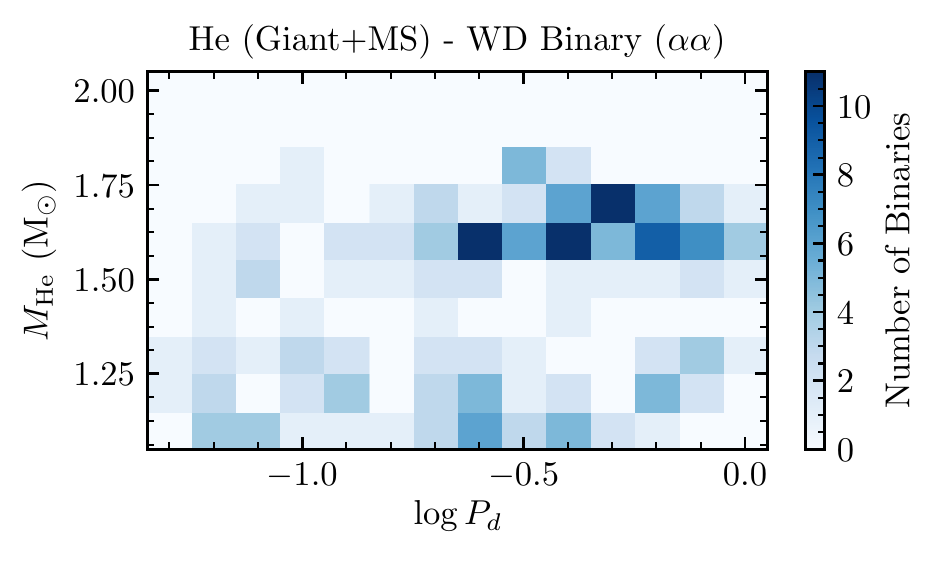}{0.5\textwidth}{(a)}
          \fig{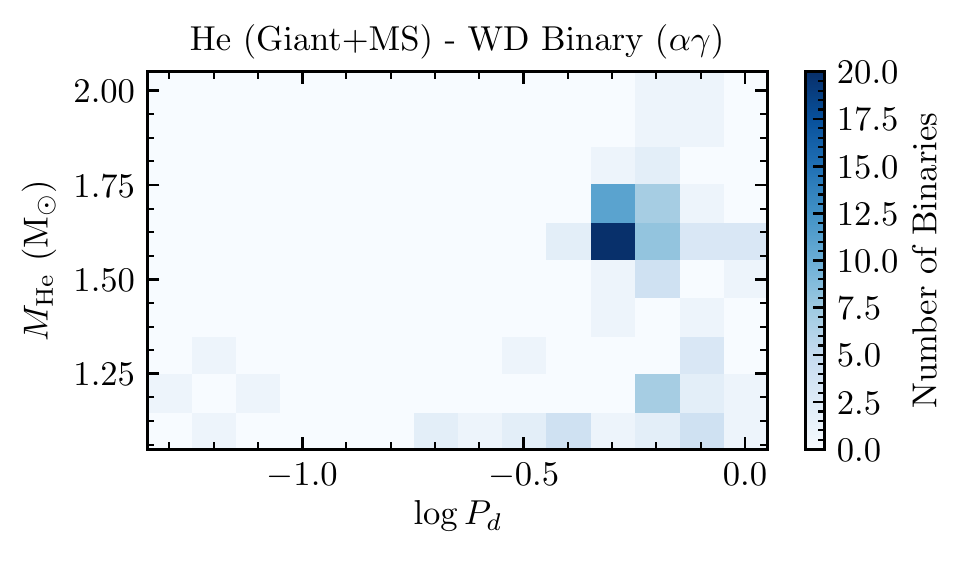}{0.5\textwidth}{(b)}
          }
\caption{Population synthesis results from \citet{2012A&A...546A..70T} of He star - WD systems resulting from two common envelope events. We choose systems with $\mwd$ of 0.95-1.05 $\msun$ which may inform the properties of the primordial systems in our work. Panel (a) uses the $\alpha$ formalism in both events, whereas panel (b) uses the $\alpha$ formalism followed by the $\gamma$ formalism. The latter appears to favor longer period systems. 
\label{fig:toonen}}
\end{figure*}

\replaced{One may also be curious about the distributions in the He star mass and binary orbital period when the He star - CO WD binary forms.}{The various formation channels affect the likelihood of forming a He star-WD system for a given $\inipara$, which then informs the contribution of that particular $\inipara$ to TN SN rates.}
We refer the reader to Figure 5 of \cite{2014A&A...563A..83C} as one source. In Figure \ref{fig:toonen} we show the distributions in He star mass and binary orbital period with a WD mass of $0.95$ $\msun$ $\leqslant \mwd \leqslant$ $1.05$ $\msun$ from the binary population synthesis calculations in \cite{2012A&A...546A..70T}, which is aimed at investigating double WD populations. This means the He star - CO WD binaries in the distributions shown in Figure \ref{fig:toonen} have undergone two common envelope episodes. The difference between panels (a) and
(b) lies in the common envelope prescriptions being used in the calculations. The first common envelope calculation is computed with the $\alpha$ formalism, based on energy conservation by \cite{1984ApJ...277..355W}. The $\alpha$ formalism assumes that the change in orbital energy, $\Delta E_{\mathrm{orb}}$ is expended, with efficiency $\alpha$, in unbinding the common envelope, which has binding energy $GMM_{\mathrm{env}}/\lambda R$, where $M$ and $R$ are the mass and radius of the donor and $\lambda$ depends on the structure of the donor. Panel (a) assumes the $\alpha$ formalism again in the second common envelope episode, whereas panel (b) assumes the  $\gamma$ formalism proposed by \cite{2000A&A...360.1011N}. The $\gamma$ formalism is based conservation of angular momentum instead, assuming that the specific angular momentum lost, $\Delta J/ \Delta M$, where $\Delta J = J_{i} - J_{f}$, the change in binary angular momentum, is proportional to the initial binary specific angular momentum, $J_{i}/(M+m)$, where $M$ and $m$ are the masses of the donor and the companion respectively. The values adopted in \cite{2012A&A...546A..70T} are $\alpha \lambda = 2$ and $\gamma = 1.75$, based on the optimization by \cite{2000A&A...360.1011N}. 

The result of using different common envelope prescriptions can be seen in Figure \ref{fig:toonen}. Panel (a), which uses the $\alpha$ formalism in both common envelope episodes, results in a more even distribution in $\iniMhe$ and $\iniP$. There is a cluster of binaries for donor mass 1.6-1.7 $\msun$ and $\iniP$ between $-0.6$ and $-0.1$. On the contrary, panel (b), which uses the $\gamma$ formalism in the second common envelope episode, results in a very concentrated distribution of binaries at donor mass 1.6-1.7 $\msun$ and $\iniP$ between $-0.4$ and $-0.3$. Longer periods are more favored in panel (b). 

\added{The common envelope ejection efficiency is another important parameter that enters these population synthesis studies. The parameter $\alpha \lambda = 2$ used by \cite{2012A&A...546A..70T} implies a highly efficient common envelope ejection, which leads to higher formation rates of long period systems. In contrast, the population synthesis study by \cite{2009ApJ...701.1540W} adopts $\alpha \lambda = 0.5$ in one case, which leads to a higher contribution to TN SNe by short period systems of $\iniP \leqslant -1.2$. }

Given the outcomes shown in Figure~\ref{fig:grids}, it appears that the scenario 
in the Figure \ref{fig:toonen}, panel (a) would predict a fair fraction of core ignitions (and hence TN SNe) whereas in Figure \ref{fig:toonen}, panel (b) almost all of the predicted systems
are at periods where we would predict the formation of detached double WDs.
It would be useful to better characterize the properties of the He star - WD binaries,
as there are few known systems with the properties of the binaries
modeled here.  The best example, HD 49798, is still not a direct analogue due to likely hosting a more massive ONe WD \citep{2018MNRAS.474.2750P}.

\subsection{Observational Constraints}
\label{subsec: observational constraints}

It remains an important task to observationally distinguish the different scenarios that may contribute to TN SNe. We discuss the several properties that may be important in identifying the systems emanating from the helium donor channel.

\subsubsection{Delay Times}
Studies have suggested that the helium donor channel may only be a sub-channel to SNe Ia (e.g., \citealt{2003A&A...412L..53Y}), contributing to a galactic rate of $\sim 0.2 \times 10^{-4}$ yr$^{-1}$ \citep{2017MNRAS.472.1593W}. Nevertheless the helium donor channel is an important channel to SNe Ia for short delay times (e.g.,  \citealt{2009ApJ...699.2026R}). It is therefore likely that thermonuclear supernovae produced by this channel may be observed in late-type galaxies, possibly offering an explanation for the preference of SNe Iax for late-type galaxies \citep{2013ApJ...767...57F} and their delay times of
50-100 Myr \citep{2019arXiv190105461T}.

\subsubsection{Progenitor System Evolutionary Phases}

Helium donor channel systems spend 
time in several evolutionary phases in advance of explosion.
For an initial donor mass higher than $\approx 1.3$ $\msun$, the system may undergo an optically-thick wind phase which lasts about $\approx 10^{4}$ years and lose a total mass ranging from $0.01$ $\msun$ for the $\approx1.3$ $\msun$ donors to more than $0.1$ $\msun$ for higher mass donors. These systems will then undergo a phase of stable mass transfer for another $\approx 10^{4}$ years, where they appear as supersoft x-ray sources (SSS; \citealt{1992A&A...262...97V}). The systems lower than $\approx 1.3$ $\msun$ may always appear as SSS, for up to $\sim 10^{5}$ years. The circumstellar material originating from the helium flashes, or the wind material during the optically-thick wind phase, may obscure the supersoft x-ray from the underlying stably accreting WD. It is still under debate whether circumstellar material may be sufficient to obscure SSS systems (e.g., \citealt{2013A&A...549A..32N}, \citealt{2013ApJ...762...75W}, \citealt{2015MNRAS.453.2927N}), and further investigations may look into non-solar composition materials, such as the helium-rich material in our systems and carbon-enriched materials as seen in the helium nova V445 Pup \citep{2003A&A...409.1007A}. 

\subsubsection{Pre-explosion}
Some of the constraining pre-explosive properties the helium donor systems are the luminosity and color of the He star. In particular, the blue point source in the Hubble pre-explosion images for the Type Iax SN2012Z has been suggested to be a $\approx 2$ $\msun$ He star (SN 2012Z-S1; \citealt{2014Natur.512...54M}). To relate to SN 2012Z-S1, we examine the likely system properties of our models when the WD reaches $\mch$. In Panel (a) of Figure \ref{fig:final_he}, we report the He donor mass by the end of the simulation $\finalMhe$, which represents an upper limit since many models (i.e., those undergoing helium flashes) terminate before the WD reaches $\mch$. The black thick contour delineates the likely TN SN progenitors on the $\iniP-\iniMhe$ space. It is likely that any progenitors from this channel have a He star of mass $\approx 0.9 - 1.1$ $\msun$ at the time of the WD explosion. The likely He star luminosity is $\log( \Lhe / \lsun) \approx 3.4 - 4$, as can be seen in Panel (b). The luminosities reported there are likely to be lower limits, since the He star will gradually evolve to higher luminosities due to the continued evolution of the He star. Comparing this with Panel (a), Figure 2 of \cite{2015ApJ...808..138L} which stacks the pre-explosion model properties of the helium donor channel, our models are situated near the upper end of the luminosity range spanned by their models. Our models also span roughly the same range in effective temperature as the models of \cite{2015ApJ...808..138L}, $\log(T_{\mathrm{eff}}/{\rm K}) \approx 4.5 - 5.0$.\deleted{, but to convert into the colors observed by Hubble of SN 2012Z-S1 one needs proper modeling of the He star atmospheres.}

\added{Note however, that the luminosity and colors of the source in 2012Z require a cooler, $\log(T_{\mathrm{eff}}/{\rm K}) \approx 4.2$ object, which in the models of \citet{2015ApJ...808..138L} corresponds to those that assume an initially more massive WD (1.2 or 1.3 $\msun$).  A higher initial WD mass fits naturally into the hybrid CONe WD+He star scenario for SNe Iax \citep{2014ApJ...794L..28W,2016A&A...589A..38B} and has been specifically invoked as the explanation for 2012Z \citep{2015ApJ...808..138L}.  Since we restricted this study to CO WDs with initial masses $\le 1.05\,\msun$,
none of the specific models presented here are an exact match the source associated with 2012Z.
The non-detection of a progenitor in SN 2014dt \citep{2015ApJ...798L..37F}, which has a pre-explosion image that reaches a comparable depth to that of 2012Z, could be consistent with either a less luminous or hotter progenitor than in 2012Z, and so our models are compatible with that event.}

\subsubsection{Companion Interaction with Supernova}

Theories predict that the impact of supernova ejecta onto the companion should produce a shock and excess emission in the early light curve (e.g., \citealt{2010ApJ...708.1025K}). A stronger constraint on the helium donor channel comes from the detection of helium in the spectra. The helium comes from entrainment of companion material in the ejecta. \cite{2010ApJ...715...78P} and \cite{2013ApJ...774...37L} have simulated the supernova impact onto a He star companion. The latter have found a stripping of 2\% to 5\% of the initial companion mass. In relating to these works, our He star models where the WD grows to $\mch$ have very similar structures to model He02 of \cite{2013ApJ...774...37L}, but are slightly more evolved than the models of \cite{2010ApJ...715...78P,2012ApJ...750..151P} (closest to their He-WDc). The entrainment of companion material may be related to the presence of He I lines in the spectra of 2 Type Iax supernovae, SNe 2004cs and 2007J \citep{2013ApJ...767...57F}.  However, note that \citet{2018arXiv181211692J} report non-detections of He lines in late-time Iax spectra corresponding
to upper limits comparable to the theoretically-predicted stripped masses.

Furthermore, \cite{2013ApJ...773...49P,2014ApJ...792...71P} predict that after the supernova explosion, the remnant He stars would release the energy deposited by the supernova impact, expand and become luminous helium OB stars for $\approx 10-30$ years and later sdO-like stars. These may inform searches for the companion shortly after the supernova explosion, or within galactic supernova remnants.

\subsubsection{Ejected Companions}
In the aftermath of the TN SN, the He donor will likely survive and the WD may even leave a bound remnant. Either of these components may be ejected from the system, at the orbital velocity if the system loses roughly more than half of the total mass. 
\replaced
{Our models predict that at the moment of the supernova, the orbital velocities of both components are no more than $\approx 400$ $\mathrm{km}$ $\mathrm{s^{-1}}$. In addition, \cite{2013ApJ...774...37L} have suggested that the He star would receive a small kick of $\approx 60$ $\kms$. Similarly, models He-WDc or He-WDd of \cite{2013ApJ...773...49P} which are the closest models to our He star models at TN SN, predicts a linear velocity of $\approx 400-500$ $\kms$ for the remnant He star. These suggest that the remnant He star in our channel will not likely result in hypervelocity He stars like US 708 \citep{2005A&A...444L..61H,2015Sci...347.1126G}.}
{Our models predict that at the moment of the supernova, the orbital velocity of the He star is in the range of $\approx 200 - 450$ $\mathrm{km}$ $\mathrm{s^{-1}}$, and that of the WD is about $\approx 100 - 350$ $\kms$. In comparison, \cite{2009A&A...508L..27W} have found He star pre-explosion orbital velocities in the range $300-500$ $\kms$ (their Fig. 1). The upper limit in their He star orbital velocities is slightly higher than ours possibly since their binary evolution code allows a shorter period system to form (see Section \ref{subsec:comparison with Wang 2017}).

In addition, interaction with the supernova may introduce a kick velocity to the He star \citep[e.g.,][]{2000ApJS..128..615M}. After accounting for the kick velocity using momentum conservation, \cite{2009A&A...508L..27W} found spatial velocities ranging from $400$ to $700$ $\kms$. The more recent hydrodynamic simulations by \cite{2013ApJ...774...37L} have suggested that the He star would receive a small kick of $\approx 60$ $\kms$. Thus, we suggest that the spatial velocities of ejected He stars are about $\approx 300 - 600$ $\kms$, slightly lower than those in \cite{2009A&A...508L..27W}. Our results are in agreement with models He-WDc or He-WDd of \cite{2013ApJ...773...49P}, which are the closest models to our He star models at TN SN and predict a linear velocity of $\approx 400-500$ $\kms$ for the remnant He star.  Thus, the surviving donors from this channel can
produce a population of high velocity He stars, though the channel likely cannot produce the $\approx 1000\,\kms$ hypervelocity sdO star US 708 \citep{2005A&A...444L..61H,2015Sci...347.1126G}. }

\subsubsection{He Nova Luminosities \& Colors}
The helium donor channel also gives rise to the phenomenon of helium novae, as in V445 Pup \citep{2003A&A...409.1007A}. An exciting possibility for V445 Pup is that it may grow up to $\mch$, which may be evaluated from the component masses and the binary orbital period. Based on pre-explosion plate archives, and a distance derived from expansion of the nova nebula, \cite{2009ApJ...706..738W} have derived a pre-explosion He star luminosity of $\log( \Lhe / \lsun) \approx 3.3 - 4.3$. The large uncertainty is based on whether a large circumstellar reddening is to be corrected for, since the color from optical V band and near-infrared K band appears too red for a He star \citep{2009ApJ...706..738W}. \cite{2010PZ.....30....4G} have suggested that a pre-explosion He star luminosity of $\log( \Lhe / \lsun) \approx 3.0$, and derived a variability period  of $\approx 0.65$ days through constructing light curves from digitized plates.

\begin{figure*}
\gridline{\fig{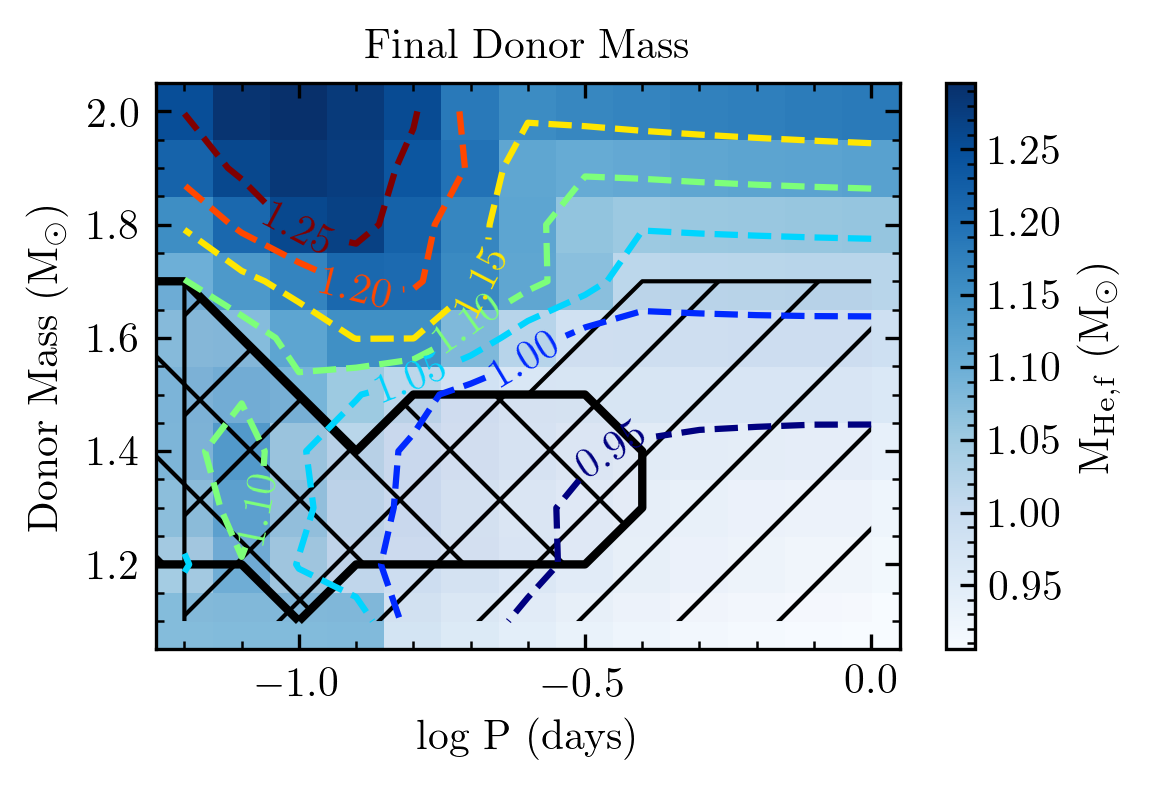}{0.5\textwidth}{(a)}
          \fig{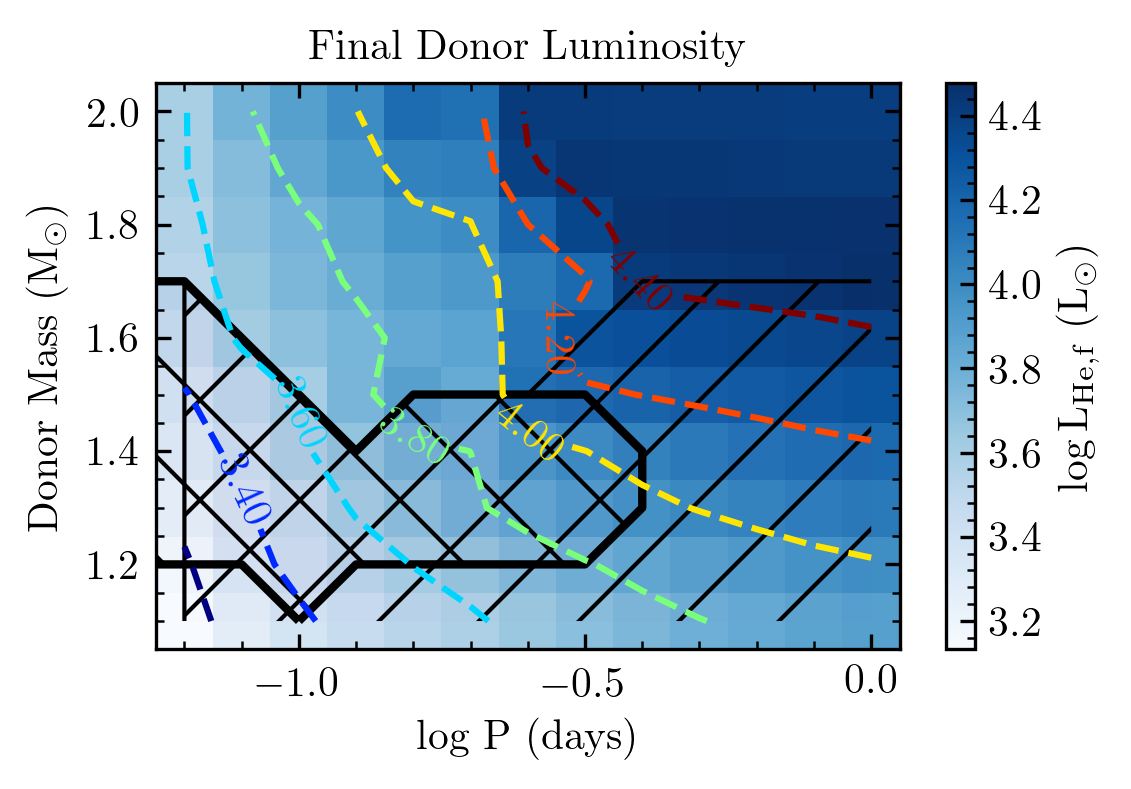}{0.5\textwidth}{(b)}
          }
\caption{Masses (panel a) and luminosities (panel b) of the He stars for each system in our fiducial grid ($\mwd =1.0$ $\msun$) by the end of our binary run.  As discussed in the text, the masses are upper limits and the luminosities are lower limits to the properties at the time of explosion. The black thick contour delineates the systems where the WD eventually reaches $\mch$, whereas the black thin contour includes the systems that eventually enter the helium flash regime. These two regions overlap since most WDs that are able to grow to $\mch$ ultimately need to do so through helium flashes. 
\label{fig:final_he}}
\end{figure*}

We discuss the relation of the V445 Pup system to our parameter space based on these observations. The black thin contour in each panel of Figure \ref{fig:final_he} delineates the systems which undergo helium flashes. In our simulations, the final period does not deviate much from the initial period, so the long period suggested by the above studies places the V445 Pup system on the right side of the contours. The systems on the right have such high initial mass transfer rates that the WD starts helium flashes when the donor envelope is almost depleted; this may be said of the He star of V445 Pup. Panel (a) indicates that the He star of V445 Pup is likely to have a mass of $1.0$ $\msun$ or lower; whereas the bolometric luminosity is likely to be $\log( \Lhe / \lsun) \approx 3.8$ and above. Furthermore, the nova light curve fitting by \cite{2008ApJ...684.1366K} under the optically-thick wind framework and assumption of free-free absorption suggests a WD mass of $\geqslant 1.35$ $\msun$. However, given the long period, for the WD to have a mass of $\geqslant 1.35$ $\msun$ when it undergoes helium flashes, we suggest that the initial WD is more massive than $1.0$ $\msun$, and the helium flashes have high retention efficiencies. If the initial WD is indeed more massive than 1.0 $\msun$, it is possible that the WD is in fact a massive ONe WD, although \cite{2008ApJ...684.1366K} have disfavored this noting that there was no indication of neon during the nebular phase of the nova. Alternatively, a downward revision of the current WD mass may be required. If the initial WD mass is $\approx 1.05$ $\msun$, the WD may barely grow up to the Chandrasekhar mass according to our grid. 

\subsubsection{Environment Densities}
Finally, the environment properties of both TN SNe and helium novae from the helium donor channel can be tested from observations. In particular, the fast wind emanating from the WD during the optically-thick wind phase will likely form a wind-blown cavity around the system \citep{2007ApJ...662..472B}. This may inform inferences from supernova remnants. On the other hand, inferences about environmental density profiles have been made during the first $\sim$year after the supernova explosion through radio and x-ray observations for example (see \citealt{2016ApJ...821..119C} and references therein). In the helium donor channel, the WD wind will likely have ceased for $\approx 10^{4}$ years before TN SN. The source of any inferred circumstellar material would thus likely be nova shells ejected more recently before the TN SN.

\section{Conclusion}
\label{sec: conclusion}

Using the stellar evolution code $\mesa$, we study the time-dependent mass transfer history and binary evolution of a $1.1-2.0\,\msun$ non-degenerate He star and a $0.9-1.05$ $\msun$ CO WD in a $0.05-1$ day orbit. 
We characterize the possible outcomes:  either a core ignition, off-center ignition, helium flashes, or formation of a detached double WD binary.
We identify the region of this parameter space (i.e., the core ignitions) that can contribute to thermonuclear supernovae 
when the WD approaches the Chandrasekhar mass.
We model the full WD structure throughout the mass transfer history, and so
can self-consistently account for the occurrence of an off-center carbon ignition in the WD.
In the systems in which this occurs, it likely precludes the occurrence of thermonuclear supernova.
The results of our work are in agreement with similar previous work by \cite{2017MNRAS.472.1593W} which accounted for off-center ignitions via a simpler procedure.

We also critically investigate several important modeling
assumptions for these systems that have not previously been 
systematically explored.
At mass transfer rates above the upper stability line $\mdotup$, the WD cannot burn material as efficiently as mass is accreted and so rapidly expands.
This material is typically assumed to be lost from the system in a fast isotropic wind
that carries the specific angular momentum of the WD.
We quantitatively discuss the possibility of this wind launching
and construct simple wind models that generally confirm the physical plausibility of these winds.
However, in the case of inefficient wind driving the wind speed may not necessarily be fast compared to the orbital speed, and hence the wind may gravitationally torque the binary. We parameterize the wind specific angular momentum loss and re-calculate our model grids.
We find that although increased wind angular momentum loss may significantly alter individual mass transfer histories and lead many modelled systems to undergo a common envelope, that the shift in the region of parameter space that leads to thermonuclear supernovae is not significant. 

Overall, our work predicts the evolutionary outcome He star - WD binaries as a function of mass and period.  This is of utility for future population synthesis calculations, for associating observed binary systems with their final fates, for characterizing He nova systems, and for confronting observations of supernova progenitors.


\acknowledgments
We thank Jared Brooks for providing the input files from his 2016 study and for many helpful communications at the start of this project.
We thank Silvia Toonen for helpful discussions and for providing the results of her population synthesis calculations.
\added{We thank Ryan Foley for discussions of SN Iax observations.}
We thank Enrico Ramirez-Ruiz for helpful comments
on the manuscript and for his support and co-supervision of T.L.S.W.
\added{We thank the anonymous referee for a constructive report.}
Support for this work was provided by NASA through Hubble Fellowship
grant \# HST-HF2-51382.001-A awarded by the Space Telescope Science
Institute, which is operated by the Association of Universities for
Research in Astronomy, Inc., for NASA, under contract NAS5-26555.
\added{T.L.S.W. thanks the Koret Scholars program, the Ron Ruby award and the Lamat summer REU program for financial support.}
The simulations were run on the Hyades supercomputer at UCSC, purchased
using an NSF MRI grant.  This research made extensive use of NASA's
Astrophysics Data System.

\added{%
\software{%
\texttt{MESA} \citep[v10108;][]{2011ApJS..192....3P, 2013ApJS..208....4P, 2015ApJS..220...15P, 2018ApJS..234...34P},
\texttt{ipython/jupyter} \citep{perez_2007_aa,kluyver_2016_aa},
\texttt{matplotlib} \citep{hunter_2007_aa},
\texttt{NumPy} \citep{der_walt_2011_aa}, and
\texttt{Python} from \href{https://www.python.org}{python.org}.}}


\appendix


\section{Mass Loss Prescription} \label{sec:mass loss prescription}

As we have described in Section \ref{subsec:mesa setup}, our binary simulations used of the built-in implicit scheme for the mass transfer rate in $\mesa$ as well as an implicit scheme for the wind mass loss rate (hereafter the $\beta$-scheme) of our own design. In an explicit scheme, wind mass loss at one step may remove too much mass such that the WD shrinks significantly, leading to a small mass loss at the second step, which in turn leads to rapid expansion and hence large mass loss at the third step, etc. For us to obtain converged mass loss rates, we prefer to implement an implicit scheme instead of an explicit scheme which requires very fine time steps. The $\beta$-scheme is intricately tied to the implicit mass loss scheme and piggybacks on the latter within the $\tt\string binary\_check\_model$ procedure. Essentially the $\beta$-scheme performs a bisection search for the wind mass loss fraction $\beta = \mdotw/|\mdothe|$  as follows.

At the start of a step, $\mesa$ evolves both stellar components and the binary system with some value of $|\dot{M}_{\mathrm{He,current}}|$ and $\betacurrent$. Then in the $\tt\string binary\_check\_model$ procedure, $\mesa$ evaluates the value of $\mdothe$ and $\beta$ from some explicit function \textit{if} the current step were to be accepted, where we call the latter  $\betaexplicit$. The procedure for solving $\mdothe$ implicitly is described in \citep{2015ApJS..220...15P}  briefly summarized in Section \ref{subsec:mesa setup}. We only comment that we usually start the implicit $\beta$-scheme only after $|\mdothe|$ is already bounded within some fraction by the implicit mass transfer scheme. The explicit function for $\beta$ is given as $1-x$, where $x$ is the retention efficiency of $|\mdothe|$ onto the WD. This depends on the expansion of the WD, and is quantified as $r = \Rwd/R_{\mathrm{RL}}$. We apply two limits for the wind mass loss, the maximum $r_{\mathrm{max}}=\text{min}(0.6,10 \Rcwd/R_{\mathrm{RL}})$ sets a zero retention efficiency $x=0$ and the minimum $r_{\mathrm{min}}=2 \Rcwd/R_{\mathrm{RL}}$ sets a full retention $x=1$. In between the wind mass loss increases increasingly as a function of $r$:

\begin{displaymath}
x_{0} = 1 - \frac{r-r_{\mathrm{min}}}{r_{\mathrm{max}}-r_{\mathrm{min}}} \\
\end{displaymath}
\begin{displaymath}
x = \frac{1}{2} \left[ 1 - \cos \left( \pi x_{0} \right) \right] \\
\end{displaymath}

We compare the current and proposed next-step values as $\fbeta = \betaexplicit - \betacurrent$. If $f_{\beta}$ is within the tolerance $\xi$, we accept the step. If not, we retake the step and adjust the value of $\betacurrent$, solving for the root of $\fbeta$ iteratively by bisection. The upper and lower bounds for the value of $\beta$, $\betahi$ and $\betalo$, to be solved for are given by checking the sign of $\fbeta$. If $\fbeta = \betaexplicit - \betacurrent > 0$, then the current $\beta$ is too low that the WD accumulates mass and expands (which is why $\betaexplicit>\betacurrent$). This suggests $\betacurrent$ to be a lower bound. Thus we establish $\betalo=\betacurrent$ and scale-up $\betacurrent$ in the next iteration. If $\fbeta = \betaexplicit - \betacurrent < 0$,
we perform the analogous procedure with $\betacurrent$ as an upper bound
and scale-down $\betacurrent$ in the next iteration. Until both bounds $\betahi$ and $\betalo$ are established, we will scale $\betacurrent$ to find the next guess. Once both $\betahi$ and $\betalo$ are established, with the corresponding function values $\fbetahi$ and $\fbetalo$, we use a quadratic solve in $\mesa$ to find the next value of $\beta$. In general the combination of the implicit mass transfer scheme and our $\beta$-scheme lead to between 3 and 9 iterations before a step is finally adopted. 


\section{Carbon burning rate} \label{carbon burning rate}

During this work, we became aware of an error in the \mesa\ implementation of the \citet{2005PhRvC..72b5806G} carbon fusion rate ($^{12}{\rm C}$ + $^{12}{\rm C}$), which we used in the evolutionary calculations presented in this paper.
The formulae given in \citet{2005PhRvC..72b5806G} are for pure carbon, but were being applied in a carbon-oxygen mixture.  This rate is not subject to the default \mesa\ screening treatment because their expressions already include the effects of screening.
When evaluating this rate, the carbon ion density was being used in quantities where the total ion density would be more appropriate, effectively evaluating various dense plasma corrections as if they were in a lower density medium.  This led to a significant underestimate of the rate in the pycnonuclear regime.  A workaround was applied in \mesa\ revision 10792.  (The correct extension of the \citet{2005PhRvC..72b5806G} results to multicomponent plasmas is given by \citet{2006PhRvC..74c5803Y}, but such an option for the carbon fusion rate is not presently implemented in \mesa.)

\begin{figure}
\centering
\includegraphics[]{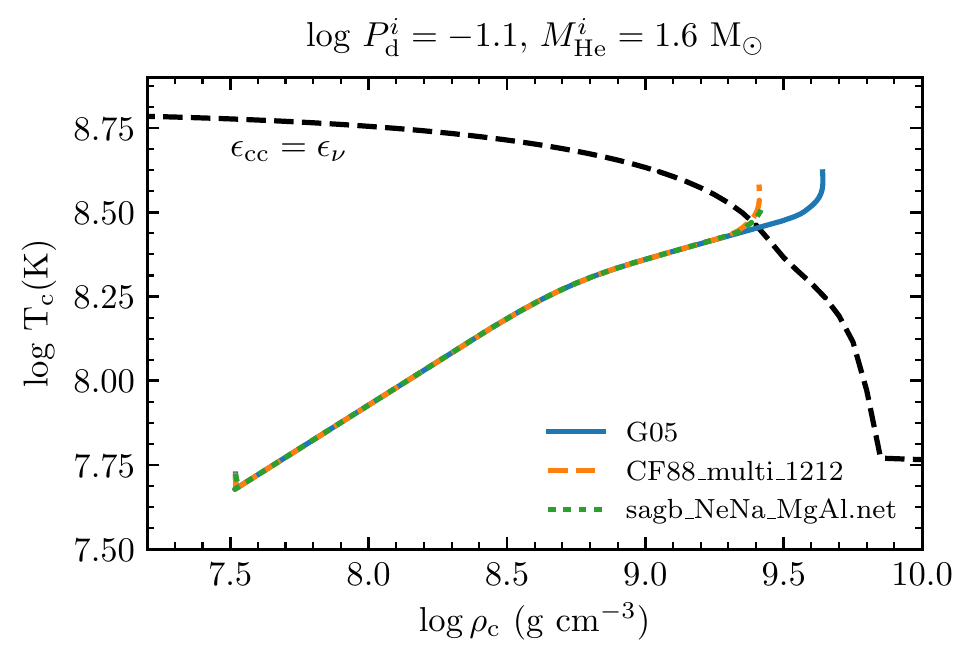}
\caption{Comparison of the central evolution of a representative model that undergoes core carbon ignition.  The solid line uses the options adopted throughout this paper.  Due to an error, this leads to ignition at slightly higher densities than it should, as illustrated by comparison to the other two lines (see text for the specific meaning of the labels).  In the text, we argue that the size and direction of this shift means that it does not meaningfully affect our results or conclusions.
\label{fig:cburn}}
\end{figure}

As such, central carbon ignitions were delayed to artificially high densities in our models. In Figure~\ref{fig:cburn}, we show the central evolution of three models run using different choices for the \mesa\ network and input reaction rates. 
One way to control the carbon burning rate is via the option \texttt{set\_rate\_1212}.  Importantly, this option applies only to reaction networks that include carbon burning via the compound reaction \texttt{r1212}, in which the exit channels for the reaction are combined.
Figure~\ref{fig:cburn} shows a model run with the options
used throughout the paper (the network \texttt{co\_burn.net} and a choice for
\texttt{set\_rate\_1212} of \texttt{G05}) as the solid line.  The dashed line shows the same model, but with \texttt{set\_rate\_1212} set to \texttt{CF88\_multi\_1212}.  This uses the \citet{1988ADNDT..40..283C} rate, with screening applied via the default \mesa\ treatment. 
The comparison of these two lines illustrates the erroneous ignition shift to higher densities.
For completeness, the dotted line shows the result using a network (\texttt{sagb\_NeNa\_MgAl.net}) 
that explicitly treats $^{23}{\rm Na}$ and thus includes carbon burning via the reactions \texttt{r\_c12\_c12\_to\_he4\_ne20} and \texttt{r\_c12\_c12\_to\_h1\_na23}; in such a case the option \texttt{set\_rate\_1212} is irrelevant and thus this agrees with the CF88 line.

We are confident that this shift does not significantly affect our results or conclusions.
Since the shift is in the direction of higher density, all models that we identify as core ignitions would  remain core ignitions.  Overall, the net effect if we re-ran the model grids might be to shift some of the hybrid ignitions to core ignitions.  However, because of the rapid rise of central density with increasing mass near the Chandrasekhar mass, this corresponds to a shift in the WD mass at ignition of only $\sim 3 \times 10^{-3}\,\msun$. This argues that the binary evolution remains very similar.
Because of the computational cost of our model grids, we opt not to re-run them using a corrected rate. 

\section{Influence of \mesa\ \mltpp} \label{uncertainties resulting from stellar evolution controls}

Convective energy transport can become inefficient in radiation-dominated, near-Eddington stellar envelopes.  When convection fails to make the convective regions nearly adiabatic, this can lead to the formation of a steep entropy gradient near the base of the convection zone.  Especially
when this region is moving Lagrangianly (for example, due to the growth of the core or the shrinking of the envelope due to mass loss), this steep gradient can lead to a strong
timestep constraint.  Section 7 of \cite{2013ApJS..208....4P} describes
a capability in \mesa\ (referred to as \mltpp) that artificially
enhances the energy transport in these regions, thus reducing the superadiabaticity
and alleviating the numerical issues.
Physically, additional energy transport could be due to three-dimensional effects that are not captured in standard
mixing length theory \citep[e.g.,][]{2015ApJ...813...74J}.

We employ \mltpp\ in both the He star and the WD.  In the WD, it is sometimes helpful during early He flashes or when the WD envelope is near its maximal extent during the red giant
accretion regime.  In the He star, 
it is particularly helpful as the systems begin to come out of contact, when the CO core mass is the largest (and the luminosity is highest) and the He envelope is small. 
Since \mltpp\ does change the envelope structure of the WD and He star, it can influence the
rates at which mass is donated and accepted.
Given that \mltpp\ is an {\it ad hoc} prescription, it is important to demonstrate that our results do not significantly dependent on its usage.

\begin{figure*}
\gridline{\fig{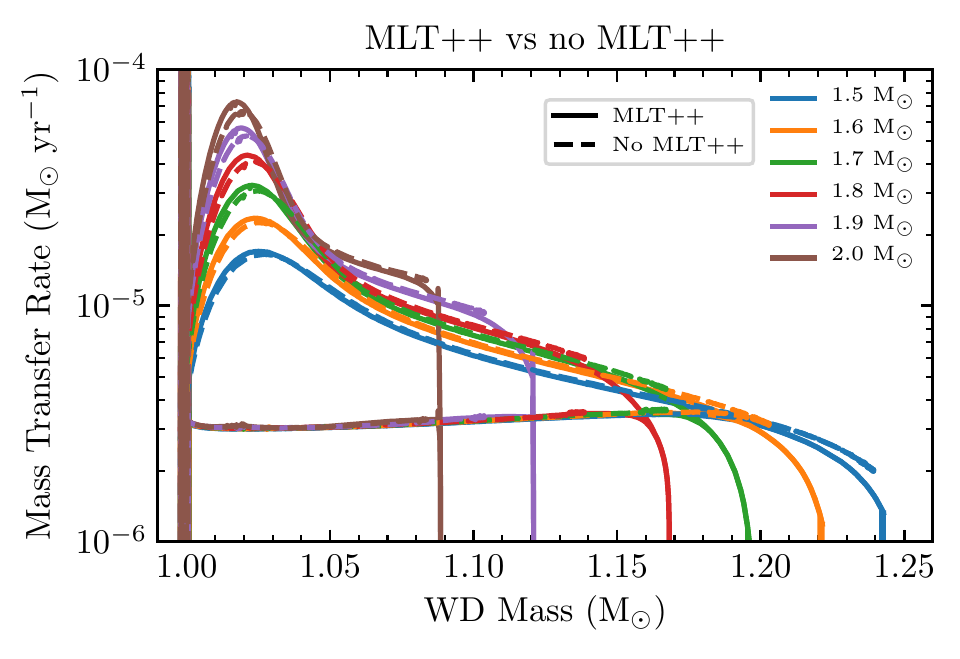}{0.5\textwidth}{}
          }
\caption{Comparison of systems with and without \mltpp.  Models
using \mltpp\ are numerically easier to evolve to a detached
double WD state.  The mass transfer histories show only small ($\la 10\%$)
differences.
\label{fig:mlt}}
\end{figure*}

Figure \ref{fig:mlt} compares two sequences of models with and without {\tt\string MLT++}.
These begin with $\mwd = 1.0\,\msun$ and at $\iniP = -0.1$, with He stars ranging from 1.5 to 2.0 $\msun$.  We found \mltpp\ was particularly needed for these
longer period and higher donor mass systems (that make detached double WD binaries), 
where the use of standard MLT severely limited the timesteps.
It is apparent that the donor comes out of contact more easily when using \mltpp,
as only these models were able to reach a phase of steeply falling $|\mdothe|$
in the allowed runtime.
The figure shows that the difference in $|\mdothe|$ is smaller than 10\%. 
The difference is even smaller for shorter period and lower mass systems
that lie within the TN SN region,
so we conclude that the usage of \mltpp\ has little influence on our overall results.

\section{Convergence Test}

\begin{figure*}
\gridline{\fig{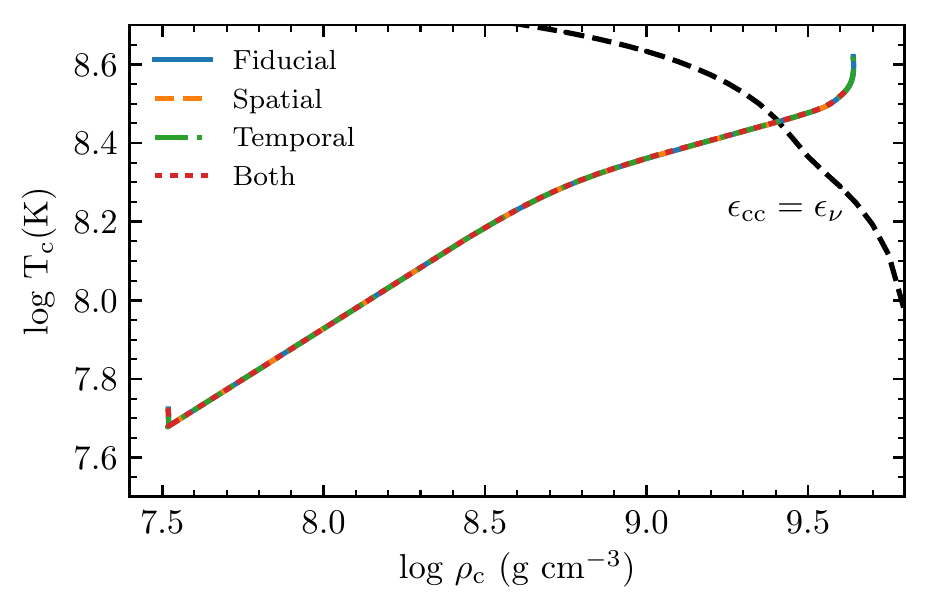}{0.5\textwidth}{(a)}
          \fig{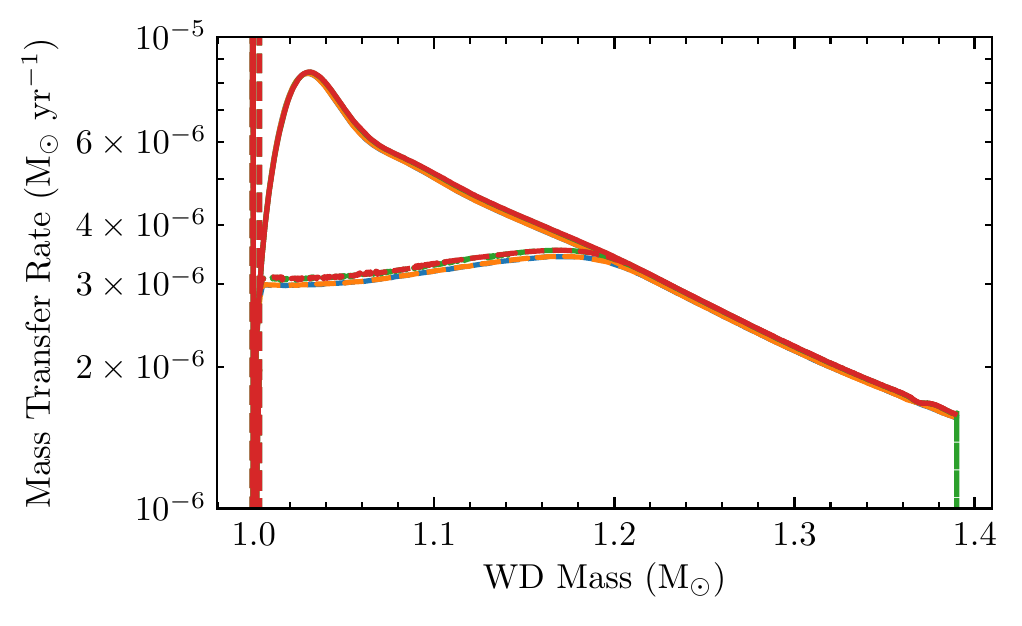}{0.5\textwidth}{(b)}
          }
\caption{Comparison of the fiducial model with $\inipara=(1.6,1.0,-0.9)$ run at different temporal and spatial resolutions.  Panel (a) shows the central evolution of the WD model and panel (b) shows the mass transfer rates.  These key quantities are essentially independent of our resolution choices, indicating these models are numerically converged.
\label{fig:convergence}}
\end{figure*}

We performed 3 tests to confirm that our results are independent of the adopted temporal and spatial resolution. Figure \ref{fig:convergence} shows our fiducial case $\inipara=(1.6,1.0,-0.9)$ along with 3 other runs with higher spatial/temporal resolution. Higher spatial resolution is achieved by increasing the number of zones in the WD via the control {\tt mesh\_delta\_coeff} and higher temporal resolution by limiting the time step based on fractional changes in the He star via {\tt varcontrol\_target}. The values adopted are shown in Table~\ref{tab:convergence}. The ``spatial'' model has almost twice as many zones around the helium-burning shell of the WD ($\approx 800$ zones) as the ``fiducial'' model ($\approx 400$ zones). Figure \ref{fig:convergence} shows the evolution of the WD core in temperature-density space and the mass transfer history in each case. It shows that the models in our fiducial case are indeed converged. 

\begin{deluxetable*}{ccCrlc}[b!]
\tablecaption{Convergence Test\label{tab:mathmode}}
\tablecolumns{6}
\tablenum{1}
\tablewidth{0pt}
\tablehead{
\colhead{Run Name} &
\colhead{${\tt\string varcontrol\_target}$} &
\colhead{${\tt\string mesh\_delta\_coeff}$} &
\colhead{Steps} &
\colhead{Max Zones} \\
 & (He star) & $\text{(WD)}$ & & 
}
\startdata
Fiducial & $1\times10^{-3}$ & 0.4 & 138105 & 4680 \\
Temporal & $4\times10^{-4}$ \tablenotemark{a} & 0.4 & 375680 & 4652 \\
Spatial & $1\times10^{-3}$ & 0.2 & 131320 & 9196 \\
Both & $4\times10^{-4}$ \tablenotemark{a} & 0.3 & 396900 & 6094 \\
\enddata
\tablenotetext{a}{Later lowered to $3\times10^{-4}$.}
\tablecomments{Table showing the stellar controls used in each model for testing spatial and temporal convergence. ${\tt\string varcontrol\_target}$ controls the time step, and ${\tt\string mesh\_delta\_coeff}$ controls the number of zones. 
\label{tab:convergence}}
\end{deluxetable*}

\end{document}